\documentclass{issue-lpr}
\usepackage[english]{babel}
\usepackage{times}
\newcommand{\one}{\mathrm{I} \! \! 1}

\graphicspath{{./}{figs/}}

\setpages{1}
\setvolume[1]{1}

\setyear{2007}%

\setdoi{20070001}%

\begin{document}

%%% Here, the file name of the title figure is to be given in the form
%%%   \titlefigure[width=<length>, ...]{<filename>}
%%% If you don't want a figure here, just omit or comment out this line.
\titlefigure[width=225pt]{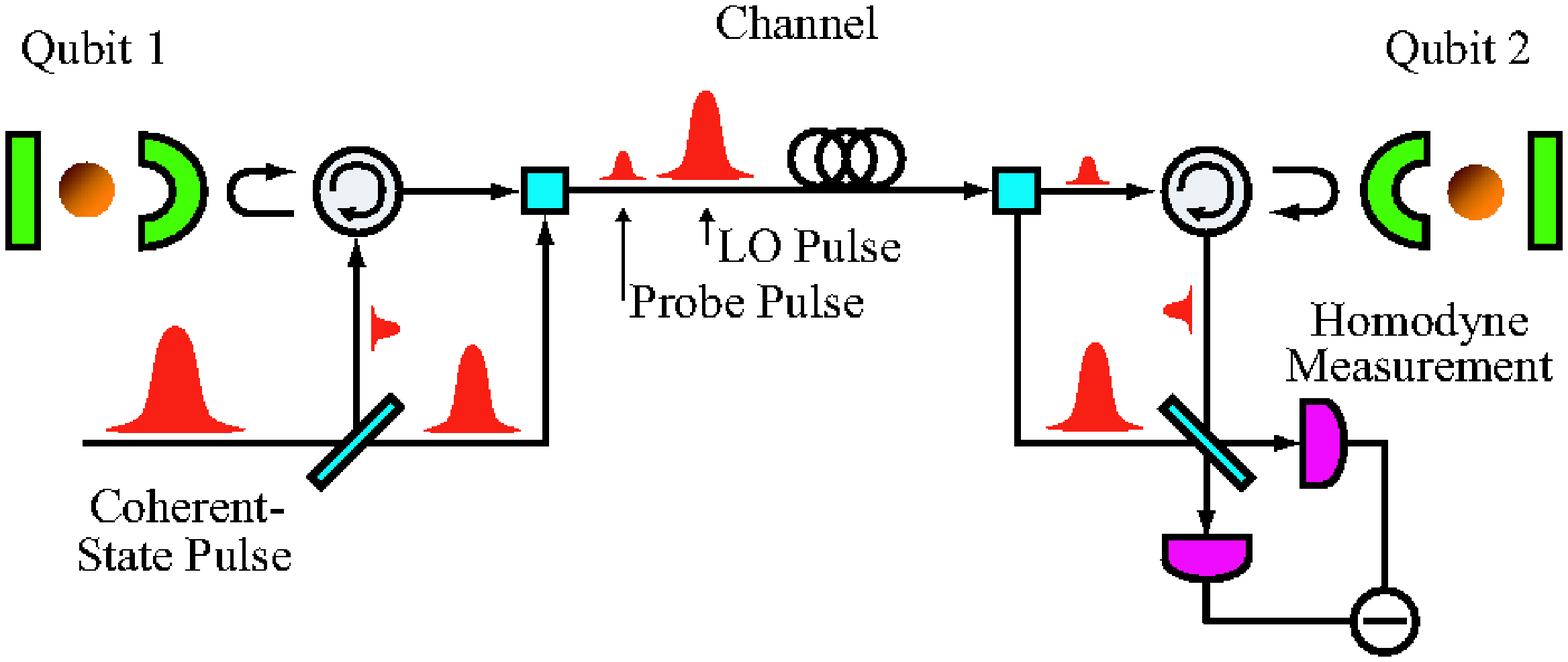}

%%% This is the caption for the title figure. If there is no title
%%% figure (see above) please delete or comment this line, too.
\titlefigurecaption{Schematics for the optical
implementation of entanglement distribution between two stations
in a hybrid quantum repeater.}

%%% Give an abstract text:
\abstract{This article reviews recent hybrid approaches to optical quantum information
processing, in which both discrete and continuous degrees of freedom are exploited.
There are well-known limitations to optical single-photon-based qubit and
multi-photon-based qumode implementations of quantum communication and quantum
computation, when the toolbox is restricted to the most practical set
of linear operations and resources such as linear optics and
Gaussian operations and states.
The recent hybrid approaches aim at pushing the feasibility, the efficiencies,
and the fidelities of the linear schemes to the limits, potentially
adding weak or measurement-induced nonlinearities to the toolbox.}

%%% In the following lines, title and authoring informaton are to be
%%% given.  First the article's main title:
\title{Optical hybrid approaches to quantum information}

%%% Please name all authors and tag them with their respective
%%% institute(s) using the \inst directive.  The corresponding author
%%% gets an additional asterisk.
\author{Peter van Loock
%\inst{1,*}
}

%%% Here the institutes are to be given.  If there is more than one,
%%% the entries have to be separated by \and tags.
\institute{%
  Optical Quantum Information Theory Group,
Max Planck Institute for the Science of Light, \\
Institute of Theoretical Physics I, Universit\"{a}t Erlangen-N\"{u}rnberg,
Staudtstr.7/B2, 91058 Erlangen, Germany
}

%%% Give the email address of the corresponding author:
\mail{e-mail: peter.vanloock@mpl.mpg.de}

%%% If the article has a very long title or lots of authors, then
%%% please give shorter versions of that for the column title:

%\titlerunning{Shorter title}
%\authorrunning{F. Author and S. Author}

%%% If there are more than two authors, please indicate that in the
%%% form ``F. Author et al.''

%%% Give some keywords and PACS code(s) for the article, if you have:

%%% The following will be completed by the publisher:
\received{\ldots}
\published{\ldots}
\maketitle
\markboth{thedoi}{thedoi}
\tableofcontents

%%% Here, the main text starts.

\section{Introduction}

Quantum information is a relatively young area of interdisciplinary research.
One of its main goals is, from a more conceptual point of view,
to combine the principles of quantum physics with those of
information theory. Information is physical is one of the key messages,
and, on a deeper level, it is quantum physical.
Apart from its conceptual importance, however, quantum information
may also lead to real-world applications for communication (quantum communication)
and computation (quantum computation) by exploiting quantum properties
such as the superposition principle and entanglement.
In recent years, especially entanglement turned out to play the most prominent role,
representing a universal resource for both quantum computation and quantum
communication. More precisely, multi\-partite entangled, so-called cluster states
are a sufficient resource for universal, measurement-based quantum computation
\cite{RaussendorfBriegel01}. Further, the sequential distribution of many copies
of entangled states in a quantum repeater allow for extending quantum communication
to large distances, even when the physical quantum channel is imperfect
such as a lossy, optical fiber \cite{Briegel98,Duer99}.

Many if not most experiments related to quantum information are conducted
with quantum optical systems. This includes the preparation, manipulation,
and measurement of interesting and useful quantum optical states, in particular,
entangled states; possibly supplemented by additional atomic systems
for storing and processing quantum states.

Why is quantum optics the preferred
field for quantum information demonstrations? Based on
mature techniques from nonlinear optics for state preparation such as
parametric down conversion, together with the most accessible means for
manipulating optical states with linear elements such as beam splitters,
there is a long list of optical proof-of-principle demonstrations of
various quantum information processing tasks. Some of these experiments
are performed with single-photon states, leading to a {\it discrete-variable}
(DV) encoding of quantum information, where, for instance, a qubit space
is spanned by two orthogonal polarizations (`photonic qubits')
\cite{Bouwmeester97}.
In other experiments,
{\it continuous-variable} (CV) states, defined in an infinite-dimensional
Hilbert space, are utilized, for example, expressed in terms of the
quadrature amplitudes of an optical, bosonic mode (`photonic qumodes')
\cite{Furusawa98}.
Typically, the DV experiments involve some heralding mechanism, rendering them
conditional, and hence less efficient; nonetheless, fidelities in the
DV schemes are fairly high \cite{Kok07}. Conversely, in the CV regime,
unconditional operations and high efficiencies are at the expense of lower
fidelities \cite{BraunsteinPvL05}.

Beyond experimental small-scale demonstrations, towards potential
real-world applications, light is clearly the optimal choice for
communication. Moreover, in order to mediate interactions between
distant matter qubits on a fixed array (for instance, in a solid-state
system), individual photons or intense light pulses may be utilized.
In such a scenario, the light field acts as a kind of quantum bus
(`qubus')
for applying entangling gates to the matter qubits.
This kind of optical, qubus-mediated quantum logic could become part
of a full quantum computer. In addition, it is exactly this
qubus approach which can be exploited for quantum communication
in a quantum repeater, where the optical qubus propagates through a fiber
(or even through free space) between neighboring repeater stations and
locally interacts with each matter qubit for nonlocal,
entan\-gled-state preparation of the two distant qubits.
In analogy to classical, optical/electronic hybrid computers,
the above schemes may be referred to as {\it `hybrid'} approaches, as they combine
the useful features of both light and matter; the former as an ideal medium
for communicating, the latter well suited for storing quantum information.

In addition to the hybrid notion mentioned in the preceding paragraph,
there is a related, but somewhat different definition of (optical) hybrid
quantum information protocols.
These are inspired by practical as well as fundamental limitations
of those optical quantum information schemes, which are solely based upon either
discrete or continuous degrees of freedom. A hybrid scheme, similar
to a classical, digital/analog hybrid computer, would then exploit
at the same time both DV and CV states, encodings, gates, measurements,
and techniques, in order to circumvent those limitations.

\subsection{CV versus DV}

It is well known that a very strong version of
universal quantum computation
(`CV universality'), namely the ability to simulate any
Hamiltonian, expressed as an arbitrary polynomial of the bosonic mode
operators, to arbitrary precision, is not achievable with only
linear transformations, i.e., Gaussian transformations \cite{Lloyd99}.
Gaussian transformations
are rotations and translations in phase space, as well as
beam splitting and squeezing unitaries, all transforming
Gaussian states into Gaussian states. Similarly, a fully Gaussian
qumode quantum computer can always be efficiently simulated by a classical computer
\cite{Bartlett02}. A single non-Gaussian element such as a cubic
Hamiltonian would be sufficient to both achieve universality
and prevent classical simulability.
In general, however, non-Gaussian transformations
are difficult to realize on optical Gaussian sta\-tes; nonetheless
deterministic protocols for approximating such gates exist \cite{GKP01}
(the `GKP' scheme).

Although the physical states in quantum optical approa\-ches,
representing quantized harmonic oscillators,
would always live in infinite-dimensional
Fock space, there is a weaker, but possibly more useful notion of universality
(`DV universality'). It refers to the
ability to approximately simulate any DV multi-qubit unitary with a finite
set of gates, logically acting on a finite subspace of the infinite-dimensional
optical Fock space, spanned by states with only a few photons (photonic qubits).
In this case, a universal set must also contain a
nonlinear interaction, unless we accept probabilistic operations \cite{KLM01}
(the `KLM' scheme).
Similarly, exactly fulfilling finite tasks
supposedly simpler than universal quantum computation,
such as a complete, photonic two-qubit Bell measurement, would be impossible with only
linear elements (including squeezers)
\cite{Luetkenhaus99,PvLLutkenhaus04,Raynal04}.
The problem is that nonlinear interactions on the level
of single photons are hard to obtain. It is very difficult
to make two photons ``talk'' to each other.

\subsection{Going hybrid}

In a hybrid scheme, where DV and CV degrees of freedom are exploited
at the same time and the goal is to circumvent the limitations
of the linear approaches, it can be useful to consider two special notions
of nonlinearity: one is that of weak nonlinearities, the other one
is that of measurement-induced nonlinearities. The former one would then
be effectively enhanced through the use of sufficiently intense light fields;
an approach first utilized for quantum nondemolition measurements
with CV states \cite{Grangier98}.
More recently, weak nonlinearities were applied to
combined DV-CV, i.e., hybrid systems, where the weak nonlinear interaction
is not only enhanced, but also mediated between the DV components
through a bright CV qubus state.
This enables one to perform various tasks from projecting onto the
complete, photonic DV Bell basis
to implementing universal, photonic two-qubit entangling gates, using either
DV (threshold) photon detectors \cite{Paris00}, CV homodyne detectors
\cite{Nemoto04,Barrett04,Munro05_2,Munro05}, or no detectors at all
\cite{Spiller06,vanloock07}.

The concept of a measurement-induced nonlinearity was initiated
by the seminal works of KLM \cite{KLM01} and GKP \cite{GKP01}.
The KLM protocol is a fully DV scheme, using complicated, entangled,
multi-photon ancilla states, measurements of photon number,
and feedforward, whereas the GKP scheme may be considered
one of the first hybrid protocols. It relies upon
DV photon number measurements, to create non-Gaussian states
from CV Gaussian resources, and to eventually realize non-Gaussian
(cubic) CV gates. The GKP proposal also contains the hybrid
concept of encoding logical DV states into physical CV states.

At this point, before going into more detail in the following sections,
let us summarize the key elements of the optical hybrid approaches
to quantum information processing reviewed in this article.
The goal of circumventing the limitations of the most practical, linear optical
schemes, maintaining to some extent their feasibility and efficiency,
and still achieving a true quantum advantage over classical schemes
may be reached through all or some of the following ingredients:

\begin{itemize}
     \item hybrid states and operations, i.e., a combination of DV and CV elements,
     \item qubus systems for mediating entangling gates,
     \item weak nonlinearities,
     \item measurement-induced nonlinearities.
\end{itemize}

The paper is organized as follows.
In Section 2, we give a brief introduction to optical
quantum information. This includes a description
of how to encode quantum information into
photonic qubits and qumodes, how to process such quantum information
using linear and nonlinear transformations, and how these
tools may be exploited to achieve (theoretically and experimentally)
efficient (scalable?) quantum computation and communication.
Section 3 then presents the concept of optical hybrid protocols,
discussing various hybrid schemes for both quantum computation
and communication.
Finally, we summarize and conclude in Section 4.

\section{Optical quantum information}

How can we encode quantum information, for instance,
a qubit, into optical states? Do quantum optical states
naturally allow for different types of encoding?
These questions, together with the issue of
processing quantum information encoded in optical states
in an efficient way, will be addressed in this section.
In particular, we discuss that there are qualitatively
different levels of `efficient' optical quantum information
processing. These depend on the type of encoding,
and on the scalability and the feasibility
of the optical resources necessary for their implementation.

\subsection{Linear versus nonlinear operations}

In Fig.~2.1.1,
%~\ref{fig2}
a table is shown summarizing possible
optical interactions and transformations for state preparation
and manipulation.

The most accessible and practical interactions are those described
by linear and quadratic Hamiltonians. Here, qua\-dratic refers to the order
of a polynomial of the mode operators $\hat a_k$ for the modes
that participate in the interaction.
Each $\hat a_k$ ($\hat a_k^\dagger$) corresponds to the annihilation/lowering
(creation/raising) operator of a quantized harmonic oscillator, each representing
a single mode from a discrete set of (spatial, frequency, polarization, etc.) modes
into which the electromagnetic field is most
conveniently expanded \cite{MilburnWalls}.

\begin{center}
  \begin{columnfigure}
    \centering%
    \includegraphics[width=225pt]{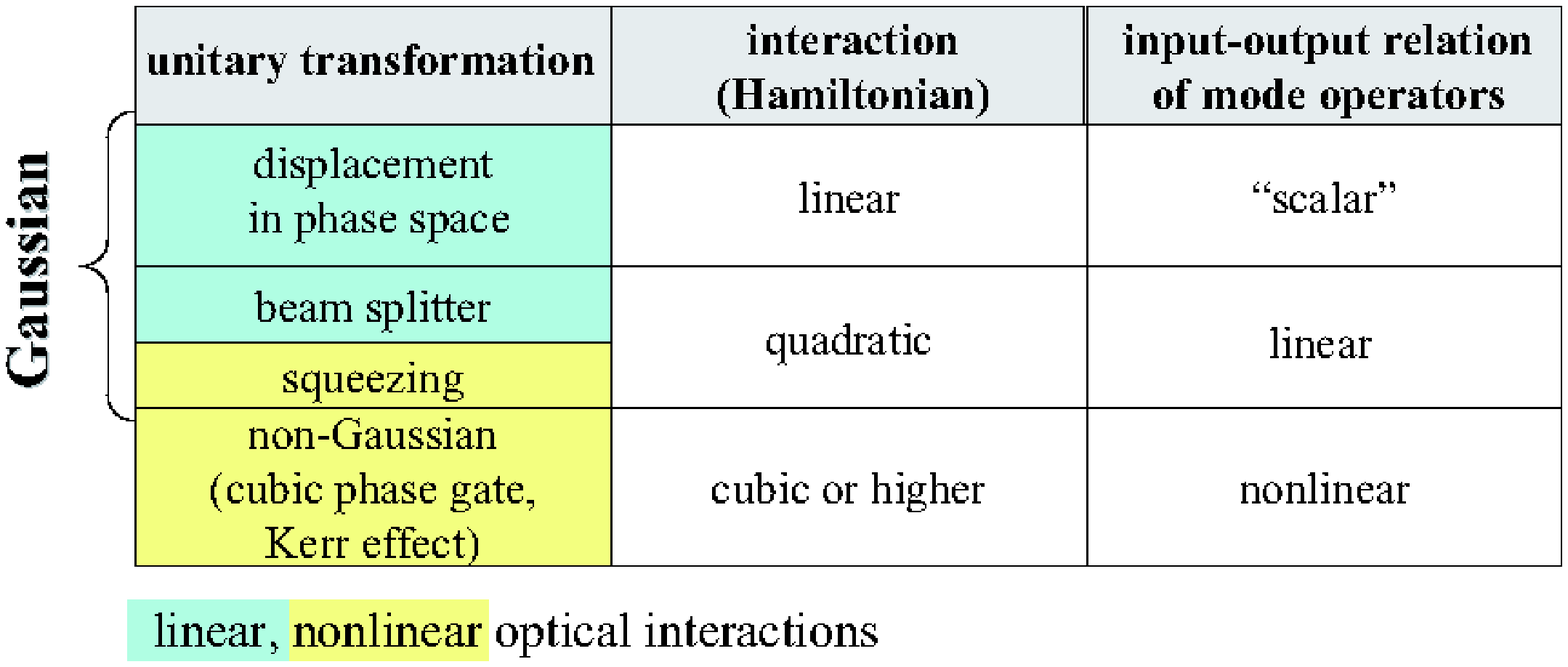}
  \figcaption{{\bf 2.1.1} Optical interactions and transformations in terms of the
  annihilation and creation operators representing a discrete set of modes
  of the optical field.} \label{fig2}
  \end{columnfigure}
\end{center}

The unitary transformations generated from quadratic Hamiltonians
are linear. An arbitrary quadratic Hamiltonian transforms
the mode operators as
\begin{equation}\label{LUBO}
\hat{a}_k'=\sum_{l} A_{kl} \hat{a}_l + B_{kl} \hat{a}_l^{\dagger}
+\gamma_l\;.
\end{equation}
Here, the matrices $A$ and $B$ satisfy the conditions
$AB^T=(AB^T)^T$ and $AA^{\dagger}=BB^{\dagger}+
\mbox{1$\!\!${\large 1}}$ according
to the bosonic commutation relations $[\hat{a}_k',\hat{a}_l'^{\dagger}]=
[\hat{a}_k,\hat{a}_l^\dagger]=\delta_{kl}$. This transformation is also referred to
as linear unitary Bogoliubov transformation (LUBO) \cite{LUBO}.
It combines passive and
active elements, i.e., beam splitters and squeezers, respectively;
the $\gamma_l$'s in Eq.~(\ref{LUBO}) describe phase-space displacements
and come from the linear terms in the Hamiltonian.
As an example, in a single-mode squeezer,
$\hat a'=
\hat a \cosh r - \hat a^{\dagger} \sinh r $,
the $x$-quadrature would be `squeezed', $\hat x' = e^{-r}\hat x$,
and the $p$-quadrature
correspondingly `antisqueezed', $\hat p' = e^{+r}\hat p$,
with
$\hat a = \hat x + i \hat p$ and $\hat a' = \hat x' + i \hat p'$.
The quadratures here are dimensionless variables playing the roles
of position and momentum with $[\hat x, \hat p] = i/2$.

Comparing the LUBO to a purely passive (photon number preserving),
linear transformation,
\begin{equation}\label{passive}
\hat{a}_k'=\sum_{l} U_{kl} \hat{a}_l\;,
\end{equation}
with an arbitrary unitary matrix $U$, we observe that
there is no mixing between the annihilation and creation
operators in the passive transformation.
Despite this difference, also the active, more general
LUBO is only linear in the mode operators.
Therefore, general linear optical transformations are here referred
to as LUBOs, including squeezers. As squeezing, however, typically
involves a nonlinear optical interaction (such as $\chi^{(2)}$)
\footnote{for which the actual, fully quantum mechanical Hamiltonian
is cubic; with the usual parametric approximation, considering
the so-called pump field classical, the pump mode operator
becomes a complex number which is then absorbed into the squeezing parameter
of the resulting quadratic Hamiltonian.},
it may as well be excluded from the `linear-optics' toolbox
(see
%Fig.~\ref{fig2}
Fig.~2.1.1).

Initially, squeezing was not really considered a useful tool
for DV quantum information processing. Moreover, it is hard to apply
squeezing to an arbitrary state such as a photonic qubit (see next section)
`online' in a controlled and efficient way.
Hence, usually, squeezing will be explicitly
excluded from the linear-optics toolbox for DV quantum information
\cite{Kok07,KLM01}. In the CV approaches, however, optical squeezed states
are the essential resource for creating Gaussian CV entangled states
\cite{BraunsteinPvL05}. In this case, squeezed states are first created
`offline' and then linearly transformed, according to the passive transformation
in Eq.~(\ref{passive}). In more recent experiments, it was demonstra\-ted
that even online squeezing may be shifted offline using squeezed ancilla states
\cite{Yoshikawa07.pra,Yoshikawa08.prl,Filip05.pra}. Only these very new
approaches would allow for efficient online squeezing of photonic DV states,
as potentially needed in hybrid schemes.

In the hybrid context, and in the view of the recent offline-squeezing
experiments \cite{Yoshikawa07.pra,Yoshikawa08.prl,Filip05.pra},
it is sensible to define the complete set of {\it linear resources and operations}
as all offline prepared, optical Gaussian states and all general LUBOs
which are equivalent to Gaussian transformations
mapping a Gaussian state back to a Gaussian state, see Fig.~2.1.1.
%Fig.~\ref{fig2}.
Particularly practical resources here are coherent states, as these
are readily and directly available from a laser source. To these coherent-state
sources, we then add deterministic, online squeezing, and as a Gaussian,
linear measurement, homodyne detection.

The link between the elementary quantum optical
devices such as phase shifters, beam splitters, and single-mode squeezers
on one side and an arbitrary LUBO as in Eq.~(\ref{LUBO}) on the other side
is provided through two important results:

\begin{itemize}
     \item any active, multi-mode LUBO as in Eq.~(\ref{LUBO})
           can be decomposed into
           a three-step circuit consisting of a passive, linear optical
           multi-mode transformation, sin\-gle-mode squeezers, and another
           passive, linear optical
           multi-mode transformation \cite{Braunstein2005}
     \item any passive, linear optical multi-mode transformation
           described by an arbitrary unitary matrix
           as in Eq.~(\ref{passive}) can be realized through a sequence of two-mode
           beam splitters and single-mode phase shifters \cite{Reck}
\end{itemize}

The former result, the so-called `Bloch-Messiah reduction', can be derived
through singular value decomposition, with  $A=UA_DV^{\dagger}$ and
$B=UB_DV^T$, a pair of unitary matrices $U$ and $V$,
and non-negative diagonal matrices
$A_D$ and $B_D$, $A_D^2=B_D^2+\mbox{1$\!\!${\large 1}}$
\cite{Braunstein2005}.
The two results together imply that any multi-mode LUBO, i.e., any linear
multi-mode transformation as in Eq.~(\ref{LUBO}),
can be implemented with single-mode phase shifters, single-mode
squeezers, and two-mode beam splitters. The displacements in Eq.~(\ref{LUBO})
(the $\gamma_l$'s) can be also realized using highly reflective beam splitters.

As shown in Fig.~2.1.1,
%Fig.~\ref{fig2},
going beyond the regime of linear resources and operations means
to include cubic or higher-order interactions leading to nonlinear
transformations. Such a nonlinear interaction
would normally map a Gaussian state onto a non-Gaussian state,
described by non-Gaussian Q and Wigner functions, see Fig.~2.1.2.
%Fig.~\ref{fig3}.
The\-se interactions are typically very weak; an example would be the
extremely weak Kerr effect in an optical fiber. Therefore,
for sufficiently long interaction times,
unwanted photon losses will normally dominate over the desired
nonlinear transformation.

Rather than performing nonlinear transformations online,
we may first create offline nonlinear resources \cite{GKP01,KLM01}
such as photon number (Fock) states as well
as other non-Gaussian states such as `cat states'
(i.e., superposition states of coherent states, see
%Fig.~\ref{fig3}
Fig.~2.1.2 on the right).
Typically, this offline preparation would be probabilistic, i.e.,
conditional, depending
on, for instance, certain photon number measurement outcomes for
a subset of modes \cite{Ourjoum06,Neergaard06,Takahashi09}.

\begin{center}
  \begin{columnfigure}
    \centering%
    \includegraphics[width=200pt]{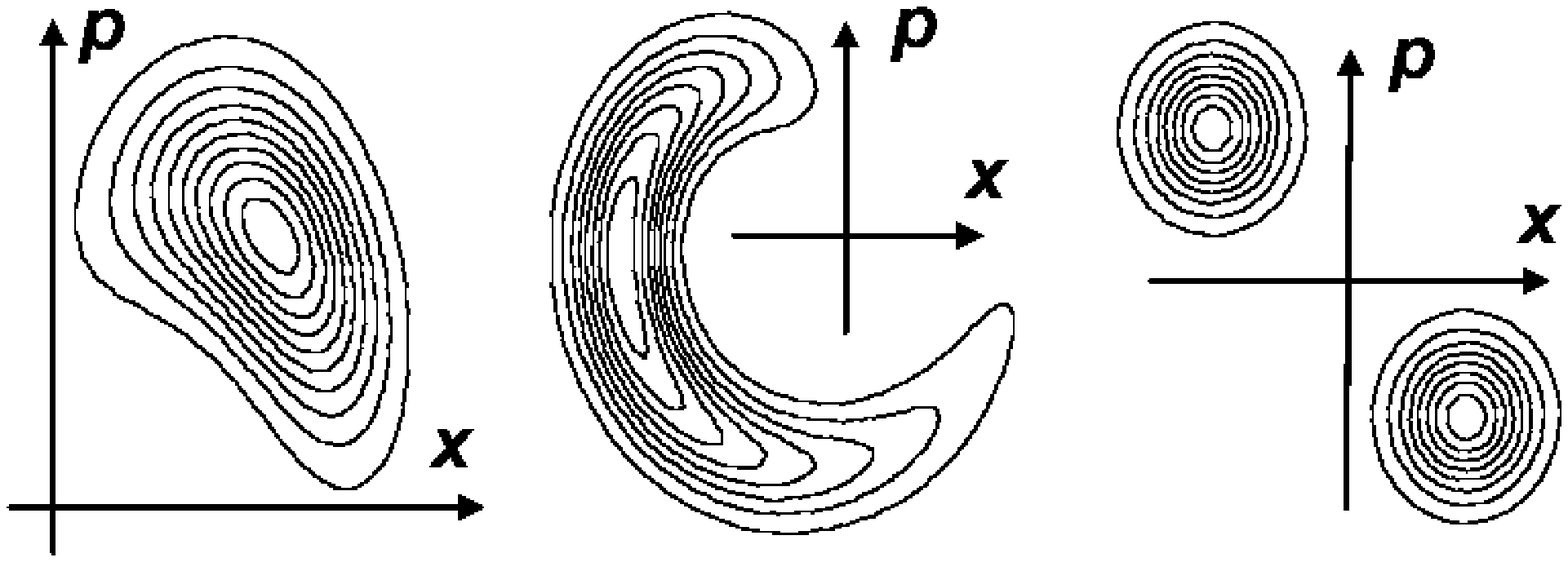}
  \figcaption{{\bf 2.1.2} Examples of single-mode states generated through
  highly nonlinear interactions. Shown is the Q function $Q(x,p)$ for a Gaussian coherent state
  evolving into various non-Gaussian states subject to a quartic self-Kerr
  interaction with $\hat H\propto \hat a^\dagger \hat a (\hat a^\dagger \hat a - 1)$.}
  \label{fig3}
  \end{columnfigure}
\end{center}

Let us now explicitly consider the encoding and processing
of quantum information using optical resources and linear/nonlinear optical
transformations.

\subsection{Qubits versus qumodes}

The information being processed through a quantum computer
is most commonly represented by a set of
two-level systems (qubits), in analogy to classical, digital
bit encoding. These qubit states could be superpositions
of two different, electronic spin projections, or, in the
optical context, superpositions of two orthogonal
polarizations.
As the bosonic Fock space is infinite-dimensional,
however, there is much more room for encoding quantum information
into optical states. Allowing for states with more than a single
photon per mode is a possible way to represent multi-level systems.
Alternatively, a multi-level state can be expressed through
sufficiently many modes with at most one photon in total.

Apart from discrete photon numbers, an optical state may be
described by its amplitude and phase. The corresponding quantum
phase-space variables could be considered the quantum analogues
of classical, analog encoding. Such phase-space representations
would completely determine the state of a quantized optical mode
(qumode).

\subsubsection{Photonic qubits}

Consider the free electromagnetic field with a Hamiltonian
$\hat H = \sum_k \hbar \omega_k (\hat a_k^\dagger \hat a_k + 1/2)$
with photon number operator $\hat a_k^\dagger \hat a_k$ for mode $k$.
Note that the sum includes the zero-point energy `1/2' for every mode.
Then the number (Fock) states $|n_k\rangle$,
eigenstates of $\hat a_k^\dagger \hat a_k$,
form a complete, orthogonal basis for each mode.
Dropping the mode index, we have the well-known
relations for annihilating and creating photons,
$\hat a |n\rangle = \sqrt{n} |n-1\rangle$ and
$\hat a^\dagger |n\rangle = \sqrt{n+1} |n+1\rangle$, respectively.
The vacuum state, containing no photons, is defined as
$\hat a |0\rangle = 0$.

Using this number basis, there are now (at least)
two different ways to encode
an optical qubit. The first encoding is called
`single-rail' (or `occupation number'), as it relies upon just a single
optical mode,
\begin{equation}\label{singlerail}
\cos(\theta/2) |0\rangle + e^{i\phi} \sin(\theta/2) |1\rangle\;.
\end{equation}
This encoding, however, is rather inconvenient,
because even simple single-qubit rotations would require
nonlinear interactions. For example, the Hadamard gate,
acting as $|k\rangle \rightarrow (|0\rangle + (-1)^k |1\rangle )/\sqrt{2}$,
transforms a Gaussian state (the vacuum) into a non-Gaussian state
(a superposition of vacuum and one-photon Fock state).

In contrast, for the so-called `dual-rail' encoding,
\begin{equation}\label{dualrail}
\cos(\theta/2) |10\rangle + e^{i\phi} \sin(\theta/2) |01\rangle\;,
\end{equation}
single-qubit rotations become an easy task (see
%Fig.~\ref{fig4}
Fig.~2.2.1).
A 50:50 beam splitter, for instance, would turn
$|10\rangle = \hat a_1^\dagger |00\rangle$ into
$(1/\sqrt{2}) (\hat a_1^\dagger + \hat a_2^\dagger) |00\rangle
= (1/\sqrt{2}) (|10\rangle + |01\rangle)$, and similarly
for the other basis state. The linear transformation here is
a simple, special case of the general passive transformation
in Eq.~(\ref{passive}), and the two modes are spatial.

\begin{center}
  \begin{columnfigure}
    \centering%
    \includegraphics[width=80pt]{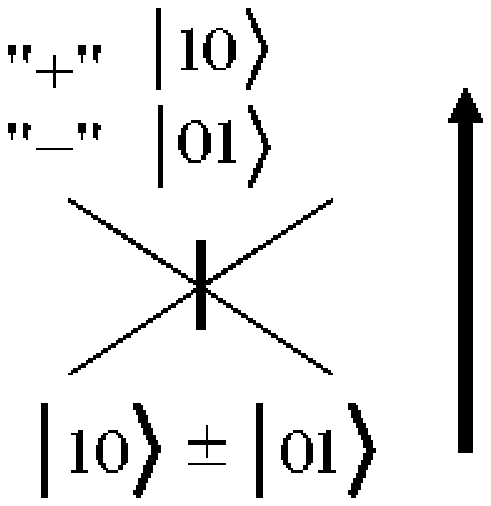}
  \figcaption{{\bf 2.2.1} Using a beam splitter to switch between the computational
  and the Hadamard transformed, conjugate basis. Here, measuring the photons at
  the output of the beam splitter would project the input state
  onto the conjugate basis.}\label{fig4}
  \end{columnfigure}
\end{center}

The most common dual-rail encoded, photonic qubit is a polarization
encoded qubit,
\begin{equation}\label{dualrailpolarization}
\cos(\theta/2) |H\rangle + e^{i\phi} \sin(\theta/2) |V\rangle\;,
\end{equation}
for two polarization modes, where one is horizontally, the other one vertically polarized.
So polarization encoding is by no means different from dual-rail encoding;
it is rather a specific manifestation of dual-rail encoding.
Single-qubit rotations are then particularly simple, corresponding to polarization
rotations.

The drawback of the dual-rail encoding is that for realizing two-qubit
entangling gates, it is necessary to make two photons (each representing
a dual-rail qubit) `talk' to each other. This kind of interaction
between two photons would require some form of nonlinearity.
Later we will discuss various possibilities for such two-photon entangling
gates. We will also discuss an extension of dual-rail to multiple-rail
encoding, where every logical basis state is represented by a single photon
that can occupy any one of sufficiently many, different modes (not just two as for
dual-rail encoding).

Towards photonic qubit processing,
the most accessible, optical resources and their optical manipulation
can be summarized as follows.

{\it Resources:} single-photon states
(i.e., conditionally pre\-pared Fock states, superpositions of Fock states),
produ\-cible through $\chi^{(2)}$ nonlinear interactions
(i.e., parametric down conversion); polarization-encoded states;
single-pho\-ton states
approximated by weak coherent states.

{\it Processing:} passive linear optics (beam splitters, phase shifters,
polarization rotators).

{\it Measurements:} photon counting; on/off detectors.

However, creating a Fock state with many photons is hard, as well as
counting large photon numbers. For this purpose,
the on/off detector is more realistic, as it does not discriminate between
different photon numbers, but only between the vacuum state (`no click')
and the non-vacuum state (`click').

For efficiently generating and measuring
states with many photons, it is more practical to enter the regime
of Gaussian states with CV homodyne detections (see next section).
We may further add squeezing to the above toolbox. The truly nonlinear
regime for processing is attainable by including, for instance,
measurement-induced nonlinearities plus feedforward operations \cite{KLM01}.

Despite the difficulty for realizing a two-photon entangling gate,
there is a clear advantage of the single-photon encoding.
Single photons are fairly robust against noise.
Therefore, typically, processing single-photon states
can be achieved with high fidelity, though, in most cases,
only conditional operations are possible,
at potentially very low success probabilities.
As extra resources for processing DV quantum information,
the atomic counterpart of the photonic polarization (spin) states
are the electronic spin states whose qubit representations we shall
introduce and utilize later.

We will now introduce a kind of complementary way to encode
quantum information, in terms of qumodes. This type of encoding
leads to states which are rather sensitive to noise,
but which can be processed in an unconditional fashion;
even entangling gates can be achieved through deterministic,
linear optics.

\subsubsection{Photonic qumodes}

Consider again the free electromagnetic field with a Hamiltonian
$\hat H = \sum_k \hbar \omega_k (\hat a_k^\dagger \hat a_k + 1/2)$
with photon number operator $\hat a_k^\dagger \hat a_k$ for mode $k$.
This time we shall rewrite the Hamiltonian as
$\hat H = \sum_k (\hat p_k^2 + \omega_k^2 \hat x_k^2)/2$,
with the position and momentum operators
$\hat x_k = \sqrt{\hbar/2\omega_k} (\hat a_k + \hat a_k^\dagger)$
and $\hat p_k = -i \sqrt{\hbar \omega_k/2} (\hat a_k - \hat a_k^\dagger)$
for each oscillator (mode) $k$. The zero-point energy `1/2' is still present
and becomes manifest in the vacuum fluctuations of the position and
momentum of every mode.

Further, we have the commutators $[\hat a_k, \hat a_l^\dagger]
= \delta_{kl}$ and $[\hat x_k, \hat p_l] = i \hbar \delta_{kl}$.
After rescaling $\hat x_k$ and $\hat p_k$ into dimensionless variables,
we arrive at $\hat a_k = \hat x_k + i \hat p_k$ with
$[\hat x_k, \hat p_l] = i \delta_{kl}/2$,
corresponding to $\hbar = 1/2$, as before.
The position and momentum (quadrature) eigenstates may serve as a CV basis
to represent the infinite-dimensional state of an optical qumode.

The vacuum state can now be written, for example, in the position basis,
$|0\rangle = \int dx\, \psi_0(x) |x\rangle_x$, with the wave function
$\psi_0(x) = (2/\pi)^{1/4} \exp(-x^2)$.
It must not be confused with the (unphysical, unnormalized)
zero-position eigenstate $|x=0\rangle_x$.
The position probability distribution of the vacuum state is
a normalized Gaussian,
$|\psi_0(x)|^2 = \sqrt{2/\pi} \exp(-2 x^2)$.
The first and second moments of the vacuum state
are easily calculated as (dropping the mode index)
$\langle 0|\hat x|0\rangle = 0$ and
$\langle 0|\hat x^2|0\rangle = 1/4$, and similarly for the momentum.
Thus, the quadrature vacuum variances are $1/4$.
The Wigner function of the vacuum state is
$W(x,p) = (2/\pi) \exp[-2(x-x_0)^2-2(p-p_0)^2]$,
with $x_0=p_0=0$. For finite $x_0$ and $p_0$, with $\alpha = x_0 + i p_0$,
we obtain the Wigner function for a displaced vacuum, i.e.,
a coherent state $|\alpha\rangle$.

\begin{center}
\begin{columnfigure}
    \centering%
    \includegraphics[width=80pt]{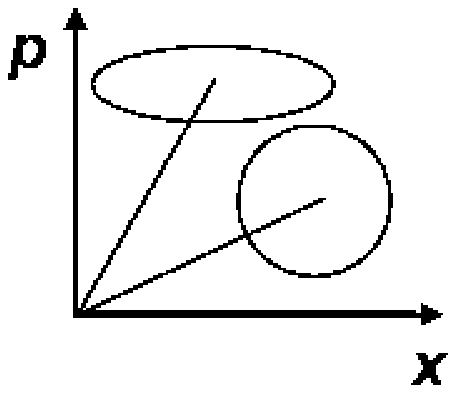}
  \figcaption{{\bf 2.2.2} Encoding of qumodes into CV Gaussian squeezed and coherent states.
  Largely squeezed states would approximately resemble position or momentum
  eigenstates.}
  \label{fig4a}
  \end{columnfigure}
\end{center}

The coherent state is also the eigenstate of the non-Hermitian
annihilation operator, $\hat a |\alpha\rangle = \alpha |\alpha\rangle$,
with, correspondingly, complex eigenvalues $\alpha$.
Its mean photon number is therefore
$\langle\alpha|\hat a^\dagger\hat a|\alpha\rangle = |\alpha|^2$.
As a displaced vacuum, it may be written as
$\hat D(\alpha)|0\rangle = |\alpha\rangle$, with the well-known
displacement operator \footnote{whenever
there is no ambiguity we may drop operator hats.}
\begin{equation}\label{displacement}
\hat D(\alpha) = \exp(\alpha\hat a^\dagger
- \alpha^*\hat a)\;.
\end{equation}
Further, with regard to photon number,
the coherent state obeys Poissonian statistics, and can be expanded as
\begin{equation}\label{coherentstate}
|\alpha\rangle = \exp(-|\alpha|^2/2)
\sum_{n=0}^\infty \frac{\alpha^n}{\sqrt{n!}} \,|n\rangle\;.
\end{equation}
As opposed to the linear displacement operator, the squeezing operator,
$\hat S(\zeta) = \exp[(\zeta^*\hat a^2 - \zeta \hat a^{\dagger 2})/2]$,
with $\zeta = r\exp(i\phi)$,
is quadratic in $\hat a$ and $\hat a^\dagger$. Together with the qua\-dratic
single-mode phase rotations (by an angle $\theta$),
\begin{equation}\label{rotation}
\hat R(\theta) = \exp(-i\theta\hat a^\dagger \hat a) \;,
\end{equation}
and the quadratic two-mode
beam splitter interactions, this completes the linear regime
of Gaussian transformations. Note that
\begin{equation}\label{coherentstate2}
\hat R(\theta)|\alpha\rangle =
|\alpha \exp(-i\theta)\rangle \;,
\end{equation}
as can be easily understood from Eq.~(\ref{coherentstate}).

\begin{center}
  \begin{columnfigure}
    \centering%
    \includegraphics[width=150pt]{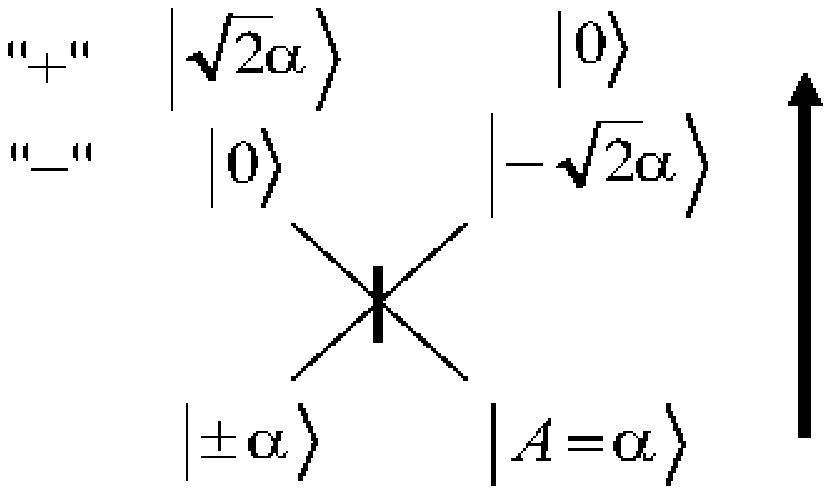}
   \figcaption{{\bf 2.2.3} Optimal unambiguous state discrimination of equally likely,
  binary coherent states
  $\{|\pm \alpha\rangle\}$ using a beam splitter, an ancilla coherent state,
  and on/off detectors. The inconclusive event here corresponds to the detection of
  the two-mode vacuum $e^{-\alpha^2}|00\rangle$ for the two output modes
  with probability $e^{-2\alpha^2}$.}
  \label{fig7}
  \end{columnfigure}
\end{center}

In the linear, CV Gaussian regime, the optical encoding into qumodes
(see
%Fig.~\ref{fig4a}
Fig.~2.2.2)
is achieved either through approximate $x/p$-eigenstates
(largely squeezed states), for which projection measurements
are well approximated by homodyne detections;
alternatively, the overcomplete and nonorthogonal set of coherent states
may serve as a basis for qumodes. Perfectly projecting onto this basis
is only possible for sufficiently large amplitudes $|\alpha|$, for which
the coherent states become near-orthogonal. Nonetheless, two coherent states
can also be unambiguously discriminated in the regime
of small amplitudes using a beam splitter, an ancilla coherent state,
and on/off detectors
(see
%Fig.~\ref{fig7}
Fig.~2.2.3, $\alpha$ real). This unambiguous state
discrimination (USD) is probabilistic, but error-free.

Remarkably, the linear optical
scheme for the USD of two arbitrary coherent states \cite{Banaszek99}
such as $\{|\pm \alpha\rangle\}$
achieves the quantum mechanically optimal USD
for two pure non\-orthogonal states $\{|\psi_1\rangle,|\psi_2\rangle\}$,
where the success probability for a conclusive result equals
$1 - |\langle \psi_1|\psi_2\rangle |= 1 - |\langle \alpha|-\alpha\rangle |=
1 - \exp(-2\alpha^2)$ (assuming $\alpha$ real) \cite{IDP1,IDP2,IDP3}.
For discriminating a larger set of coherent states,
optimal USD becomes more subtle. The quantum mechanical optimum
for more than two symmetrically distributed coherent states can be approached
using linear optics and feedforward \cite{vanEnk02}.

Let us summarize the most common optical resources and potential optical
manipulations of qumodes.

{\it Resources:} Gaussian states; squeezing by means of
$\chi^{(2)}$ nonlinear interactions
(optical parametric amplification);
non-Gaussian states (e.g., `cat states'),
producible, in principle, directly from
$\chi^{(3)}$ nonlinear interactions.

{\it Processing:}
Gaussian transformations (LUBOs:
phase-space displacements,
passive linear optics, active squeezers).

{\it Measurements:} Gaussian measurements (e.g., homodyne detection);
non-Gaussian measurements (e.g., on/off detectors).

The creation of non-Gaussian resource states such as `cat states'
becomes more feasible when conditional state preparation is allowed.
In this case, a hybrid approach is useful, as discussed later.
A drawback of the CV qumode encoding is that these states
are fairly sensitive to losses and noise.
The quality of CV Gaussian entangled states, unconditionally
and efficiently producible
from squeezed light through beam splitters, is fundamentally
limited by the constraint of finite squeezing (energy).
Fidelities drop quickly in the presence of excess thermal noise,
or simply when photons leak into the environment. Nonetheless,
most operations are unconditional and homodyne detection
is possible with near-unit efficiency.

Finally, once again we note that additional atomic systems
and their degrees of freedom may be utilized. In the case
of sufficiently large atomic ensembles, the collective spin
variables can play the analogous roles of the CV qu\-mode phase-space
variables.

\subsection{Implementing efficient quantum computation efficiently?}

A necessry criterion for a quantum computer to give a true
advantage over classical computers is that its realization does not
require exponential resources. In other words,
the exponential `speed-up' quantum computation is usually associated with
must not be at the expense of an exponential increase of physical resources.
The exponentially large dimension of the Hilbert space of $N$ logical qubits,
$2^N$, should be exploited with a number of physical resources scaling as
$\sim N$ (or a polynomial of $N$) rather than $\sim 2^N$. If this criterion
is satisfied, the quantum computation is considered to be `efficient'.

Besides this theoretical, in-principle `efficiency', an actual physical
implementation of a quantum computation should be experimentally `efficient' as well.
While the former type of efficiency is fundamental, the latter one depends
on current technology, and an in-principle efficient quantum computation protocol
may be infeasible and impractical today, but implementable in the future.
Before we shall assess some of the existing proposals for optical quantum computation
with regard to these criteria, let us first recall the most commonly used
quantum gate sets for DV as well as for CV universal quantum computation.

\subsubsection{Universal sets}

For several qubits, a combination of arbitrary single-qubit rotations
with one fixed two-qubit entangling gate is known to be sufficient for
universality such that any unitary gate can be exactly realized on
any given multi-qubit state \cite{Nielsen}. This universal set, however, is too large
(in fact, it is infinitely large) to be implemented in an error-resistant fashion.
Therefore, a discrete, finite set of elementary gates must be chosen which no longer
achieves exact multi-qubit gates (as the set of unitary gates is continuous),
but rather an approximate realization to arbitrary precision. To be efficient,
a sufficiently good approximation must not require an exponential
number of elementary gate applications. A convenient universal set of gates is
\begin{equation}\label{DVset}
\{H, Z_{\pi/2}, Z_{\pi/4}, C_Z\}\,,
\end{equation}
where we omitted the operator hats. Here, $H$ is the Hada\-mard gate,
$H |k\rangle = |0\rangle +  (-1)^k |1\rangle$,
and $C_Z$ acts as an entangling gate, with
\begin{equation}\label{DVC_Z}
|k\rangle\otimes |l\rangle\rightarrow
(-1)^{kl}|k\rangle\otimes |l\rangle\,,\; k,l=0,1\,.
\end{equation}
The gates $Z_\theta \equiv
\exp(-i \theta Z/2)$ describe single-qubit rotations about the $Z$-axis
by an angle $\theta$ with $Z$ being one of the usual Pauli operators $X,Y,Z$,
and $Z|k\rangle = (-1)^k |k\rangle$, $X |k\rangle = |k\oplus 1\rangle$.
Note that removing the gate $Z_{\pi/4}$ from the elementary gate set
means that only so-called Clifford unitaries can be realized which are known
to be insufficient for a quantum computational speed-up over classical computation.
For both universality and speed-up when computing with computational basis states,
the non-Clifford phase gate $Z_{\pi/4}$ must be included here.

Universality for qumodes can be defined and achieved similarly to the qubit case.
Arbitrary single-qumode transformations, together with beam splitters for two qumodes,
are sufficient to simulate any Hamiltonian expressed as an arbitrary polynomial
of the qumode position and momentum operators \cite{Lloyd99}.
As a discrete, elementary gate set for approximate simulations
to any precision, one may choose \cite{Bartlett02}
\begin{equation}\label{CVset}
\{F, Z(\tau), D_2(\kappa), D_3(\lambda), C_Z\}\,,
\end{equation}
with $\tau$, $\kappa$, and $\lambda$ real.
\footnote{commonly, both in the DV and the CV case,
the CNOT or SUM gates are used for the canonical entangling gate
instead of $C_Z$; however, these are equivalent up to local
DV-Hadamard or CV-Fourier transformations.}
In this case, $F$ represents the Fourier transform operator
to switch between the position $|\; \rangle_x$ and momentum
$|\; \rangle_p$ basis states,
$F|x\rangle_x = |x\rangle_p$.
The entangling gate $C_Z$ is an $x$-controlled $p$-displacement,
$C_Z = \exp(2i \hat x\otimes\hat x)$, with
\begin{equation}\label{CVC_Z}
C_Z |x\rangle_x |p\rangle_p =
|x\rangle_x |p + x\rangle_p\,,
\end{equation}
while the roles of the Pauli gates are now played
by the Weyl-Heisenberg (WH) momentum and position shift
operators, $Z(\tau) = \exp(2 i \tau \hat x)$ and
$X(\tau)=\exp(-2 i \tau \hat p)$, respectively,
$Z(\tau)|p\rangle_p= |p+\tau\rangle_p$
and
$X(\tau)|x\rangle_x = |x+\tau\rangle_x$.
Finally, the phase gates $D_l(\kappa_l) = \exp(i\kappa_l \hat x^l)$
are included in order to simulate any Gaussian (CV Clifford)
transformation ($l=2$) and to achieve full CV universality including
non-Gaussian (CV non-Clifford) gates ($l=3$).
Recall that the Gaussian transformations map Gaussian states
onto Gaussian states; they correspond to quadratic Hamiltonians
with linear input-output relations for the qumode operators
as in Eq.~(\ref{LUBO}). Gaussian operations on Gaussian states
can be efficiently simulated classically \cite{Bartlett02}.

Both for the DV and the CV case, for those encodings discussed so far,
there is always at least one universal gate that is not realizable
through linear transformations alone. In single-photon single-rail encoding,
even a single-qubit Hadamard gate, transforming a Gaussian vacuum state
into a non-Gaussian superposition of vacuum and one-photon Fock state
would be highly nonlinear. The hardest part of universally processing
dual-rail encoded qubits would be the entangling gate which has to
act upon at least two photons.
Ultimately, the universal processing of even a single qumode requires
some form of nonlinearity.

\subsubsection{Nonlinear versus linear optics}

The most obvious approach now to optically implement an entire
set of universal quantum
gates would be directly through nonlinear interactions.
The two-qubit $C_Z$ gate
is accomplished by applying a quartic cross-Kerr interaction
on two photonic occupation number qubits,
\begin{equation}\label{crossKerr}
\exp(i\pi\, \hat a_1^\dagger\hat a_1\otimes \hat a_2^\dagger\hat a_2)
|k\rangle\otimes |l\rangle =
(-1)^{kl}|k\rangle\otimes |l\rangle \;.
\end{equation}
The same interaction leads to a $C_Z$ gate
for two photonic dual-rail qubits, with the cross-Kerr interaction
acting on the second rail (mode) of each qubit such that
only the term $|01\rangle\otimes |01\rangle$ acquires a sign flip.
This is the conceptually simplest and theoretically most efficient
(only one optical device needed!) method to complete the set
of universal gates in dual-rail encoding. In the
low-photon number subspace here, we may even decompose
the cross-Kerr two-mode unitary into a beam splitter,
two self-Kerr one-mode unitaries,
\begin{eqnarray}\label{selfKerr}
\exp\left[i\frac{\pi}{2}\, \hat a_1^\dagger \hat a_1 (\hat a_1^\dagger \hat a_1 - 1)\right]
\otimes\,
\exp\left[i\frac{\pi}{2}\, \hat a_2^\dagger \hat a_2 (\hat a_2^\dagger \hat a_2 - 1)\right],
\nonumber\\
\end{eqnarray}
and another beam splitter
%(see Fig.~\ref{fig5b}).
(see Fig.~2.3.1).

Thus, a sufficiently strong one-mode self-Kerr interaction would
be enough to fulfill the criteria for DV universality \cite{Nielsen} on
the finite-dimensional multi-qubit subspace of the infinite-dimensional,
multi-mode optical Fock space.
At the same time, the quartic one-mode self-Kerr interaction, together
with Gaussian, linear transformations (LUBOs), would be also sufficient
for the strong notion of full (asymptotically arbitrarily precise)
CV universality \cite{Lloyd99}.

\begin{center}
  \begin{columnfigure}\label{fig5b}
    \centering%
    \includegraphics[width=180pt]{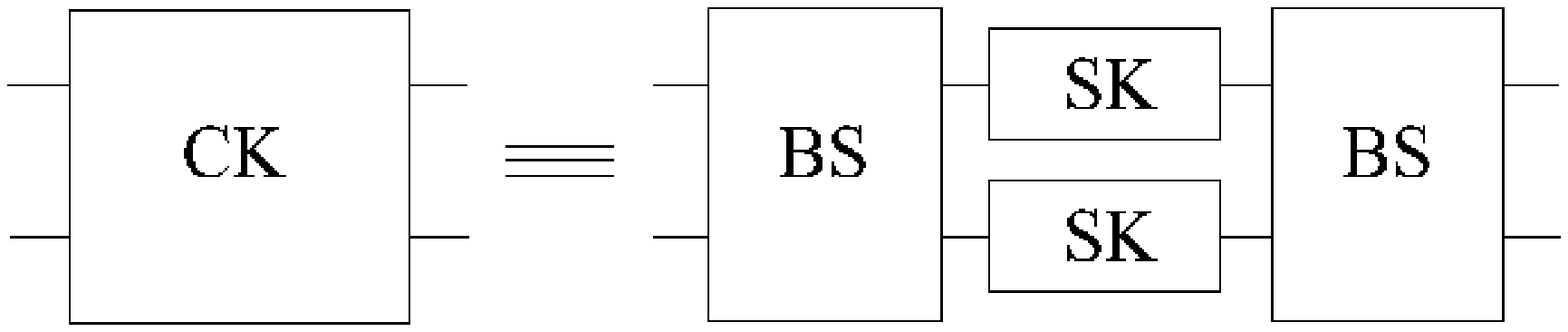}
  \figcaption{{\bf 2.3.1} Implementing a controlled sign gate ($C_Z$) on two single-rail
  qubits using cross-Kerr (CK) or self-Kerr (SK) nonlinearities.
  The first beam splitter (BS) transforms
  the term $|11\rangle$ into $|20\rangle - |02\rangle$, while the
  other terms stay in the vacuum and one-photon space.
  As the SK interactions affect sign flips only
  for the two-photon components, only the term $|11\rangle$
  acquires a sign flip.}
  \end{columnfigure}
\end{center}

The problem of this approach, however, is that an effective
coupling strength of $\sim\pi$ for the self/cross-Kerr interactions is totally
infeasible on the level of single photons. Therefore,
it is worth examining carefully if there is a way to
implement universal
quantum gates through linear optical elements, ideally
just using beam splitters and phase shifters.
A very early proposal for linear-optics-based quantum computation
indeed does work with only linear elements \cite{AdamiCerf98}.
It is based upon so-called multiple-rail encoding, where a $d$-level system
is encoded into a single photon and $d$ optical modes, with the basis states
$\hat a_k^\dagger |00\cdots 0\rangle$, $k=1,2,...,d$.
Any unitary operator can be realized in the space spanned by this basis,
as we only need $\hat U \hat a_k^\dagger |00\cdots 0\rangle =
\sum_{l=1}^{d} U_{kl} \hat a_l^\dagger |00\cdots 0\rangle$, $\forall k$.
This linear transformation, as in Eq.~(\ref{passive}),
is easily achieved through a sequence of beam
splitters and phase shifters \cite{Reck}.

As a result, universal quantum computation is, in principle, possible
using a single photon and linear optics. This kind of realization
would be clearly efficient from an experimental point of view.
In fact, implementing a universal two-qubit gate in a
$d=2^2 = 4$-dimensional Hilbert space, would only require the modest set
of resources of an optical `ququart', in `quad-rail' encoding
corresponding to a single photon and four optical modes.
In fact, for small quantum applications, by adding to the polarization
of the photons (their spin angular momenta)
extra degrees of freedom such as orbital angular momenta,
this kind of approach can be useful \cite{Kwiat98,Schuck,Barbieri,Wilde}.

Nonetheless, the drawback of the multiple-rail-based linear-optics
quantum computer \cite{AdamiCerf98} is its bad scaling.
In theory even, this type of quantum computer is inefficient.
Scaling it up to computations involving $N$ qubits, we need
$2^N$ basis states, and hence $2^N$ optical modes.
All these modes have to be controlled and processed in a linear
optical circuit with an exponentially increasing number of optical
elements. For example, a 10-qubit circuit would only require
10 photons and 20 modes in dual-rail encoding, while it consumes
$2^{10} = 1024$ modes (for just a single photon) and at least as many
optical elements in multiple-rail encoding.

\subsubsection{Teleportation-based approaches}

There are now indeed a few proposals that aim at circumventing
the in-principle (scaling) inefficiency of the linear multiple-rail protocol
and the experimental infeasibility of the direct nonlinear
optical approach. Certainly, the two most important and
conceptually unique proposals are the KLM \cite{KLM01} and the
GKP \cite{GKP01} schemes. In both schemes,
a new concept of {\it measurement-induced nonlinearities} is exploited.
The KLM scheme is a fully DV-based protocol, demonstrating that,
in principle, passive linear optics and DV photonic auxiliary states
are sufficient for (theoretically) efficient, universal DV quantum computation.
Inducing nonlinearity through photon counting measurements
renders the KLM scheme non-deterministic. However,
the probabilistic quantum gates can be made asymptotically
near-deterministic by adding to the toolbox feedforward and complicated,
multi-photon entangled auxiliary states with sufficiently high
photon numbers, and by employing quantum teleportation \cite{GottesmanChuang}.
KLM is `in-princi\-ple efficient', as the number
of the ancillary photons grows only polynomially with the success rate.
Fidelities are always, in principle, perfect
in the KLM approach.

We shall briefly describe the basic elements
of KLM. The GKP scheme combines linear CV resources with linear operations
and nonlinear measurements; therefore, a discussion of GKP, which achieves
both fault-tolerant DV universality and (non-fault-tolerant) CV universality,
in the spirit of Ref.~\cite{Lloyd99},
is postponed until the section on hybrid schemes.
Similarly, the alternate concept of weak nonlinearities
relies on hybrid systems, and will also be discussed later.

\begin{center}
  \begin{columnfigure}\label{fig6b}
    \centering%
    \includegraphics[width=180pt]{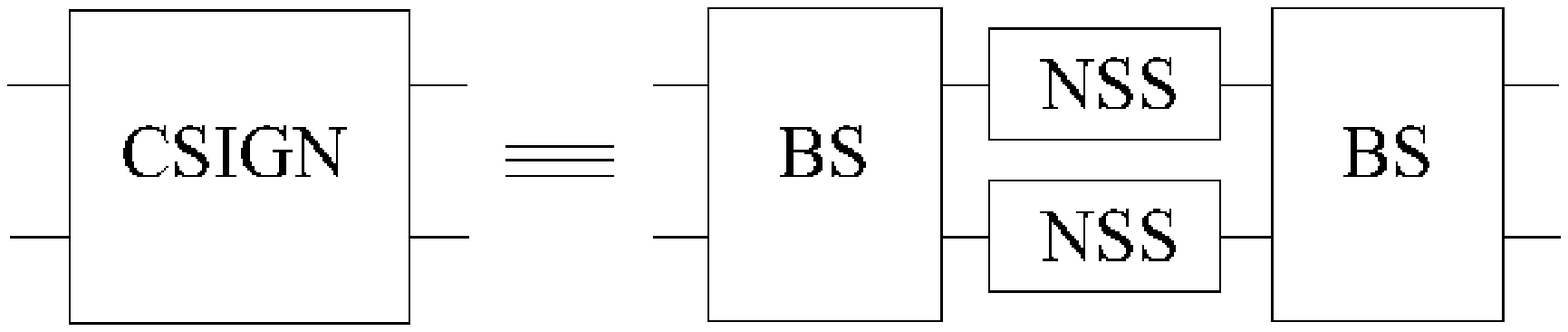}
  \figcaption{{\bf 2.3.2} Implementing a probabilistic
  controlled sign gate (CSIGN $\equiv C_Z$) on two single-rail
  qubits using two non-deterministic NSS gates. The resulting
  two-qubit gate works in a similar way to the deterministic
  implementation described in
  %Fig.~\ref{fig5b}
  Fig.~2.3.1 using Kerr
  nonlinearities.}
  \end{columnfigure}
\end{center}

The essential ingredient for the non-deterministic realization
of a two-photon two-qubit entangling gate (in dual-rail encoding)
is the one-mode nonlinear sign shift (NSS) gate \cite{KLM01}. It acts on
the qutrit subspace $\{|0\rangle, |1\rangle, |2\rangle \}$
of the optical Fock space as $|k\rangle\rightarrow
(-1)^{k(k-1)/2}|k\rangle$. Placing two such NSS gates in the middle
between two beam splitters will then act as a
controlled sign gate, $|k\rangle\otimes |l\rangle\rightarrow
(-1)^{kl}|k\rangle\otimes |l\rangle$, on two single-rail
as well as two dual-rail qubits. In fact, we may replace
the deterministic Kerr-based circuit of Fig.~2.3.1
%Fig.~\ref{fig5b}
by the equivalent
circuit depicted in
%Fig.~\ref{fig6b}.
Fig.~2.3.2. The latter, however,
becomes non-deterministic with NSS gates operating only probabilistically.

In the original KLM proposal, the NSS gate can be realized
with $1/4$ success probability, corresponding to a success probability
of $1/16$ for the full controlled sign gate as shown in
%Fig.~\ref{fig6b}.
Fig.~2.3.2.
In subsequent works, this efficiency was slightly improved \cite{Knill03}.
There are also various, more general treatments of these
non-deterministic linear-optics gates deriving bounds on their efficiencies
\cite{Scheel04,Eisert05,Uskov09}. Experimental demonstrations
were reported as well \cite{obrien1}, even entirely in an optical fiber
\cite{obrien2}.

Probabilistic quantum gates cannot be used directly for
quantum computation. The essence of KLM (see
%Fig.~\ref{fig8})
Fig.~2.3.3)
is that near-unit success probabilities
are attainable by combining non-deterministic gates on offline entangled states
with the concept of quantum gate teleportation \cite{GottesmanChuang}.
As the necessary Bell measurements for quantum teleportation succeed
at most with $1/2$ probability, if only fixed arrays of beam splitters  are used
\cite{Calsamiglia01}, entangled ancilla states and feedforward must be added
to boost efficiencies beyond $1/2$ to near $1$.

\begin{center}
  \begin{columnfigure}\label{fig8}
    \centering%
    \includegraphics[width=225pt]{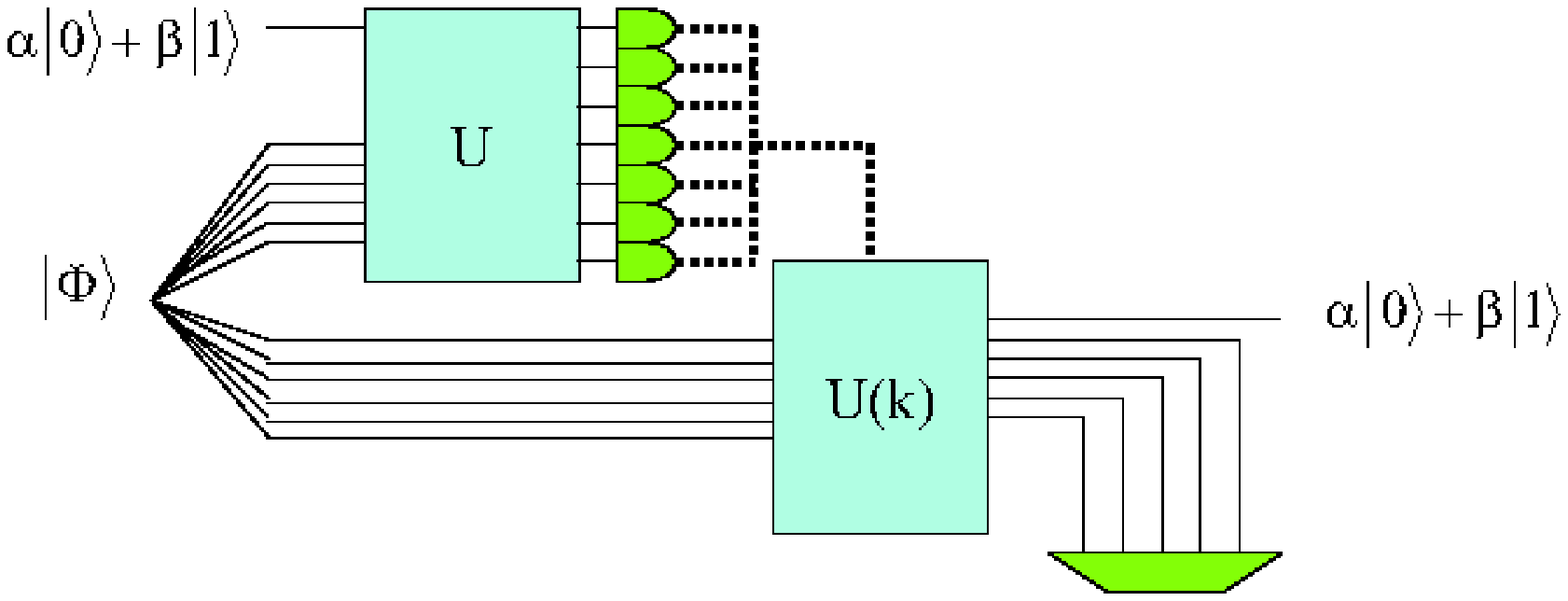}
  \figcaption{{\bf 2.3.3} Making non-deterministic gates near-deterministic
  through single-rail quantum teleportation. The Bell measurement is performed by means of
  the linear-optics circuit $U$ plus photon counting.
  For an entangled two-mode state $|\Phi\rangle\propto
  |10\rangle + |01\rangle$ with one ancilla photon, teleportation succeeds
  only in one half of the cases. For larger ancillae with sufficiently many photons,
  teleportation can be made almost perfect. In order to teleport a gate
  near-deterministically onto an input state, the corresponding gate must be first applied
  offline and probabilistically to the multi-photon entangled ancilla state.}
  \end{columnfigure}
\end{center}

Even though KLM is `in-principle efficient', it is still highly impractical,
as near-deterministic operations would require ancilla states too complicated
to engineer with current experimental capabilities. It is therefore extremely
important to further enhance the efficiencies of linear-optics quantum computation
with regard to the resource scaling. Steps into this direction have been made already
by merging the teleportation-based KLM approach with the fairly recent
concept of one-way (cluster) computation \cite{RaussendorfBriegel01}.

\subsubsection{Cluster-based approaches}

In the preceding section, we introduced the notion of
meas\-urement-induced nonlinearities, which, combined
with the more general (and implementation-independent) concept
of measurement-based quantum computation, enables one to
obtain the necessary nonlinear element in (linear) optical
approaches to quantum information processing.
As opposed to the standard, circuit model of quantum computing,
where any computation is given by a sequence of reversible, unitary
gates, in measurement-based quantum computing, universal quantum gates
are encoded `offline' into an entangled-state resource; suitable
measurements, performed `online' on this resources state, and, typically,
some form of feedforward will then lead to the desired unitary evolution.
Feedforward may sometimes be postponed until the very end of the computation,
or even totally omitted through `reinterpretation' of the `Pauli-frame';
nonetheless, in some form it will be needed in order to render measurement-based quantum
computation determinis\-tic despite the randomness induced by the measurements.

There are now further subcategories of measurement-based quantum computing.
First, that based on full quantum teleportation \cite{GottesmanChuang}
involving online {\it nonlocal} measurements such as Bell measurements, as described
in the preceding section. Secondly, there is an ultimate realization of measurement-based
quantum computing that requires all entangling operations be done offline and allows only
for {\it local} measurements applied on the offline resource state -- the cluster state
\cite{RaussendorfBriegel01}.
In such a cluster computation, a multi-party entangled
cluster state is first prepared
offline. The actual quantum computation is then performed solely through
single-party projection measurements on the individual nodes of the cluster state.
By choosing appropriate measurement bases in each step, possibly depending on earlier
measurement outcomes, any unitary gate can be applied to an input state
which typically becomes part of the cluster at the beginning of the
computation, see Fig.~2.3.4.
%Fig.~\ref{fig13}.

\begin{center}
  \begin{columnfigure}\label{fig13}
    \centering%
    \includegraphics[width=170pt]{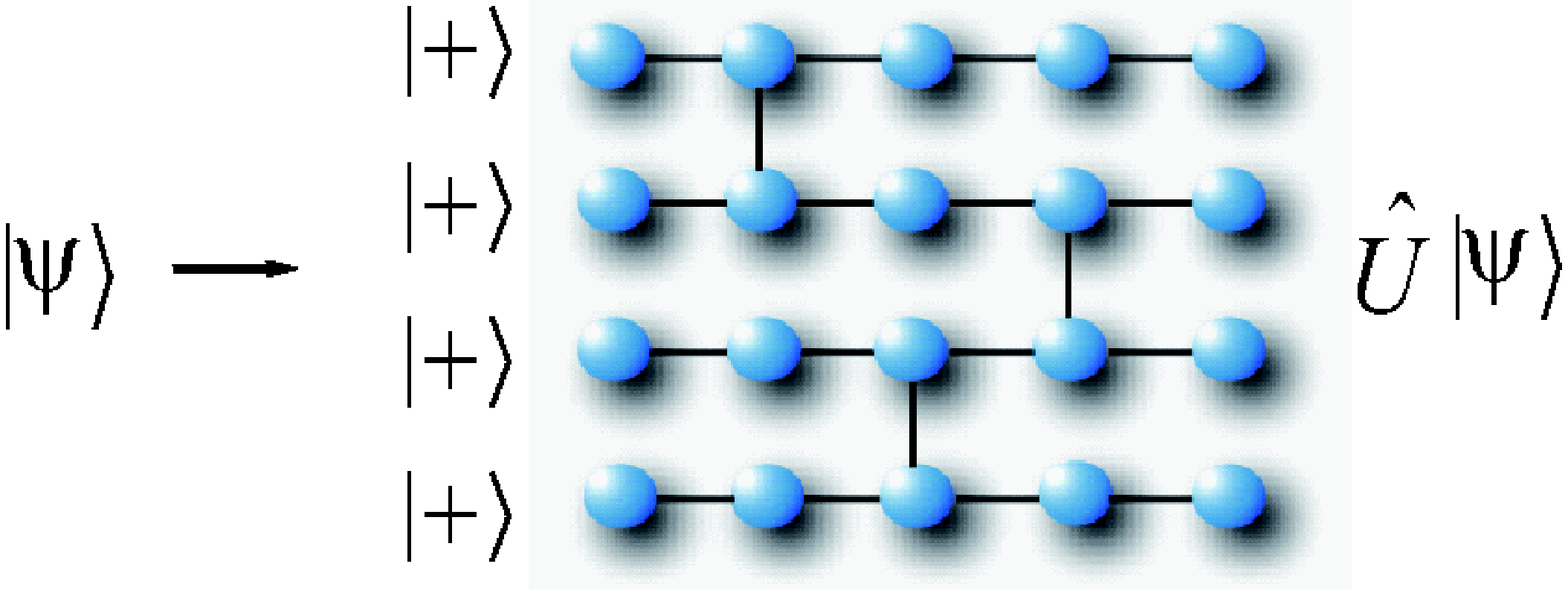}
  \figcaption{{\bf 2.3.4} One-way (cluster) computation for qubits. Certain single-qubit basis states
  become pairwise entangled to form a multi-qubit cluster state. Local projection
  measurements on the individual qubits (potentially including feedforward with
  a measurement order going from left to right) are then enough to realize universal
  quantum computation. A multi-qubit input state $|\psi\rangle$ attached to the left end
  of the cluster could, in principle, be universally processed with the output state
  occurring at the right end of the cluster. The vertical edges allow for two-qubit gates.
  }
  \end{columnfigure}
\end{center}

\begin{center}
  \begin{columnfigure}\label{figqubitcluster}
    \centering%
    \includegraphics[width=179pt]{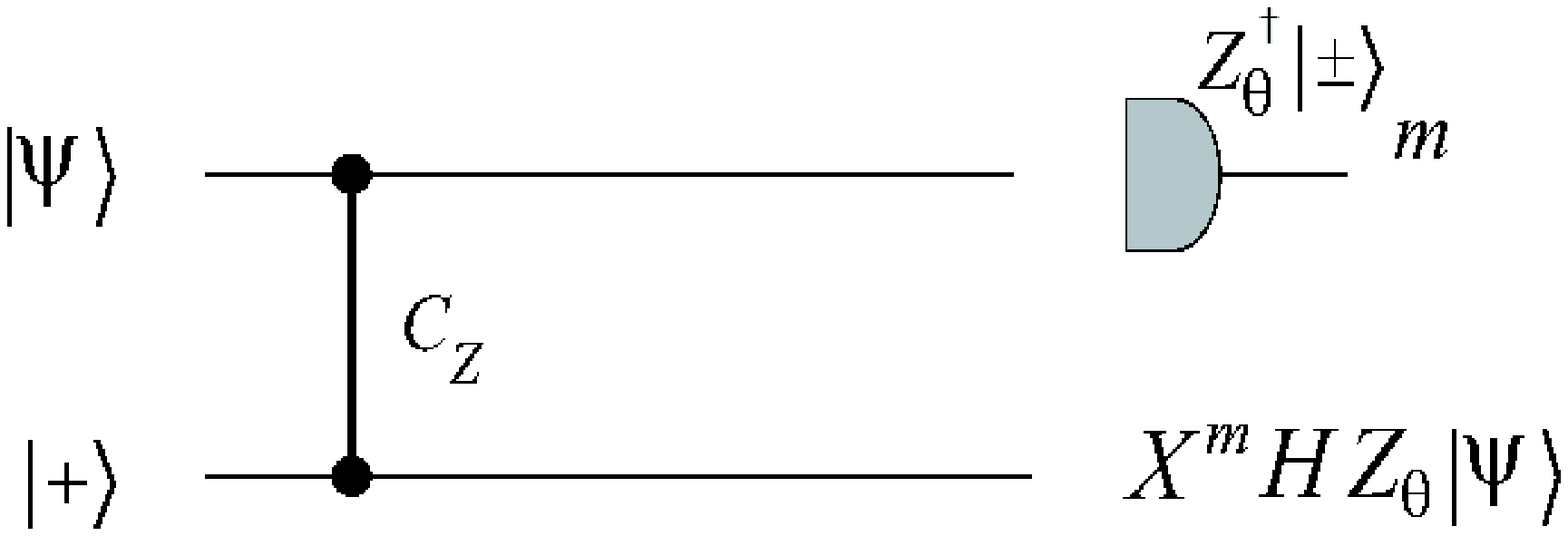}
  \figcaption{{\bf 2.3.5} Elementary `one-bit' teleportation circuit for qubit cluster computation.
  The $C_Z$ gate represents a horizontal edge connecting two nodes of the cluster state. An input state $|\psi\rangle$
  is teleported into the left node, and a subsequent, local single-qubit measurement in the binary basis
  $\{Z_\theta^\dagger |\pm\rangle \}$ leaves the second node, up to a Hadamard gate $H$ and a Pauli correction
  $X$ depending on result $m=0,1$, in the unitarily evolved state $Z_\theta|\psi\rangle$, with the rotation angle
  $\theta$ for a $Z$-rotation $\exp(-i \theta Z/2)$ controlled by the actual choice $\theta$ for the measurement basis.}
  \end{columnfigure}
\end{center}

The essence of cluster computation as described by the one-way model
of quantum computation \cite{RaussendorfBriegel01} can be summarized as follows:
the cluster state is {\it independent of the computation};
universality is achieved through {\it choice of measurement bases}.
This is illustrated in Fig.~2.3.5 for an elementary `one-bit' teleportation circuit
between just two nodes of a qubit cluster state; once an input state is part of the
cluster (after it got teleported into the cluster), a local single-qubit measurement
is then sufficient to apply an arbitrary $Z$-rotation. At the very beginning of a cluster computation,
the cluster computer may be initiated in a product of `blank' basis states $|+\rangle$
such that no extra encoding teleportation step is needed; an arbitrary state can be anyway
prepared within the cluster through local measurements. For example, an arbitrary
single-qubit rotation requires just three elementary steps as shown in Fig.~2.3.5.
In such a concatenation of elementary steps, for a given desired evolution,
the later choice of the measurement bases depends on earlier measurement outcomes
whenever non-Clifford gates are involved.

Apart from the conceptual innovation, it turned out that, in particular,
with regard towards linear optical quantum computation proposals,
the cluster approach helps to reduce the resource costs significantly \cite{Nielsen04,Browne05}.
As there are no more nonlocal Bell measurements needed in cluster-based
computation, but only local projections, the problem of the nondeterministic
linear-optics Bell measurements can be circumvented. Such Bell measurements
may only be used for preparing a DV optical cluster state offline \cite{Browne05},
whereas the online computation is perfectly deterministic.
Eventually, scalability and resource costs in linear-optics quantum computation are determined
by the efficiency with which cluster states can be grown using
probabilistic entangling gates \cite{Kieling1,Kieling2}.

In a very recent extension of the DV cluster model, the analogous CV cluster computation
approach is considered \cite{Menicucci06}. In the CV model, full CV universality
can be approached by applying linear Gaussian and nonlinear non-Gaussian measurements to a Gaussian, approximate CV
cluster state. A discussion of universal CV cluster computation
combining CV and DV measurements will be presented in the section on hybrid schemes.

We conclude this section by noting that typically there is a trade-off
between the DV and CV optical approaches. The DV schemes are necessarily probabilistic
and only at the expense of special extra resources can they be made neardeterministic;
fidelities are usually quite high, near-unit fidelities.
Conversely, in the CV schemes, fidelities tend to be modest and are necessarily below unity;
nonetheless, CV operations are typically deterministic.
These characteristic features were already present in the earliest experiments
of DV and CV quantum information processing, namely those demonstrating
quantum teleportation of an unknown quantum state between two parties
\cite{Bouwmeester97,deMartini,Furusawa98}.

Let us now consider the possibility of optically realizing efficient quantum
communication. Even for this supposedly simpler task than universal
quantum computation, similar constraints and nogo results exist,
when the toolbox is restricted to only linear operations.
Especially when efficient and reliable quantum communication is to be
extended over large, potentially intercontinental distances,
it turns out that this is, in principle, possible, but
would require not much less resources than needed for doing
optical quantum computation.

\subsection{Implementing efficient quantum communication efficiently?}

The goal of quantum communication is the reliable transfer of arbitrary quantum states,
possibly drawn from a certain alphabet of states. Quantum communication is ``the art to transfer quantum states"
\cite{Gisin07}. This may then lead to various applications such as the secure distribution of a classical key
(quantum key distribution \cite{GisinRMP,NJPfocus,Stebila}) or the connection of spatially separated quantum computers for distributed quantum computing
or a kind of quantum internet \cite{Kimble}. As light is an optimal information carrier for communication,
one may send quantum states encoded into a stream of single photons or a multi-photon pulse through an optical channel.
However, quantum information encoded in fragile superposition states,
for example, using photonic qubits or qumodes, is vulnerable against losses and other sources of excess noise
along the channel such that the fidelity of the state transfer will exponentially decay with the length of the channel.

For instance, the $|1\rangle = \hat a^\dagger |0\rangle$ term of a single-rail qubit would partially leak into
the vacuum modes of the channel, $\hat a^\dagger |0\rangle\otimes |0\rangle_{\rm ch} \rightarrow
\sqrt{\eta} |10\rangle + \sqrt{1-\eta} |01\rangle$, such that tracing over the channel mode leads to the final signal
state $\eta |1\rangle\langle 1| + (1-\eta) |0\rangle\langle 0|$, with the transmission parameter $\eta = \exp(-L/L_{\rm att})$
and the attenuation length $L_{\rm att}$. If a photon that did make it through the channel
is to be detected and, in particular, resolved against detector dark counts, this will become exponentially harder
for longer channels. Similarly, a qumode in a coherent state would be transformed
as $|\alpha\rangle\otimes |0\rangle_{\rm ch} \rightarrow |\sqrt{\eta}\alpha\rangle |\sqrt{1-\eta}\alpha\rangle$,
corresponding to the signal map $|\alpha\rangle \rightarrow |\sqrt{\eta}\alpha\rangle$. Although the coherent state remains
pure, its transmitted amplitude would be decreased by an exponential factor. Any nonclassical qumode state
would exponentially decohere into a mixed state.

\begin{center}
  \begin{columnfigure}\label{figqrepeater}
    \centering%
    \includegraphics[width=225pt]{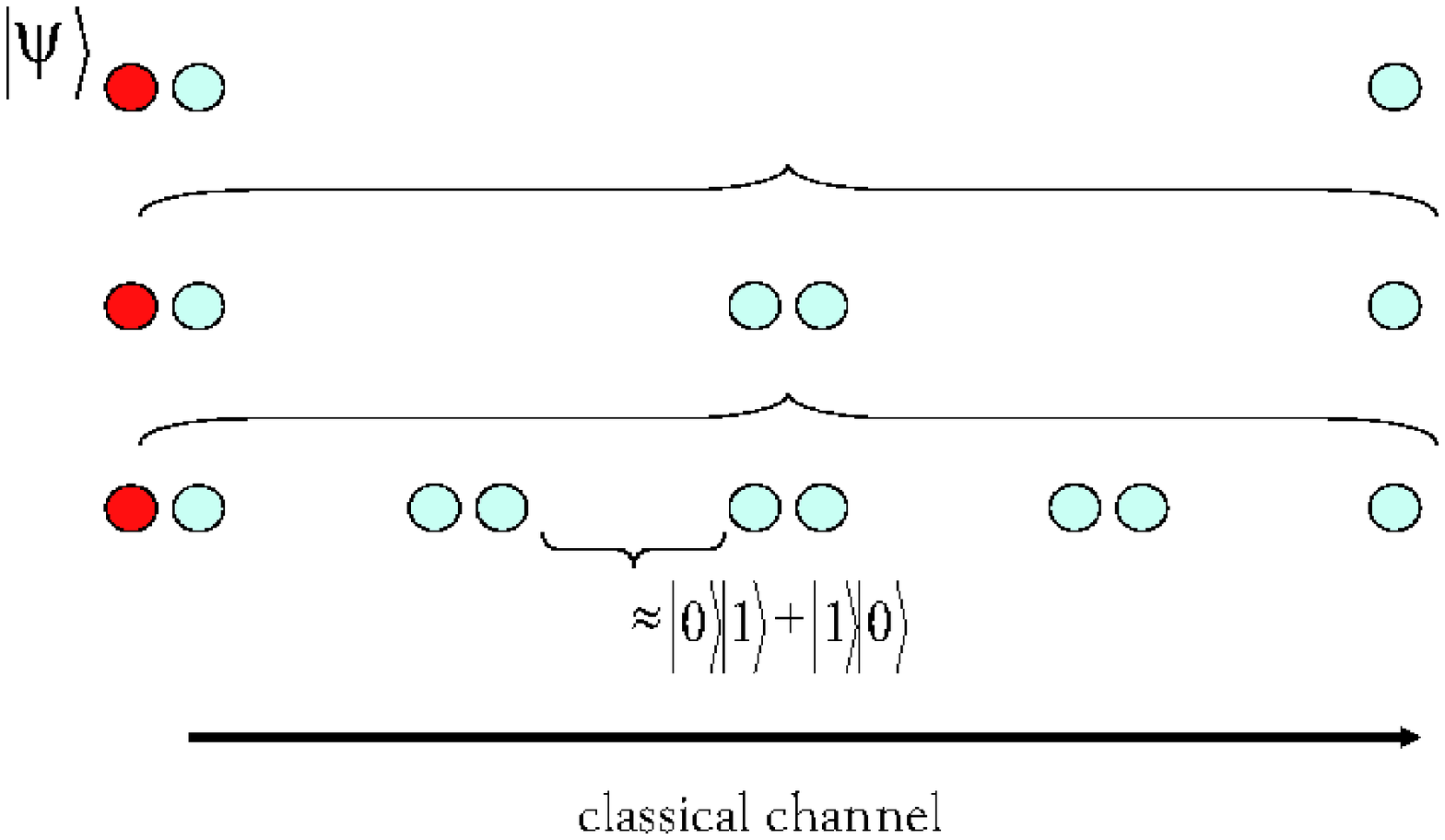}
  \figcaption{{\bf 2.4.1} Concept of a quantum repeater. The state $|\psi\rangle$ is teleported
  to the remotest station after a long-distance entangled pair is created over the total channel.
  For this purpose, a supply of short-distance pairs is distributed over sufficiently short
  channel segments such that high-fidelity entangled pairs can be distilled and connected
  through entanglement swapping. Only this combination of entanglement
  distillation and quantum teleportation in a fully nested protocol enables one to suppress
  the exponential decay of the transfer efficiencies or fidelities as obtained in direct state transmission.
  As the full protocol typically contains probabilistic elements, sufficient local memories are required.
   }
  \end{columnfigure}
\end{center}

In long-distance, classical communication networks, signals that are gradually
distorted during their propagation in a channel are repeatedly recreated through
a chain of intermediate stations along the transmission line.
For instance, optical pulses traveling through a glass fiber and being subject to
photon loss can be reamplified at each repeater station. Such an amplification is
impossible, when the signal carries quantum information. If a quantum bit is encoded
into a single photon, its unknown quantum state cannot be copied along the line
\cite{WoottersZurek,Dieks}; the photon must travel the entire distance with an exponentially
decreasing probability to reach the end of the channel.

The solution to the problem of long-distance quantum communication
is provided by the so-called quantum repeater \cite{Briegel98,Duer99}, see Fig.~2.4.1.
In this case, prior to the actual quantum-state communication, a supply of
standard entangled states is generated and distributed among not too distant
nodes of the channel. If sufficiently many of these imperfect entangled states
are shared between the repeater stations, a combination of entanglement
purification and swapping extends this shared entanglement over
the entire channel. Through entanglement swapping \cite{Zukowski},
the entanglement of neighboring pairs is connected,
gradually increasing the distance of the shared
entanglement. The entanglement purification \cite{Bennett96,Deutsch}
enables one to distill (through local operations) a high-fidelity entangled pair
from a larger number of low-fidelity entangled pairs,
as they would emerge after a few rounds of entanglement swapping with imperfect
entangled states and at the very beginning after the initial, imperfect
entanglement generation and distribution between two neighboring repeater stations.

The essence of long-distance quantum communication as realized through the quantum
repeater model \cite{Briegel98,Duer99} can be summarized as follows:
provided {\it sufficient local quantum memories} are available and {\it some form of
quantum error detection} is applied, quantum communication over arbitrary distances is
possible with an increase of (spatial or temporal) resources scaling only
subexponentially with distance. Similar to what we concluded for efficient quantum computation,
an in-principle efficient realization depends on a non-exponential resource scaling.
Otherwise, without fulfilling this criterion, we could as well choose
to directly transmit quantum states, similar to performing exponentially many gate
operations in an inefficient quantum computer.
\footnote{the naive approach of dividing the total channel into several segments
that are connected through quantum teleportation without incorporating any form of quantum
error detection and without using quantum memories
is not enough to render quantum communication efficient with regard
to resource scaling; however, it may still help to enhance practicality of a scheme,
for instance, in order to resolve single-photon signals against detector dark counts
(``quantum relay" \cite{GisinRMP,qrelay1,qrelay2,qrelay3}).}

The main distinction between the
communication and the computation scenario is that in the former case we may now
use probabilistic operations; in particular, quantum error correction may be replaced
by quantum error detection. However, this supposed advantage becomes a real advantage
only provided that quantum states can be reliably stored during the waiting times
for classical signals communicating successful events. So eventually, there is a trade-off
between the requirements on memory times and local quantum gates,
with the rule of thumb that only less efficient (in terms of fidelity)
and more complex quantum error detection/correction
schemes would lead to a reduced need for efficient memories \cite{Hartmann07,Aschauer}.

When it comes to turning the in-principle solution to scalable quantum communication
over arbitrary distances in form of a quantum repeater into a realistic implementation,
what are the currently available resources regarding quantum memories and gates?
For a repeater segment of the order of $L_0\sim L_{\rm att}\sim 20$ km, with an optical fiber
at minimal absorption of 0.17 dB/km corresponding to $L_{\rm att}\approx 25.5$ km,
we have a classical communication time of at least $L_0/c \approx 0.066$ ms
(ideally considering the vacuum speed of light $c$),
in order to verify successful entanglement creation, swapping, or distillation events between
two neighboring stations. In order to extend this to a total distance of $L \sim 2000$ km, even if
no intermediate stations are involved such that only a single heralded entanglement creation
event is to be confirmed over the distance $L$ ,
a classical communication time of
about 6.6 ms would be needed. Although memory times approaching 60 ms are achievable in electron spin systems
\cite{Tyryshkin},
and single excitations may be stored and retrieved over a time scale of 1-10 ms \cite{Zhao1,Zhao2},
longer memory times would have to rely upon nuclear spins.
However, even in this case, currently available memories ($\sim 1$ s \cite{Morton08,LaddPRB05})
do not match those
needed in a fully nested repeater protocol. Moreover, the local quantum logic for error detection
(requiring, effectively, a small quantum computer at each station), would further
increase the actual memory requirements, for instance, when using probabilistic
linear-optics gates \cite{obrien1,obrien2}, as discussed in the preceding section.

The maximum distance for experimental quantum communication is currently about 250 km
\cite{stucki09,GisinRMP}. Although extensions to slightly larger distances may be possible
with present experimental approaches \cite{scheidl09}, there are also various proposals
for actually implementing a quantum repeater.
The most recent proposals
are based on the nonlocal generation of atomic (spin) entangled states, conditioned upon
the detection of photons distributed between two neighboring repeater stations.
The light, before traveling through the communication channel and being detected,
is scattered from either individual atoms, for example, in form of solid-state single photon
emitters \cite{Childress1,Childress2}, or from an atomic ensemble,
i.e., a cloud of atoms in a gas \cite{Duan01} (the `DLCZ' scheme).

In these heralded schemes, typically,
the fidelities of the initial entanglement generation are quite high, but the heralding mechanism
leads to rather small pair generation rates.
Other complications include interferometric phase stabilization
over large distances \cite{Simon07,Zhao07,Chen07,Jiang07} and the purification of atomic ensembles.
Yet some elements towards
a realization of the DLCZ protocol have been demonstrated already
\cite{Chou05,Chaneliere05,Eisaman05,Chen08}.

The DLCZ scheme is initially based upon ``hybrid" CV Gaussian entangled states,
where hybrid here means that the entanglement is effectively
described by a two-mode squeezed state with the two modes being a (symmetric) collective atomic mode
for a large ensemble of $N_a$ atoms,
\begin{eqnarray}\label{collectivespin}
\hat S_a = \frac{1}{\sqrt{N_a}}\sum_{i=1}^{N_a} |g\rangle_i \langle s| \,,
\end{eqnarray}
and a Stokes-light mode, $\hat a_s$ \cite{Duan01,Razavi06}; here, $|g\rangle_i$, $|s\rangle_i$ are
the ground and metastable states of the $i$th atom before and after a Raman transition
in a $\Lambda$ configuration, respectively; $|g\rangle \rightarrow |e\rangle$ is induced
by a classical pump (with $|e\rangle$ being the excited state), while $|e\rangle \rightarrow |s\rangle$
produces the forward-scattered Stokes light. In the regime of only low excitations,
corresponding to short interaction times $t_{\rm int}$ and Stokes mean photon numbers less than unity,
the $n=0,1$ terms of the two-mode squeezed state,
\begin{eqnarray}
\frac{1}{\cosh r} \sum_{n=0}^{N_a}
(\hat S_a^\dagger \hat a_s^\dagger \tanh r)^n |0\rangle_a |0\rangle_s/n!\,,
\end{eqnarray}
with
$|0\rangle_a\equiv \otimes_{i=1}^{N_a}|g\rangle_i$  and ``squeezing parameter" $r$, are dominating such that
the resulting atom-light state becomes approximately
%\begin{eqnarray}
$|0\rangle_a |0\rangle_s + r\, \hat S_a^\dagger \hat a_s^\dagger |0\rangle_a |0\rangle_s$.
%\end{eqnarray}
To this order, dropping terms $O(r^2)$ and higher,
by creating the same state at the nearest repeater station equipped with atomic memories
and combining the Stokes fields coming from both stations, labeled by $s1$ and $s2$, at a central beam splitter
in order to erase which-path information,
\begin{eqnarray}
\hat a_{s1/s2}^\dagger |0\rangle_{s1}|0\rangle_{s2}
\rightarrow (\hat a_{s1}^\dagger \pm \hat a_{s2}^\dagger)|0\rangle_{s1}|0\rangle_{s2}/\sqrt{2},
\end{eqnarray}
single-photon detector events would trigger an entangled state of the form
$(\hat S_{a1}^\dagger \pm \hat S_{a2}^\dagger) |0\rangle_{a1}|0\rangle_{a2}/\sqrt{2}$
between the ensembles denoted by $a1$ and $a2$.
The initial atom-light entanglement
is completely swapped onto the atomic memories.
Additional atomic ``vacuum" contributions
originating from detector dark counts would be (to some extent) automatically removed from the final states;
a kind of purification ``built into" the entanglement swapping process.
The effect of atomic spontaneous emissions is suppressed, because a single atomic spin mode
gets collectively enhanced for sufficiently large ensembles $N_a \gg 1$.

Though, in principle, being ``fully CV" at the beginning,
the DLCZ protocol uses DV measurements with photon detectors
in order to conditionally prepare DV entangled states.
Ideally only the single excitations from the initial Gaussian states
would contribute to the final DV state. Only in this limit (corresponding to small
``squeezing" $r$), we will obtain near-unit fidelity pairs, $F\approx 1 - r^2$,
at the expense of longer average creation times, $T\approx T_0/r^2$,
with $T_0\equiv t_{\rm int} + L_0/c$.
In some sense, we could as well add DLCZ to the list of hybrid schemes according to our
definition, as it combines (very weakly excited)
CV resources with DV measurements, resulting in DV output states.

This is in contrast to the complementary approaches of the Polzik group
who use similar physical systems and interactions, i.e., atomic ensembles with Stokes light,
but remain fully CV throughout for light-matter teleportations \cite{Polzik1}
and interfaces \cite{Polzik2}, as their operations consist of CV homodyne detections
and feedforward. In their atom-light CV approach, two orthogonal components
of the optical Stokes operators and the atomic collective spin operators
are each well approximated by (rescaled) qumode phase-space variables.
To sum up, we may say now that in DLCZ effectively
``flying qumodes" become ``static qubits", while in Polzik's schemes,
``flying qumodes" become ``static qu\-modes". The advantage of the latter approach is clearly
that it is entirely unconditional with no need for any heralding element;
this, as typical in fully CV schemes, is at the expense of only imperfect non-unit
fidelities.

In either case, however, the interactions take
place in free space, with many atoms effectively enhancing the coupling
between the collective spins and the light field. Yet other approaches to quantum
communication would turn ``flying qubits" (e.g. polarization-encoded photonic
qubits) into ``static qubits" \cite{GisinRMP,Childress1,Childress2,Tanzilli05}.
Later, in the section on hybrid
schemes, we shall describe a scenario in which a single DV spin system (an atomic qubit)
is to be entangled with a CV qumode; so then the light-matter coupling is qualitatively
different, for instance, taking place and being enhanced in a cavity,
and a ``flying qumode" will mediate the entangling interaction between
two ``static qubits"; as a ``genuine" hybrid scheme
-- the CV qumode component will have to have high excitation numbers as opposed
to the single excitations of DLCZ -- such a scheme, though not being fully
unconditional, will contain an only moderate conditional element.

In summary, all those approaches discussed in this section
are still fairly demanding with current technology.
Especially, in the DV setting, with its highly conditional
en\-tangled-pair creation and connection protocols,
the minimal requirements on the necessary quantum memories \cite{Norbert,Collins,Brask}
are far from being met using state-of-the-art resources.
Even though a fully CV approach, as implemen\-ted in the Polzik experiments,
is tempting, because of its unconditionalness for entanglement creation and swapping,
entanglement distillation has been shown to be impossible with only Gaussian states
and Gaussian operations \cite{Eisert02,Fiurasek02,Giedke02}.
So it seems there is always a price to pay when certain types of resources are
replaced by supposedly cheaper ones.
In the next section, we will now discuss the very recent
concept of hybrid quantum information processing, which
could be useful for both efficient quantum computation and communication.

\section{Hybrid approaches}

Let us recall our definition of a hybrid protocol.
We refer to a quantum information scheme as hybrid, whenever it is based
upon both discrete and continuous degrees of freedom for
manipulating and measuring the participating quantum subsystems.
In the quantum optical setting, this includes in particular those
approaches that utilize light for communication and employ matter
systems for storage (and processing) of quantum information \cite{recentLukinReview},
as the optical qumodes are most naturally represented by
their quantized position and momentum (amplitude and phase quadrature)
variables, whereas the atomic spins or
any two-level structures in a solid-state system provide the natural realization
of qubits. An important ingredient of such hybrid schemes may then be
a particularly intriguing form of entanglement --
hybrid entanglement, i.e., an inseparable state of two systems of different
dimensionality, for example, between a qubit and a qumode.

We shall also remind the reader of the
motivation for combining CV and DV approaches, besides the
``natural" motivation of representing and using hybrid light-matter systems,
as stated in the preceding paragraph.
CV Gaussian resources can be unconditionally prepared and Gaussian operations
are deterministic and (experimentally) efficient. Nonetheless, there are various,
highly advanced tasks which would require a non-Gaussian element:

\begin{itemize}
     \item quantum error detection and correction for qumodes are impossible in the Gaussian regime \cite{Eisert02,Fiurasek02,Giedke02,Niset09},
     \item universal quantum computation on qumodes is impossible in the Gaussian regime \cite{Lloyd99,Bartlett02}.
\end{itemize}

The non-Gaussian element may be provided in form of a DV measurement
such as photon counting. There are also a few simpler tasks which can be performed
better with some non-Gaussian element compared to a fully Gaussian approach,
for instance, quantum teleportation \cite{Cochrane00} or optimal cloning \cite{CerfKrueger05,Demkowicz}
of coherent states.

Similarly, (efficient) universal quantum computation on photonic qubits would depend on
some nonlinear element, either directly implemented through nonlinear optics
or induced by photon measurements, as discussed in detail in Sec.~2.3.
In addition, there are even supposedly simpler tasks which are impossible
using only quadratic interactions (linear transformations)
and standard DV measurements such as photon counting.
The prime example for this is a complete photonic Bell measurement
\cite{Luetkenhaus99,PvLLutkenhaus04,Raynal04}.

It is worth pointing out that the above restrictions and nogo results apply
even when linear elements and photon detectors are available that operate with 100\%
efficiency and reliability (i.e., fidelity). In other words,
the imposed constraints are of fundamental nature and cannot be resolved
by improving the experimental performance of the linear elements, for example,
by further increasing squeezing levels.

Incorporating both nonlinear resources and nonlinear operations
into an optical quantum information protocol would enable one, in principle,
to circumvent any of the above constraints. Such an approach,
unless weak nonlinearities are employed, will most likely be more expensive
than schemes that stick to either linear resources or linear operations.
There are now many proposals for optically implementing
quantum information protocols through a kind of hybrid approach.
Among other classifications, two possible categories for such hybrid schemes
are:

\begin{itemize}
     \item those based upon nonlinear resources using linear operations,
     \item those based on linear resources using nonlinear operations.
\end{itemize}

In the latter case, for instance, DV photon number measurements may be
applied to CV Gaussian resources. The former type of
implementations would utilize, for example, CV homodyne measurements
and apply them to DV photonic qubit or other non-Gaussian states.
Some of these approaches will be presented, with regard to hybrid quantum computing (Sec.~3.3),
in Secs.~3.3.3 and 3.3.4, after a discussion on qubit-into-qumode encodings
in Sec.~3.3.1 and hybrid Hamiltonians (Sec.~3.3.2). As an additional approach for
incorporating a nonlinear element into a quantum information protocol,
in Sec.~3.3.5, we shall give a brief description of schemes using
weakly nonlinear operations. Finally, we will turn to hybrid quantum communication
in Sec.~3.4.

The notion of hybrid entanglement and some of its
applications as well as qubit-qumode entanglement transfer will be discussed in Sec.~3.2.
As a start, however, let us give a short summary of some earlier
hybrid proposals and implementations and some more recent schemes.

\subsection{Overview}

The hybrid approaches may either aim at ``simpler" tasks
such as quantum state engineering and characterization,
or at the ultimate applications of universal quantum computing
and long-distance quantum communication. Here is a short overview
on theory and experiments.

{\it Realizing POVMs:}\\
theory: optimal unambiguous state discrimination of binary coherent
qumode states $|\pm\alpha\rangle$ using a 50:50 beam splitter, a coherent-state ancilla,
and a DV measurement discriminating between zero and nonzero photons (on/off detector)
\cite{Banaszek99} (see
%Fig.~\ref{fig7}
Fig.~2.2.3);\\
theory and experiment:
near-minimum error discrimination of $|\pm\alpha\rangle$ using displaced
on/off detectors beating the optimal Gaussian homodyne-based discrimination scheme \cite{Wittmann}.

{\it Quantum state engineering:}\\
generation of
Schr{\"o}dinger-cat (coherent-state superposition, CSS) states with a ``size" of $|\alpha|^2 \sim 1.0 - 2.6$;
from Gaussian squeezed vacuum through DV photon subtraction using beam splitters and photon detectors,
theory \cite{Dak97} and experiment \cite{Ourjoum06,Wakui07,Neergaard06,Takahashi09};
from squeezed vacuum and
one-photon (two-photon) Fock states through beam splitters,
CV homodyne detection, and
postselection yielding squeezed Fock states and approximate
odd (even) CSS states (see Fig.~3.1.1),
theory \cite{Lance06} and experiment \cite{Ourjoum07};
experiments: single-mode photon-added/subtracted coherent state \cite{Zavatta}
and thermal state \cite{Parigi}.

\begin{center}
  \begin{columnfigure}
    \centering%
    \includegraphics[width=80pt]{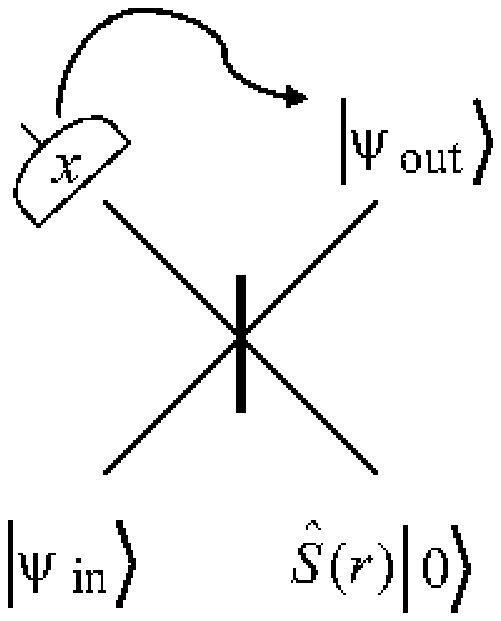}
  \figcaption{{\bf 3.1.1} Conditional preparation of an odd (even)
  CSS state using DV one-photon (two-photon) Fock states,
  $|\psi_{\rm in}\rangle = |1\rangle$ $\left(|\psi_{\rm in}\rangle = |2\rangle\right)$
  and CV squeezed vacuum resources, $\hat S(r) |0\rangle$,
  together with CV homodyne detection and postselection
  \cite{Lance06}.}\label{figJeongscheme}
  \end{columnfigure}
\end{center}

{\it Quantum state characterization:}\\
theory:
measurement of entanglement and squeezing
of Gaussian states through beam splitters and photon counting \cite{Fiurasek04};
measurement of Bell nonlocality of Gaussian
two-mode squeezed states via photon number
parity detection \cite{Banaszek01};\\
experiment:
homodyne tomography of one-photon state \cite{Lvo01},
homodyne tomography of two-photon state \cite{Ourjoum06PRL}.

{\it Quantum communication subroutines:}\\
theory:
entanglement concentration (EC) of pure Gaussian two-mode
squeezed states (TMSSs) through DV photon subtraction \cite{Opa00};
entanglement distillation (ED) of noisy Gaussian TMSSs using beam splitters and
on/off detectors \cite{Eis04} or
photon number QND-measurements \cite{Duan00a,Duan00b};\\
experiment: EC of pure TMSSs through nonlocal \cite{OurjoumGrangier}
and local \cite{TakahashiQPH} DV photon subtraction;
ED of Gaussian TMSSs
subject to non-Gaussian noise such as phase diffusion \cite{Hage08} and
random attenuation \cite{Dong08}.

{\it Quantum computational resources:}\\
theory:
conditional creation of non-Gaussian cubic
phase states from TMSSs
through DV photon counting, as a resource to implement
CV cubic phase gates (GKP, \cite{GKP01});
generalization of such cubic-gate schemes \cite{Gho06};
universal DV gates through weak Kerr-type
nonlinearities and strong CV Gaussian probes \cite{Nem04}.

More details on the GKP scheme will be presented in Sec.~3.2.
The pioneering ``hybrid" work is the experiment of Lvovsky {\it et al.}
in which CV, homodyne-based quantum tomography is
performed for the discrete one-photon Fock state \cite{Lvo01}.
The reconstructed Wigner function in this experiment has a strongly
non-Gaussian shape including negative values around the origin in phase space.

According to an even earlier, theoretical proposal,
CV quantum teleportation \cite{SamKimble} is applied
to DV entangled states of polarization-encoded photonic qubits \cite{Polkinghorne},
transferring DV Bell-type nonlocality through CV homodyne detection
and optimized displacements in phase-space for feedforward (``gain tuning").
Optical CV quantum teleportation of DV photonic states was further explored
by Ide {\it et al.} \cite{IdeHofmann01,IdeHofmann02}, combining CV tools such as gain tuning
with postselection, an ingredient inherited from the conditional DV approaches.

It is important to notice that for an experimental implementation
of a hybrid scheme in which DV and CV techniques and resources are to be combined
(for example, for CV quantum teleportation of DV states),
the standard way of applying such methods has to be generalized.
In particular, frequency-resolved homodyne detection,
as used, for instance, in CV quantum teleportation of coherent states
\cite{Furusawa98}, must be extended to time-resolved homodyning
\cite{LvovskyRMP} in order to synchronize the CV operations with
DV photon counting events. \footnote{this extension means
that, for instance, measurements on quantum states in the weak-excitation regime
are no longer restricted to time-resolved photon counting only (like in the
DLCZ repeater proposal), but would include time-resolved homodyning;
conversely, an extension from frequency-resolved homodyning to
frequency-resolved photon counting could be considered
as well \cite{RalphHuntingtoon08}.}
CV operations must act on a faster scale:
while the standard CV experiments used single-mode cw light sources
with narrow sidebands of $\sim 30$ kHz, the new generation of hybrid
experiments relies upon bandwidths of at least $\sim 100$ MHz,
corresponding to time scales of $\sim 100$ ns \cite{Akiraprivate}.

A beautiful example of a typical hybrid scheme according to our definition
is the ``offline squeezing" protocol from Ref.~\cite{Lance06} for
quantum state engineering, experimentally demonstrated in Ref.~\cite{Ourjoum07},
see Fig.~3.1.1. In this scheme, approximate CSS states are built
using linear CV measurements with outcomes within
a finite postselection window, linear CV squeezed-state, and nonlinear DV Fock-state
resources. The protocol works by squeezing the input Fock state, e.g.,
$|1\rangle \rightarrow \hat S(r)|1\rangle$, which corresponds approximately
to an odd CSS state ($\propto |\alpha\rangle - |-\alpha\rangle$)
and would be hard to achieve
``online" using the standard squeezing techniques such as optical parametric
amplification. \footnote{in the experiment of Ref.~\cite{Ourjoum07},
a two-photon state $|n=2\rangle$ was simply split at a beam splitter;
so the squeezed vacuum in Fig.~3.1.1 was just a vacuum state.
Postselection through time-resolved homodyne detection led to
an output CSS state which was squeezed by 3.5 dB. Theoretically,
the fidelity of the CSS state would approach unity for input Fock states
$|n\rangle$ with $n\to\infty$ \cite{Ourjoum07}.}
Though postselection renders the protocol probabilistic,
it enables one to preserve the non-Gaussianity of the input Fock state.
This is different from the non-hybrid ``offline squeezing" approach
of Refs.~\cite{Filip,Lam}, where postselection is replaced by a continuous
feedforward operation such that ``offline squeezing" is applied
in a fully Gaussian, deterministic fashion. Later we shall discuss
how to realize arbitrary squeezing operations on an arbitrary
input state through deterministic, homodyne-based cluster computation.

\begin{center}
  \begin{columnfigure}
    \centering%
    \includegraphics[width=220pt]{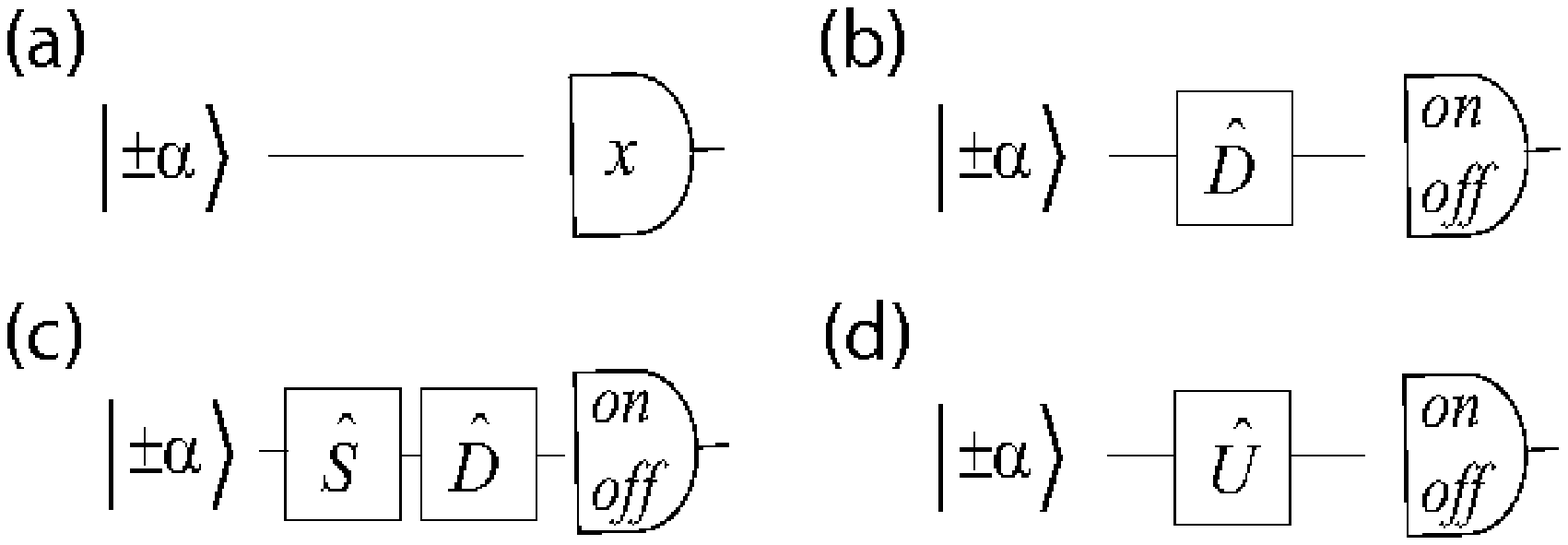}
  \figcaption{{\bf 3.1.2} Approximate discrimination of binary coherent states
  $\{|\pm \alpha\rangle\}$. Complementary to the probabilistic, error-free USD scheme (Fig.~2.2.3),
  an appropriate single-mode transformation [(b) optimal displacement $\hat D$, (c) optimal squeezing $\hat S$
  and displacement $\hat D$, (d) optimal nonlinear non-Gaussian transformation $\hat U$]
  in front of an on/off photon detector achieves close-to-optimal (b), even closer-to-optimal
  (c), and optimal (minimum-error) discrimination (d),
  and would always beat the optimal CV receiver solely based upon homodyne detection (a) \cite{Takeoka08}.
  }\label{figTakeokascheme}
  \end{columnfigure}
\end{center}

In the non-hybrid, Gaussian CV regime, it is known how useful
largely (offline) squeezed states are for engineering all kinds of multi-party
entangled states \cite{BraunsteinPvL05} including arbitrary CV graph states \cite{PvL07}.
On the other hand, typically, squeezing was explicitly excluded from
the toolbox for DV (linear-optical) quantum information processing \cite{Kok07}.
One important aspect of the more recent hybrid approaches is that
squeezing is no longer considered a resource solely for CV protocols.
The above quantum-state-engineering schemes may serve as examples for this.

Besides quantum state engineering, however, there are other tasks
in quantum information that may benefit from the use of squeezing,
especially when the squeezing transformation can be applied online
to an arbitrary state at any time during a quantum protocol. One example for this
is the near-minimum error discrimination of binary coherent-state signals
\cite{Takeoka08}, a protocol complementary to the error-free USD
as depicted in Fig.~2.2.3. In this case, squeezing is needed to obtain
the optimal Gaussian transformation that, in combination with DV
photon detection, leads to a near-optimal state discrimination,
\footnote{The actual optimal (i.e., minimum-error) discrimination
would correspond to a projection onto a CSS basis. This so-called
Helstrom bound \cite{Helstrom} is attainable by replacing the
Gaussian transformation in front of the photon detectors by a non-Gaussian one
\cite{Sasaki}, see Fig.~3.1.2. Remarkably, this highly nonlinear, optimal
measurement can also be achieved by only photon detection and real-time
quantum feedback \cite{Dolinar,Cook}.}
see Fig.~3.1.2.

The bottom line of the discussion here is that squeezing added as an online tool
to the standard linear-optics toolbox may be of great benefit,
beyond the more conventional quantum-state engineering schemes.
Further examples are given later in the section on hybrid quantum computing,
where squeezing is a necessary correction operation in order to implement
nonlinear quantum gates in a measurement-based fashion. Online squeezing
could also be used to ``unsqueeze" the squeezed CSS state emerging in the experiment of Ref.~\cite{Ourjoum07}.

Finally, concluding this overview of hybrid proposals and experiments,
let us at least mention the ultimate
application of CSS states for fault-tolerant, universal quantum computation
presented in Ref.~\cite{LundRalphCat}
and discuss yet another way to create such states \cite{Savage}.

In Section 2.3, we started using the qubit Pauli operator basis $X$, $Y$, and $Z$
as elementary gates, and rotations along their respective axes, $Z_{\theta}$, etc.,
to describe and realize arbitrary single-qubit unitaries. In analogy, we used
a similar notation for the qumode WH (displacement) operator basis, $X(\tau)$ and $Z(\tau)$.
Here, in the hybrid context, we shall exploit interactions and operations involving
combinations of DV qubit and CV qumode operators and therefore we prefer
to use unambiguous notations: for qumodes, still $X(\tau)=\exp(-2 i \tau \hat p)$
and $Z(\tau) = \exp(2 i \tau \hat x)$ for the WH group elements,
and $\hat x$ and $\hat p$ for the Lie group generators with
$\hat a = \hat x + i \hat p$; for qubits, now
$\sigma_x\equiv X$, $\sigma_y\equiv Y$, and $\sigma_z\equiv Z$
for the Pauli basis.
Now look at the effective interaction obtainable
from the fundamental Jaynes-Cummings Hamiltonian,
$\hbar g (\hat a^{\dagger} \sigma_-  + \hat a \sigma_+)$, in the
dispersive limit \cite{Schleich},
\begin{equation}
\label{Hint} \hat H_{\rm int} = \hbar \chi \sigma_z \hat a^{\dagger} \hat a\,.
\end{equation}
Here, $\hat a$ ($\hat a^{\dagger}$) refers to the annihilation (creation)
operator of the electromagnetic field qumode in a cavity and
$\sigma_z = |0\rangle\langle 0| - |1\rangle\langle 1|$ is the
corresponding Pauli operator for a two-level atom in the
cavity (with ground state $|0\rangle$ and excited state
$|1\rangle$).
The atomic system may as well be an effective
two-level system with an auxiliary level (a $\Lambda$-system),
as described earlier for the DLCZ quantum repeater.
However, this time we do not consider the collective spin
of many atoms, but rather a single electronic spin for a single atom
(in a cavity).
The operators $\sigma_+$ and ($\sigma_-$)
are the raising (lowering) operators of the qubit.
The atom-light coupling strength
is determined by the parameter $\chi=g^2/\Delta$, where $2g$ is
the vacuum Rabi splitting for the dipole transition and $\Delta$
is the detuning between the dipole transition and the cavity
field. The Hamiltonian in Eq.~(\ref{Hint}) generates a controlled
phase-rotation of the field mode depending on the state of the
atomic qubit. This can be written as ($\theta = \chi t$)
\begin{equation}\label{controlledrotation}
\hat R(\theta\sigma_z) = \exp(-i\theta\sigma_z\hat a^\dagger \hat a) \;,
\end{equation}
which, compared to Eq.~(\ref{rotation}), now describes a unitary operator
that acts in the combined Hilbert space of a single qubit and a single qumode.
We may apply this operator upon a qumode in a coherent state,
and may formally write
\begin{equation}\label{controlledcoherentstate2}
\hat R(\theta\sigma_z)|\alpha\rangle =
|\alpha \exp(-i\theta\sigma_z)\rangle \;.
\end{equation}
Compared to the uncontrolled rotation in Eq.~(\ref{coherentstate2}),
this time the qumode acquires a phase rotation depending on the state of the qubit,
see Fig.~3.1.3. As the eigenvalues of $\sigma_z$ are $\pm 1$,
applying $\hat R(\theta\sigma_z)$ to the initial qubit-qumode state $|\alpha\rangle\otimes
\left(|0\rangle + |1\rangle\right)/\sqrt{2}$ results in
\begin{eqnarray}\label{hybridentstate}
&&|\alpha \exp(-i\theta\sigma_z)\rangle \otimes \left(|0\rangle + |1\rangle\right)/\sqrt{2}\nonumber\\
&&\quad\quad\quad\quad = \left( |\alpha e^{-i\theta}\rangle |0\rangle +
|\alpha e^{i\theta}\rangle |1\rangle \right)/\sqrt{2} \;,
\end{eqnarray}
a hybrid entangled state between the qubit and the qumode.

The observation that this hybrid entangled state can be used for creating a macroscopic
superposition state of a qumode, a CSS state, by measuring the microscopic system,
the qubit, is about 20 years old \cite{Savage}. A suitable measurement
is a projection onto the conjugate $\sigma_x$ qubit basis, $\{|\pm\rangle\}$,
equivalent to a Hadamard gate applied to the qubit,
\begin{eqnarray}\label{hybridcatgeneration}
\left( |\alpha e^{-i\theta}\rangle |+\rangle +
|\alpha e^{i\theta}\rangle |-\rangle \right)/\sqrt{2} \;,
\end{eqnarray}
followed by a qubit computational $\sigma_z$ measurement. Depending on the result,
we obtain $\propto |\alpha e^{-i\theta}\rangle  \pm
|\alpha e^{i\theta}\rangle$ for the qumode. The size of this CSS state
depends on the distance between the rotated states in phase space, see Fig.~3.1.3,
scaling as $\sim\alpha\theta$ for typically small $\theta$ values. However, sufficiently
large initial amplitudes $\alpha$ still lead to arbitrarily ``large" CSS states
(at the same time increasing their vulnerability against photon losses).

\begin{center}
  \begin{columnfigure}\label{figContrRotation}
    \centering%
    \includegraphics[width=100pt]{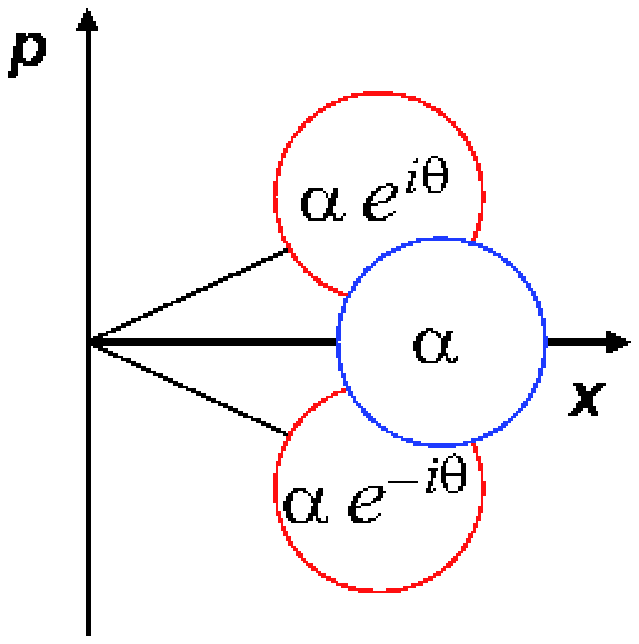}
  \figcaption{{\bf 3.1.3} Controlled phase rotation of a qumode in a coherent state,
  $\alpha$ real. Depending on the qubit state, $\sigma_z =\pm 1$,
  the phase angle of the controlled rotation will be $\mp\theta$.
  When the qubit starts in a superposition state, $\propto|0\rangle + |1\rangle$,
  we obtain a hybrid entangled state between the qubit and the qumode.}
  \end{columnfigure}
\end{center}

This is a manifestation of a weak nonlinearity which is effectively
enhanced through a sufficiently intense light field.
Although typically the ``single-atom dispersion" itself is rather weak with small phase
angles of at most $\theta \sim 10^{-2}$, it does not require strong
coupling; in a CQED setting, the only requirement is a sufficiently large
cooperativity parameter \cite{Ladd2006}.
In the following sections, we shall also consider using weak nonlinearities
for certain quantum gate constructions and, more generally,
sets of hybrid operations such as controlled displacements and rotations
sufficient for universal quantum computing.

Essential part of any such protocol is the generation of hybrid entanglement,
as expressed by Eq.~(\ref{hybridentstate}). Then measuring the qubit appropriately
yields macroscopic CSS states, as described; measurements on the qumode may be useful
for creating entangled qubit memory pairs in quantum communication,
and, for the ultimate application, no measurements at all may still give
universal quantum gates, as shown later.
Let us now look in a little more detail at the notion of hybrid entanglement.

\subsection{Hybrid entanglement}

In the preceding section, we encountered the example
of an entangled state between a qubit and a qumode.
This state, though defined in a combined qubit-qumode,
hence infinite-dimensional Hilbert space,
can be formally written in a two-qubit Hilbert space,
as we shall explain in this section.
Thus, the entanglement of this state can be conveniently quantified.

Besides entanglement measures for quantifying entanglement,
entanglement qualifiers, so-called entanglement witnesses,
have become useful tools for delineating inseparable states.
Such witnesses are well-known for CV Gaussian states \cite{Duaninsep,Simoninsep}
as well as for DV density operators \cite{HorodeckiWitness}.
Optically encoded quantum information, however, may
comprise DV photonic qubit states in a Fock subspace
or general non-Gaussian entangled states.
Using the partial transposition inseparability criteria \cite{Peres,HorodeckiWitness},
Shchukin and Vogel demonstrated that certain inseparability criteria for states
living in a physical bipartite space of two qumodes can be
unified under one umbrella in terms of a hierarchy of conditions for
all moments of the mode operators \cite{ShukinVogel}
(their results were further refined by Miranowicz and Piani \cite{Piani}).
These general conditions then include the previously known criteria
expressed in terms of second moments \cite{Duaninsep,Simoninsep} as a special case,
even those believed to be independent of partial transposition \cite{Duaninsep}.
For non-Gaussian entangled states, the second-moment criteria
would typically fail to detect entanglement, and a possible
entanglement witness would have to incorporate higher-order moments.

Another interesting aspect when comparing qubit and qumode entangled states is
whether the readily available, naturally given amounts of potentially unbounded CV entanglement
can be transferred onto DV qubit systems \cite{Son02,CiracKrausPRL}.
In the following subsection, we shall first consider qubit-qumode hybrid entangled states,
including a discussion on how to witness and possibly quantify
their entanglement. Further, we will devote another subsection to
the topic of transferring entanglement between qubit and qumode systems.

\subsubsection{Qubit-qumode entangled states}

Consider the following bipartite state,
\begin{equation}\label{generalhybridentstate}
\left(|\psi_0\rangle |0\rangle + |\psi_1\rangle |1\rangle\right)/\sqrt{2} \;,
\end{equation}
with an orthogonal qubit basis $\{|0\rangle,|1\rangle\}$ and a pair of
linearly independent qumode states $|\psi_0\rangle$ and $|\psi_1\rangle$. A specific
example of such a state and a possible way to build it
was presented in Eq.~(\ref{hybridentstate}) and the preceding discussion.

Clearly, the state in Eq.~(\ref{generalhybridentstate}) becomes
a maximally entangled, effective two-qubit Bell state when
$\langle\psi_0|\psi_1\rangle\to 0$. For $0<|\langle\psi_0|\psi_1\rangle|<1$,
the state is nonmaximally entangled, but still can be expressed
effectively as a two-qubit state. This can be seen by writing the two
pure, nonorthogonal qumode states in an orthogonal, two-dimensional basis,
$\{|u\rangle,|v\rangle\}$,
\begin{eqnarray}\label{qumodebasis1}
|\psi_0\rangle&=&\mu|u\rangle+
\nu|v\rangle,\nonumber\\
|\psi_1\rangle&=&(\mu|u\rangle -
\nu|v\rangle)
\,e^{i\phi}\,,
\end{eqnarray}
where $\nu=\sqrt{1-\mu^2}$ with
$\mu = \left[1+e^{-i\phi}\langle\psi_0|\psi_1\rangle \right]^{1/2}/\sqrt{2}$.
Then, using this orthogonal basis,
\begin{eqnarray}\label{qumodebasis2}
|u\rangle&=&(|\psi_0\rangle+e^{-i\phi}
|\psi_1\rangle)/(2\mu),\nonumber\\
|v\rangle&=&(|\psi_0\rangle-e^{-i\phi}
|\psi_1\rangle)/(2\nu)\,,
\end{eqnarray}
the hybrid entangled state of Eq.~(\ref{generalhybridentstate})
becomes
\begin{equation}\label{generalhybridentstate2}
\mu|u\rangle|0\rangle+
\sqrt{1-\mu^2}|v\rangle|1\rangle
 \;,
\end{equation}
a nonmaximally entangled two-qubit state with
Schmidt coefficients $\mu$ and $\sqrt{1-\mu^2}$,
where 1 ebit is obtained only for $\mu\to 1/\sqrt{2}$ and
$\langle\psi_0|\psi_1\rangle\to 0$.
Quantifying the entanglement is straightforward,
as the entropy of the reduced density matrix
is a function of the Schmidt coefficients.

It is interesting to compare the state
of Eq.~(\ref{generalhybridentstate}) with a bipartite qumode-qumode
entangled state of the form
\begin{equation}\label{qumodequmodeentstate}
\left(|\psi_0\rangle |\psi_0\rangle \pm |\psi_1\rangle |\psi_1\rangle\right)/\sqrt{N_\pm} \;,
\end{equation}
assuming the overlap $\langle\psi_0|\psi_1\rangle$ is real \cite{Wang02}.
First of all, in this case, a normalization constant $N_\pm$ is needed,
depending on $\langle\psi_0|\psi_1\rangle$.
Secondly, and quite remarkably, such a state may always represent
a maximally entangled two-qubit state (in the subspaces spanned by
$|\psi_0\rangle$ and $|\psi_1\rangle$),
independent of $\langle\psi_0|\psi_1\rangle$,
but depending on the relative phase \cite{Wang02,vanEnk01,Hirota01}, i.e.,
the sign in Eq.~(\ref{qumodequmodeentstate}).

The prime example for such qumode-qumode entangled states are two of the
so-called quasi-Bell states,
\begin{equation}\label{quasiBellstates}
|\Psi^\pm\rangle\equiv
\left(|\alpha\rangle |\alpha\rangle \pm |-\alpha\rangle |-\alpha\rangle\right)/\sqrt{N_\pm} \;,
\end{equation}
with $N_\pm\equiv 2 \pm 2 e^{-4 |\alpha|^2}$. The state $|\Psi^-\rangle$
is identical to the two-qubit Bell state $(|u\rangle |v\rangle + |v\rangle |u\rangle)/\sqrt{2}$
when $2 \mu = \sqrt{2 + 2 e^{-2|\alpha|^2}}$, $2 \nu = \sqrt{2 - 2 e^{-2|\alpha|^2}}$,
and $N_- = 8 \mu^2 \nu^2$,
which is maximally entangled with exactly 1 ebit of entanglement for {\it any} $\alpha\neq 0$.
In contrast, the state $|\Psi^+\rangle$ only equals the one-ebit Bell state
$(|u\rangle |u\rangle + |v\rangle |v\rangle)/\sqrt{2}$ in the limit of orthogonal
coherent states, $\langle\alpha|-\alpha\rangle\to 0$ for $\alpha\to\infty$.\footnote{similarly,
for the other two (quasi-)Bell states, we then have
$|\alpha\rangle |-\alpha\rangle - |-\alpha\rangle |\alpha\rangle\propto
|u\rangle |v\rangle - |v\rangle |u\rangle$ for {\it any} value of $\alpha$, but
$|\alpha\rangle |-\alpha\rangle + |-\alpha\rangle |\alpha\rangle\propto
|u\rangle |u\rangle - |v\rangle |v\rangle$ only when $\alpha\to\infty$.}

The amount of entanglement in the qubit-qumode and qumode-qumode states of
Eqs.~(\ref{generalhybridentstate}) and (\ref{qumodequmodeentstate}), respectively,
is bounded above by one ebit, corresponding to a maximally entangled
two-qubit Bell state. This is different from a ``genuine" CV qumode-qumode entangled state
such as a Gaussian two-mode squeezed state, which contains an arbitrary amount
of entanglement for sufficiently high levels of squeezing, see
Fig.~3.2.1 in the next section.\footnote{the entropy
of the reduced density matrix of a two-mode squeezed state, i.e., its entanglement
would exceed one ebit
at about 4.5 dB squeezing. Thus, the currently available squeezing levels of about 10 dB
\cite{Mehmet,Vahlbruch,Takeno}, corresponding to about 3 ebits (see Fig.~3.2.1),
would easily suffice to outperform
those hybrid entangled states discussed in the present section.}

Let us further mention that the quantification of entanglement
of the hybrid states in
Eqs.~(\ref{generalhybridentstate}) and (\ref{qumodequmodeentstate})
(we may also refer to the latter as hybrid in the sense that
the two physical qumodes each effectively live in a two-dimension\-al logical subspace)
becomes more subtle, when they are mapped onto mixed states.
For example, a qumode could be
subject to an imperfect channel transmission, for instance,
in a lossy fiber. Then for the special case of
$|\psi_0\rangle$ and $|\psi_1\rangle$ being coherent states do we still
obtain qubit-like expressions, as the coherent states themselves remain
pure under amplitude damping (and the resulting mixed
states would be expressible and hence quantifiable as two-qubit density operators
in the orthogonal $\{|u\rangle,|v\rangle\}$-basis, see, for example,
Ref.~\cite{pvlhybridrepeater08}). In this case,
the density operator decoheres faster for larger amplitudes $\alpha$;
an effect which will become important later in the hybrid quantum communication schemes.

Finally, it is worth pointing out that lower bounds on the entanglement of
(pure or mixed) non-Gaussian entangled states such as those
pure-state examples in Eq.~(\ref{quasiBellstates}) can be derived
from the standard measures for Gaussian entanglement \cite{EisertPlenioReview,AdessoReview}
by simply calculating
the entanglement for the Gaussian state with the same second-mo\-ment correlation matrix
as the non-Gaussian state given; in other words, {\it for a given
correlation matrix, the corresponding Gaussian-state entanglement provides a
conservative and hence safe estimate on the actual entanglement of the state in question}
\cite{GiedkeExtrem}.\footnote{note that Gaussian states also provide an upper bound on the entanglement
when only the correlation matrix is known: for given energy, Gaussian states
maximize the entanglement \cite{vanEnkEnergybound}.}

Apart from applying entanglement measures to hybrid and non-Gaussian
entangled states, it is sometimes enough to have a (theoretical and, in particular,
experimental) tool in order to decide whether a given state
is entangled or not. Such entanglement qualifying criteria are typically related
with certain observables (Hermitian operators) $\hat W$ for which the
expectation value ${\rm Tr}(\rho\hat W)$ is nonnegative for all separable states
$\rho$, whereas it may take on negative values for some inseparable states $\rho$.
Entanglement witnesses are well-known for CV Gaussian states \cite{Duaninsep,Simoninsep}
as well as for DV density operators \cite{HorodeckiWitness}.

Qumode-qumode entangled states like those in Eq.~(\ref{qumodequmodeentstate})
may be identified through the partial transposition criteria \cite{Peres,HorodeckiWitness}
adapted to the case of arbitrary CV states \cite{ShukinVogel,Piani}.
As a result, all known CV inseparability criteria, including those
especially intended for Gaussian states and
expressed in terms of second moments \cite{Duaninsep,Simoninsep}, can be
derived from a hierarchy of conditions for all moments of the mode operators
$\hat a$ and $\hat a^\dagger$. Moreover, for non-Gaussian entangled states,
for which the second-moment criteria
typically fail to detect entanglement, the higher-moment conditions
would work. The concept for these criteria is as follows.

It is known that for any positive operator $\hat P\geq 0$, we can write
$\hat P= \hat f^\dagger \hat f$ such that ${\rm Tr}(\rho\hat f^\dagger \hat f)$
is nonnegative for any operator $\hat f$ and any physical state $\rho$.
Then we may choose the bipartite decomposition
$\hat f = \sum_{ij} c_{ij} \hat A_i \otimes \hat B_j$, for which
\begin{eqnarray}\label{derivationShukin1}
0 &\leq&  {\rm Tr}(\rho\hat f^\dagger \hat f)\nonumber\\
&=& \sum_{ij,kl}c_{ij}^*{\rm Tr}(\rho\hat A_i^\dagger \hat A_k
\otimes \hat B_j^\dagger \hat B_l) c_{kl}\nonumber\\
&\equiv&  \sum_{ij,kl}c_{ij}^* \,[M(\rho)]_{ij,kl}\, c_{kl}\;,
\end{eqnarray}
for any coefficients $c_{ij}$. Hence the matrix $M(\rho)$ is positive-semidefinite
for any physical state $\rho$. Now any separable state $\rho$ remains
a physical state after partial transposition of either subsystem such that
$M(\rho^{T_A})$ and $M(\rho^{T_B})$ remain positive-semidefinite matrices,
where $T_A$ and $T_B$ denotes partial transposition for subsystem
$A$ and $B$, respectively. Then, negativity of $M(\rho^{T_A})$ or $M(\rho^{T_B})$,
and hence negativity of any subdeterminant of $M(\rho^{T_A})$ or $M(\rho^{T_B})$
is a sufficient criterion for entanglement, from which
sets of inequalities can be derived with convenient choices for the local
operators $\hat A_i$ and $\hat B_j$ \cite{ShukinVogel}.

One such choice for a qumode-qumode state
would be each qumode's position and momentum operators, eventually reproducing
Simon's criteria in terms of second-moment correlation matrices \cite{Simoninsep}.
Another choice, adap\-ted to a qubit-qumode state of the form in
Eq.~(\ref{generalhybridentstate}), is given by
$\{\hat A_i\} = \{|\phi\rangle\langle 0|,|\phi\rangle\langle 1|\}$
and $\{\hat B_j\} = \{\one,\hat x,\hat p \}$, with some generic qubit
state $|\phi\rangle$ \cite{HaukeNorbert08}.
The resulting expectation value matrix $M(\rho)$ then serves again,
using partial transposition, as a tool to detect entanglement,
this time for hybrid qubit-qumode states. This can be particularly useful
for verifying the presence of effective entanglement in a binary
coherent-state-based quantum key distribution protocol \cite{RigasNorbert06,Leuchs09}
as a necessary security requirement \cite{Curty04}.

The choices for the local
operators $\hat A_i$ and $\hat B_j$ discussed so far all lead
to second-moment conditions only. These are experimentally most accessible,
but may fail to detect some form of non-Gaussian entanglement.
For qu\-mode-qumode states,
a more general choice is the normally ordered form for $\hat f$,
$\hat f = \sum_{nmkl} c_{nmkl} \hat a^{\dagger\, n} \hat a^m
\hat b^{\dagger\, k} \hat b^l$, with mode operators $\hat a$ and $\hat b$
for the two qumodes $A$ and $B$, respectively. Inserting this into
${\rm Tr}(\rho^{T_A}\hat f^\dagger \hat f)\geq 0$ or
${\rm Tr}(\rho^{T_B}\hat f^\dagger \hat f)\geq 0$ yields a hierarchy of
separability conditions in terms of the moments of the mode operators.
For example, the quasi-Bell state $|\Psi^-\rangle$ of Eq.~(\ref{quasiBellstates})
leads to a subdeterminant of the matrix of moments
with a sufficient order of the moments such as $\langle \hat a^\dagger \hat a
\hat b^\dagger \hat b\rangle$ which becomes
$-\alpha^6 \frac{\coth(2 \alpha^2)}{\sinh^2(2 \alpha^2)}\equiv s$,
for real $\alpha$ \cite{ShukinVogel,Piani}.
This subdeterminant is negative for any nonzero $\alpha$, proving
the entanglement of the state $|\Psi^-\rangle$ for any $\alpha\neq 0$.\footnote{note
that $s \to 0$ for $\alpha\to\pm\infty$ and that $s$ is maximally
negative, $s\approx - 0.125$, for $|\alpha| < 0.4$, even though
we know that $|\Psi^-\rangle$ has constant entanglement of one ebit
for any nonzero $\alpha$.}

To conclude this section, let us summarize that it is straightforward
to quantify the entanglement of hybrid qu\-bit-qumode and non-Gaussian
qumode-qumode states, provided these states can be represented in two-dimensional
subspaces. Otherwise, in order to detect the inseparability
of such states, the partial transposition criteria expressed
in terms of matrices of moments can be used. In a recent experiment,
employing CV quantum information encoded into the spatial wavefunction
of single photons, fourth-order-moment entanglement was detected \cite{Walborn09}.

\subsubsection{Qubit-qumode entanglement transfer}

As the entanglement in Gaussian qumode states
is unconditionally available and, in principle, unbounded,
it is tem\-pting to consider protocols in which
the CV entanglement is transferred onto DV systems. Especially
the local memory nodes in a quantum repeater are typically
represented by atomic spin states (recall the discussion in Sec.~2.4).
So the unconditional
generation of CV entanglement and its efficient distribution between
two repeater stations should then be supplemented by
a transfer of the transmitted ebits (in form of
flying qumodes) onto the local, electronic or nuclear, storage
spins (in form of static qubits).

Consider the distribution of two-mode squeezed states,
$\sum_{n=0}^\infty \tanh^n r \,|n, n\rangle/\cosh r$,
with squeezing $r$. In principle, an amount of entanglement in ebits \cite{vanEnkQT},
\begin{eqnarray}
E(r) = \cosh^2 r \log_2(\cosh^2 r)\, -\, \sinh^2 r \log_2(\sinh^2 r)\,,\nonumber\\
\end{eqnarray}
could then be shared between the ends of the channel.
Realistically, of course, the entanglement distribution will be
subject to photon losses, leading to a highly degraded, shared
CV entanglement; though always a nonzero amount of entanglement remains,
unless additional thermal noise sources
are present along the channel \cite{Duaninsep}.
We shall take into account lossy channels in a later section. In this section,
let us neglect the imperfect channel transmission and consider ideal
CV entanglement distribution. In this case,
Fig.~3.2.1 shows how many ebits could be, in principle, transferred
from the CV to the DV system. Figure~3.2.2 illustrates the principle
of entanglement distribution and transfer.

\begin{center}
  \begin{columnfigure}\label{figTMSSinEbits}
    \centering%
    \includegraphics[width=190pt]{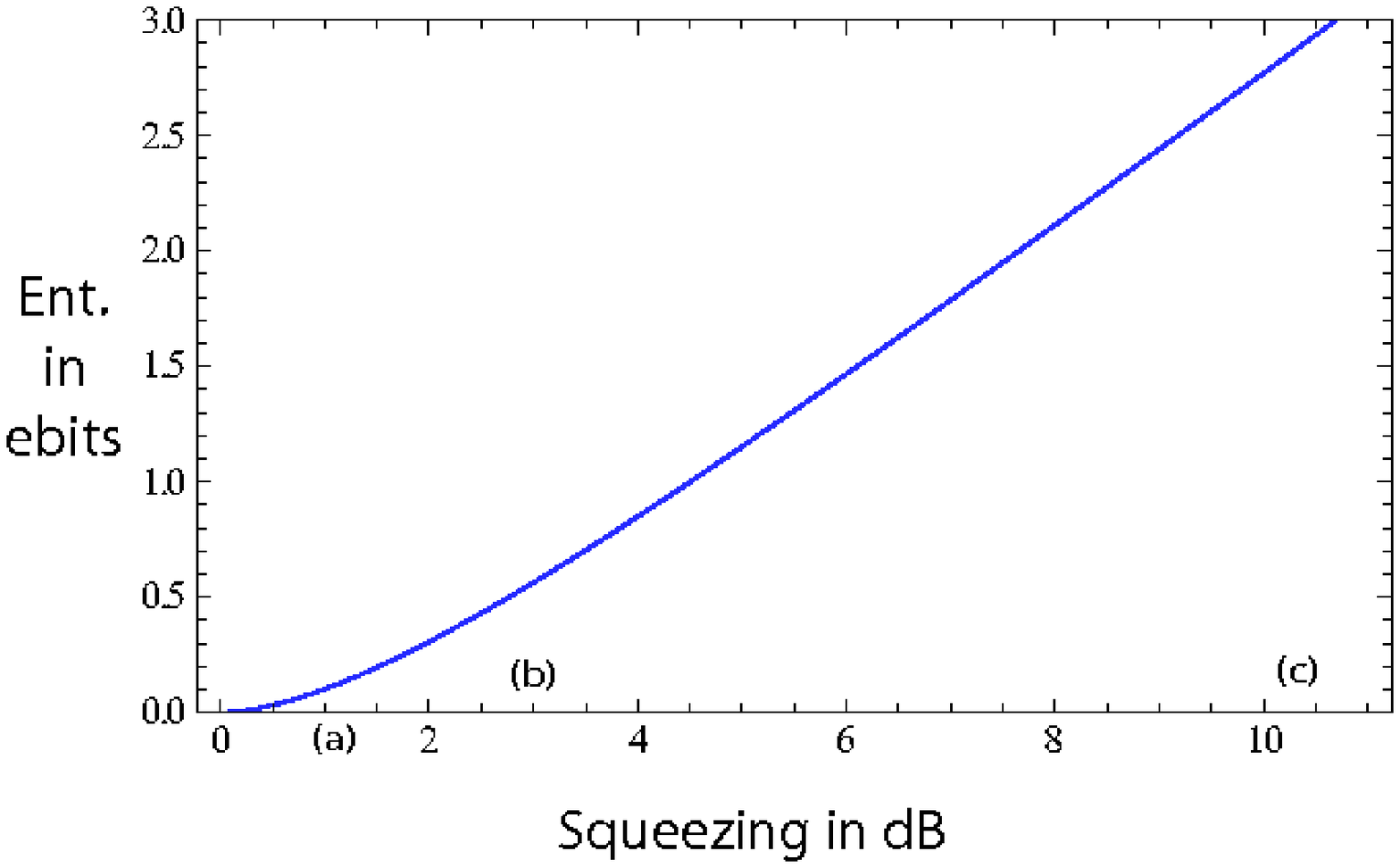}
  \figcaption{{\bf 3.2.1} Entanglement of a Gaussian two-qumode squeezed state
  in ebits (the units of maximally entangled two-qubit states) as a function
  of the squeezing values in dB. The letters indicate experiments with
  corresponding squeezing levels: (a) recent CV Shor-type quantum error correction
  scheme \cite{Aoki09},
  (b) first CV quantum teleportation \cite{Furusawa98}, and (c) most recent squeezing
  ``world records" \cite{Mehmet,Vahlbruch,Takeno}.}
  \end{columnfigure}
\end{center}

So what mechanism achieves this entanglement transfer?
Are there possibly fundamental complications owing
to the mismatch of the Hilbert spaces?\footnote{later,
for a hybrid quantum repeater, we can
avoid this mismatch from the beginning by distributing effective two-qubit
entanglement of the form of hybrid qubit-qumode states.}
These questions were addressed in Refs.\cite{SonJModOpt,KrausCirac,Paternostro2},
where it was shown that an entangled two-mode squeezed state may act as
a driving field to two remote cavities. When a two-level system is placed
in each cavity, the resonant Jaynes-Cumm\-ings Hamiltonian,
$\hbar g (\hat a^{\dagger} \sigma_-^{a}  + \hat a \sigma_+^{a})$,
can be used, describing the qubit-qumode coupling for qumode $\hat a$
and qubit $\sigma_-^{a}$,
and similarly for mode $\hat b$ and spin $\sigma_-^{b}$.
As the driving field is assumed to be an external,
broadband field, an additional interaction has to be included
that describes the desired coupling between the
external qumodes $\hat a_k e^{i \omega_k t}$
and the internal cavity qumode $\hat a$ (and similar for the internal
and external qumodes $\hat b$ and $\hat b_k e^{i \omega_k t}$, respectively),
$\sum_k \kappa_k [\hat a \hat a_k^\dagger e^{i (\omega_a - \omega_k) t}
+ \hat a^\dagger \hat a_k e^{-i (\omega_a - \omega_k) t} + a \leftrightarrow b]$
\cite{Paternostro2}.
Thus, $\kappa_k$ is the rate of this wanted coupling between each internal qumode
at frequency $\omega_{a,b}$
and a corresponding external driving qumode at frequency $\omega_k$.
In the weak-coupling regime, where $\kappa$ is much smaller
than the external bandwidth, the non-unitary dynamics of the
(internal) qumode-qubit systems is governed by a master equation.
The steady state of the cavities can then be shown to become
a two-mode squeezed state (in the bad-cavity regime using CQED language
\cite{Turchette}).

\begin{center}
  \begin{columnfigure}\label{figDistr}
    \centering%
    \includegraphics[width=225pt]{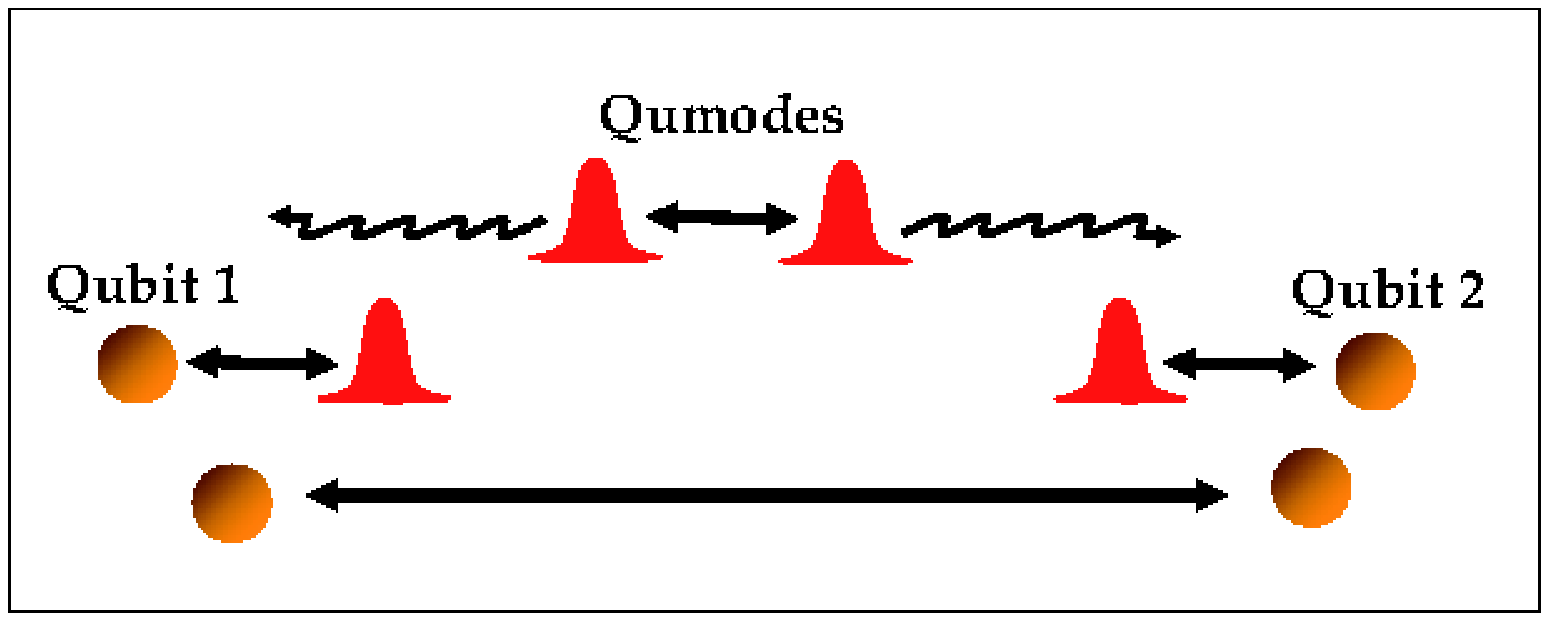}
  \figcaption{{\bf 3.2.2} Using a CV entanglement distributor for DV entanglement distribution.
  Two entangled qumode pulses travel to the opposite ends of a channel
  and interact with local atomic qubits placed, for instance, in a cavity.
  Through these interactions, the CV entanglement is dynamically transferred onto the two qubits.}
  \end{columnfigure}
\end{center}

Eventually, the CV entanglement is transferred
onto the DV systems in the steady state.
More realistically, additional dissipations have to be taken into account
such as spontaneous atomic decay. This kind of unwanted
in-out coupling will occur at a rate $\Gamma$. An important
parameter then is the so-called cooperativity, $C\propto 1/(\Gamma \kappa)$.
Only for sufficiently large $C$ do we obtain a nearly pure atomic
steady state \cite{Paternostro2}.\footnote{however,
the assumption of large $C$ may collide
with the bad-cavity/weak-coupling assumption \cite{Paternostro2,Turchette}.
Later, in the section on hybrid quantum communication,
the hybrid quantum repeater shall
also depend on sufficiently large $C$ \cite{Ladd2006}.}

The general conditions for transferring general CV entanglement
(including non-Gaussian entanglement of the form Eq.~(\ref{quasiBellstates}))
from CV driving fields
onto two qubits, both for the dynamical and the steady-state cases,
were presented in Ref.~\cite{Paternostro2}. Similar DV entanglement
generation schemes were proposed in Refs.~\cite{Benatti} and \cite{Braun}
using an indirect interaction of two remote qubits through
Markovian and non-Markovian environments, respectively.
Finally, let us mention the interesting concept
of transferring entanglement from a relativistic quantum field
in a vacuum state onto a pair of initially unentangled atoms
\cite{Reznik,ReznikCirac}.

Different from those schemes described in this section,
which are basically measurement-free and dynamical entanglement
transfer protocols utilizing an optical field as entanglement
distributor, later we shall describe how to exploit local measurements
on hybrid entangled states
in order to nonlocally prepare entangled qubit states.
The optical field will then act as a kind of quantum bus
mediating the interaction between the qubits.

\subsection{Hybrid quantum computing}

In this section, we shall now discuss various hybrid approaches
to quantum computing. This includes optical hybrid protocols
for models of universal quantum computation as well
as for certain gates from a universal gate set.
More specifically, for processing photonic quantum information,
we consider either using linear optical resources such as Gaussian
entangled cluster states and performing nonlinear operations
on them, or first creating offline nonlinear, non-Gaussian
resource states and applying linear operations such as homodyne detections.
Further, we discuss hybrid schemes for quantum computing
and universal quantum logic that do not require
any highly nonlinear resources or interactions: only
a weakly nonlinear element is needed \cite{Nemoto04}.

However, let us start this section by looking at a few proposals in which
either the encoding of quantum information \cite{GKP01,LundRalphCat} or the
unitary gate evolutions, i.e., the interaction Hamiltonians \cite{LloydHybrid},
are explicitly hybrid.

\subsubsection{Encoding qubits into qumodes}

There are various ways to encode a photonic qubit
into optical modes such as polarization or spatial modes,
as we discussed in the introductory part of this
article. In particular, the photon occupation number
in a single-rail, single-mode Fock state may serve
as a qubit or a more general DV basis. In dual-rail or,
more generally, multiple-rail encoding, a single photon
encoded into multi-mode states can be even
universally processed through linear optical elements;
though in an unscalable fashion, unless
complicated ancilla states and feedforward are employed \cite{KLM01}.

Another natural way to encode a logical DV state into a physical,
optical multi-mode state would be based upon at least two qumodes
and a constant number of photons distributed over the physical
qumodes such that, for example, a logical, $2J+1$-dimensional spin-$J$ state,
$|J,m_j\rangle$, $m_j = -J, -J+1,...,J-1, J$,
can be represented by two physical qumodes in the two-mode
Fock state $|J\equiv (n_1+n_2)/2,m_j\equiv (n_1-n_2)/2\rangle$, where
$n_1$ and $n_2$ denote the photon numbers of the two modes.\footnote{the choice
of constant total number $\hat n_1 + \hat n_2 \equiv \hat S_0$ and varying
number differences $\hat n_1 - \hat n_2 \equiv \hat S_3$ corresponds to
a specific basis in the so-called Schwinger representation. For example,
in order to faithfully represent the SU(2) algebra
by the Lie algebras of two infinite-dimensional oscillators, i.e., two qumodes
$\hat a_1$ and $\hat a_2$,
one may replace the usual Pauli matrices $\sigma_0\equiv\one$,
$\sigma_1\equiv\sigma_x$, $\sigma_2\equiv\sigma_y$, and
$\sigma_3\equiv\sigma_z$ by the so-called quantum Stokes operators
$\hat S_i = (\hat a_1^\dagger,\hat a_2^\dagger) \sigma_i (\hat a_1,\hat a_2)^{\rm T}$,
$i=0,1,2,3$, satisfying the SU(2) Lie algebra commutators
$[\hat S_1,\hat S_2] = 2 i \hat S_3$, while $[\hat S_0,\hat S_j] = 0$,
for $j=1,2,3$.}
In fact, for $n_1+n_2 = 1$, we obtain a dual-rail encoded
spin-$1/2$ qubit: $\{|J=1/2,m_j=-1/2\rangle,|J=1/2,m_j=+1/2\rangle\}
\equiv\{|n_1=0,n_2=1\rangle,|n_1=1,n_2=0\rangle\}$. Similarly,
a spin-$1$ qutrit,
$\{|J=1,m_j=-1\rangle,|J=1,m_j=0\rangle,|J=1,m_j=1\rangle\}$ corresponds to
$\{|0,2\rangle,|1,1\rangle,|2,0\rangle\}$ in the Fock basis.

These natural encodings may even automatically provide some
resilience against certain errors such as photon losses.
For instance, the dual-rail encoding serves as an error detection code \cite{KLM01,Nielsen},
and multi-photon states subject to losses may also keep high degrees of entanglement \cite{Durkin}.
However, for the purpose of fault-tolerant, universal quantum information
processing, other encodings could be preferable, though they might be much harder to realize.
One example is the fault-tolerant quantum computation proposal based upon
CSS states such as the qubit-type states $(a|\alpha\rangle + b|-\alpha\rangle)/\sqrt{N}$,
$N=|a|^2 + |b|^2 + 2 e^{-2 |\alpha|^2} {\rm Re}(a b^*)$ \cite{LundRalphCat}.
Although some universal DV gates such as the Hadamard gate
(possibly using auxiliary hybrid entangled states of the form
in Eq.~(\ref{quasiBellstates}) for a teleporta\-tion-based realization)
are hard to implement for this encoding,
the effect of photon losses on these CSS-type states corresponds to
random phase flips in the coherent-state basis such that repetition codes
known from DV qubit quantum error correction can be directly applied \cite{Glancy}.
None\-theless, these codes would still require Hadamard gates for encoding and decoding.

\begin{center}
  \begin{columnfigure}\label{fig12}
    \centering%
    \includegraphics[width=130pt]{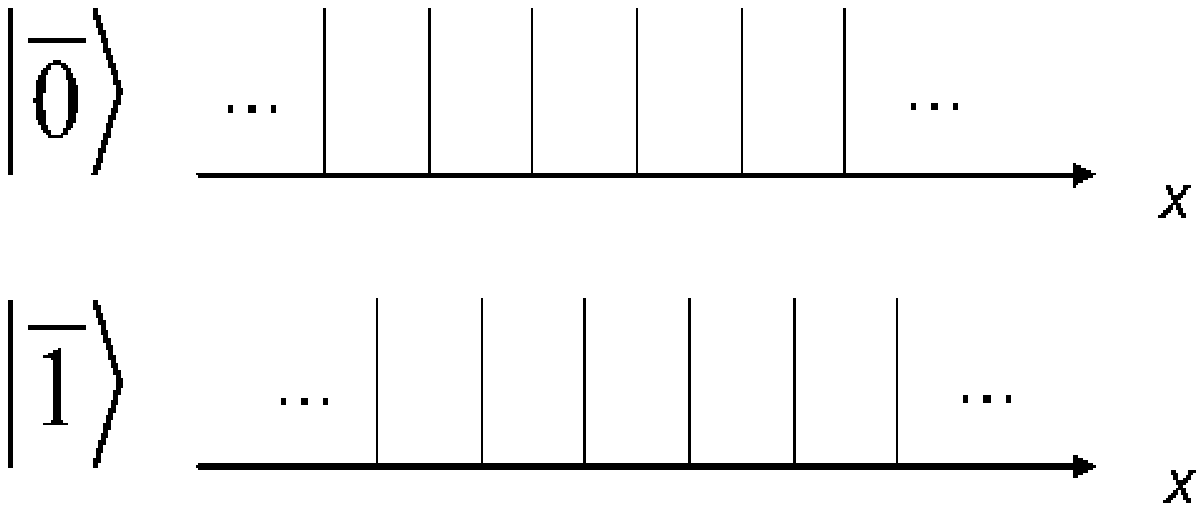}
  \figcaption{{\bf 3.3.1} Encoding a qubit into a qumode according to GKP \cite{GKP01}.
  The logical, computational qubit basis
  states, with $\bar Z |\bar k\rangle = (-1)^k |\bar k\rangle$, $k=0,1$,
  are infinite superpositions of delta peaks in position space.
  This encoding is particularly suited to protect the qubit against small, diffusive
  shift errors as arising from, for instance, weak amplitude damping of the qumode.}
  \end{columnfigure}
\end{center}

Another well-known proposal for fault-tolerant, universal quantum computation
on logical qubits encoded into a physical qumode space is the GKP scheme \cite{GKP01}.
This type of encoding (see Fig.~3.3.1) is primarily intended to
protect a logical qubit against small errors such as small shifts of the physical qumode
in phase space. Direct translations of standard qubit quantum error correction
codes against rarely occurring Pauli errors into the CV regime
\cite{LloydSlotine,Braunstein1,Braunstein2}
fail to provide any protection against such small, diffusive, typically Gaussian errors
\cite{Niset09}. Provided the error shifts are sufficiently small, i.e., smaller than
$\Delta/4$ with $\Delta$ the distance between the delta peaks in Fig.~3.3.1,
the encoded states remain sufficiently intact, and
suitable syndrome measurements enable one, in principle, to detect the errors
and correct the states.

Besides having this special error correction capability, the GKP scheme also
includes universal gate operations. The link between logical and physical gate operations
can be easily understood from Fig.~3.3.1.\footnote{in Fig.~3.3.1, we do
not show the logical eigenstates of the logical $\bar X$ Pauli operator. Similar to
the logical eigenstates of the logical $\bar Z$ Pauli operator, which are superpositions
of delta peaks along the position axis as shown,
the $\bar X$ eigenstates are delta-peak superpositions
along the momentum axis in the physical qumode's phase space. The small shift errors
can then be shifts along $x$ and $p$, and as these WH errors form a basis,
{\it any} sufficiently small error may be corrected. This reasoning is similar to that for
the standard CV qumode codes \cite{LloydSlotine,Braunstein1,Braunstein2} against
sufficiently large, stochastic shift errors in a single channel \cite{PvLNoteQEC}.}
A logical Pauli operator $U\in\{\bar X, \bar Y, \bar Z\}$
is transformed under conjugation by a logical qubit Clifford unitary
$C$ into a different logical Pauli operator, $C^\dagger U C = U'\in\{\bar X, \bar Y, \bar Z\}$.
Thus, as the logical Pauli gates correspond to WH shifts of the physical qumode,
any physical Gaussian operation, since it preserves such WH gates under conjugation, can only
lead to a logical Clifford operation. Therefore, {\it a physical non-Gaussian operation is needed
in order to achieve logical non-Clifford qubit gates and hence DV qubit universality.}
GKP demonstrate two such non-Clifford gates
of which one uses a controlled rotation through a
dispersive atom-light interaction as in Eq.~(\ref{controlledrotation}),
whereas the other one is based upon a cubic phase gate on the physical qumode,
$D_3(\kappa_3) = \exp(i\kappa_3 \hat x^3)$. Recall that this gate can also be used to complete
the universal set for full CV universality in Eq.~(\ref{CVset}), in the spirit of Ref.~\cite{Lloyd99}.

In optical quantum information processing, it is pretty natural to encode
DV quantum information into an appropriate subspace of the full infinite-dimensional
qumode space. However, in order to realize the most advanced optical quantum processors,
which achieve both universality and fault tolerance, the qubit-into-qumode encodings,
though conceptually highly interesting, may still be far from being implementable.
In this sense, the CSS-type encoding \cite{LundRalphCat} and the
position/momentum-eigenstate superposition encoding \`a la GKP \cite{GKP01}
are very similar. In both schemes, fault tolerance and universality require
complicated non-Gaussian operations or resources. Nonetheless, they do both incorporate
the necessary non-Gaussian quantum error correction steps into a DV qubit processor embedded in
the physical space of an optical qumode.

\subsubsection{Hybrid Hamiltonians}

Besides the qubit-into-qumode encodings of the preceding section,
another interesting hybrid approach is based upon
unitary evolutions, i.e., Hamiltonians which are hybrid.
These Hamiltonians contain DV qubit and CV qumode operator combinations such
as, for instance, the controlled rotation in Eq.~(\ref{controlledrotation})
corresponding to a dispersive light-matter interaction Hamiltonian.
More generally, we may consider a unitary gate of the form,
\begin{equation}
U = \exp[i \lambda f(\sigma_x,\sigma_z) \otimes g(\hat x,\hat p) ]\,,
\end{equation}
acting on the composite system of a qubit and a qumode.
In fact, in the context of combining the DV and CV approaches,
it has been pointed out \cite{wan01} that a suitable set of elementary
Hamiltonians, including the controlled rotations and additional
uncontrolled displacements,
\begin{equation}
\{\sigma_x \hat a^{\dagger} \hat a,\sigma_z
\hat a^{\dagger} \hat a,\hat x\}\,,
\end{equation}
is, in principle, sufficient for universal
quantum computation on qubits.\footnote{note that compared to
the discussion on universal sets for DV and CV quantum computation
in Sec.~2.3.1, we are now writing a universal set in terms of elementary
Hamiltonians instead of elementary gates.}
Even earlier Lloyd considered a universal set containing only
controlled displacements \cite{LloydHybrid},
\begin{equation}\label{lloydset}
\{\pm\sigma_x\hat x,\pm\sigma_z \hat x,
\pm\sigma_z \hat p\}\,.
\end{equation}
Typically, in quantum optics, controlled rotations are easier
to achieve than controlled displacements, as we started discussing
at the end of
Sec.~3.1 and shall exploit later \cite{vanloock07}.
Controlled displacements may be accessible
in other systems using ion traps \cite{mil99,mol99,sor00,mil00}
or squids \cite{Spiller}.

There are two ways to understand the universality of the above
Hamiltonian sets. One approach is based upon the decomposition,\footnote{using
the well-known Baker-Campbell-Hausdorff formula $e^{A}e^{B} = e^{A+B} e^{[A,B]/2}
+ {\rm O}([A,[A,B]],[[A,B],B])$ and so
$e^{\pm i H_2 t} e^{\pm i H_1 t} = e^{\pm i (H_1 + H_2) t} e^{-[H_2,H_1] t^2/2}
+ {\rm O}(t^3)$.}
\begin{equation}\label{lloydtrick}
e^{i H_2 t} e^{i H_1 t} e^{-i H_2 t} e^{-i H_1 t} =
e^{[H_1,H_2] t^2} + {\rm O}(t^3)\,.
\end{equation}
So by applying the Hamiltonians $H_1$ and $H_2$ for some short time,
we can also approximately implement the Ham\-iltonian $-i [H_1,H_2]$,
provided the interaction times are sufficiently short.\footnote{this is the same
asymptotic, approximate model for universal quantum computation
as it was used for discrete \cite{Lloyd95} and continuous \cite{Lloyd99}
variables on their own.}
Now first of all it can be shown that using Eq.~(\ref{lloydtrick})
and the elementary Hamiltonians of Eq.~(\ref{lloydset})
one can generate through commutation any single-qubit,
any single-qumode, as well as any qubit-qumode unitary \cite{LloydHybrid}.
The extra two-qubit and two-qumode entangling gates in order
to complete the universal sets for DV and CV universality, respectively,
are then achieved through the following commutators,
\begin{eqnarray}\label{comm1}
-i[\sigma_z^{(1)}\hat x,\sigma_z^{(2)}\hat p]
&=& \sigma_z^{(1)} \sigma_z^{(2)} /2\,,\\
\label{comm2}
-i[\sigma_z\hat x_1,\sigma_x\hat x_2] &=&
2 \,\sigma_y \,\hat x_1\, \hat x_2\,,
\end{eqnarray}
respectively. Here, the superscripts and subscripts denote
operators acting upon one of the two qubits or qumodes.
In other words, the $C_Z$-gates of the sets in
Eq.~(\ref{DVset}) and Eq.~(\ref{CVset}) can be enacted
{\it approximately} by applying some of the elementary Hamiltonians
in Eq.~(\ref{lloydset}). However, there is a crucial
difference between the above two commutators.
The commutator in Eq.~(\ref{comm1}) commutes with the
elementary Hamiltonians from which it is built,
whereas the commutator in Eq.~(\ref{comm2}) does not.
As a consequence, in the latter case, the decomposition formula
in Eq.~(\ref{lloydtrick}) is indeed only an approximation
that requires infinitesimally small interaction periods.
However, since $-i[\sigma_z^{(1)}\hat x,\sigma_z^{(2)}\hat p]$
commutes with $\sigma_z^{(1)}\hat x$ and $\sigma_z^{(2)}\hat p$,
and all higher-order commutators vanish as well,
the two-qubit $C_Z$-gate according to Eq.~(\ref{lloydtrick})
with Eq.~(\ref{comm1}) no longer depends on small
interaction times. Instead we obtain the {\it exact}
formula,
\begin{equation}\label{exact}
e^{i \sigma_z^{(2)}\hat p t} e^{i \sigma_z^{(1)}\hat x t}
e^{-i \sigma_z^{(2)}\hat p t} e^{-i \sigma_z^{(1)}\hat x t} =
e^{i \sigma_z^{(1)} \sigma_z^{(2)} t^2/2} \,.
\end{equation}
This observation leads us to an alternative way of understanding
how the hybrid gates of Eq.~(\ref{lloydset}) can be used
to achieve universal quantum computation on qubits.
In particular, a two-qubit entangling gate
of sufficient strength (i.e., $t^2\sim 2\pi$)
is then possible without direct interaction between the two qubits;
a single qumode subsequently interacting with each qubit
would rather mediate the qubit-qubit coupling -- as a kind
of {\it quantum bus} (so-called qubus \cite{Spiller,cirzol95},
see Fig.~3.3.2).
The sudden exactness of the gate sequence
can be explained by interpreting it as a controlled geometric phase gate
\cite{wan01,luis}.

So even though the original hybrid scheme of Lloyd \cite{LloydHybrid}
achieves universality using a finite gate set, it appears unrealistic
to switch between the elementary Hamiltonians
over arbitrarily short time. Accomplishing the universal gates
and hence the Hamiltonian simulation exactly over a finite number of steps,
as described by Eq.~(\ref{exact}), is thus an essential extension
of these hybrid approaches.

\begin{center}
  \begin{columnfigure}\label{fig10}
    \centering%
    \includegraphics[width=225pt]{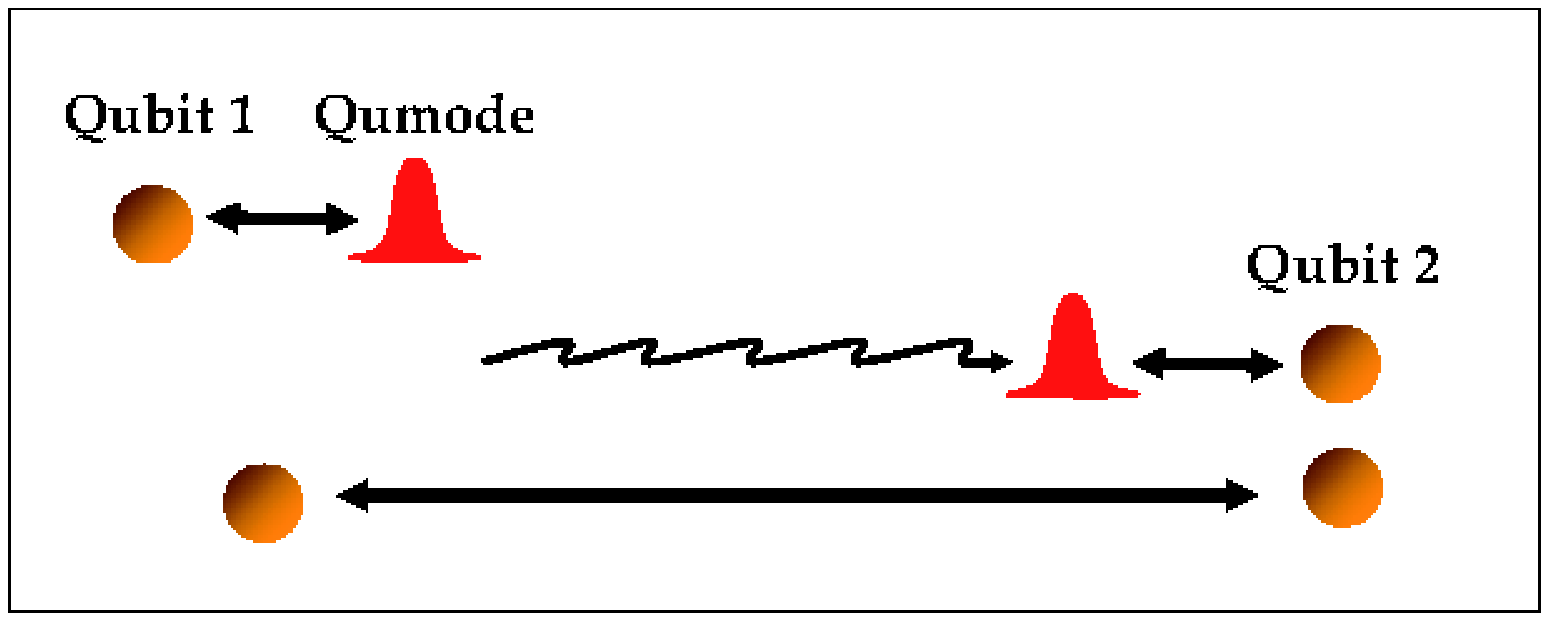}
  \figcaption{{\bf 3.3.2} A quantum optical illustration of the qubus principle.
  A CV probe pulse (qumode) subsequently interacts
  with two matter qubits. After a sequence of interactions
  an entangling gate between the two qubits
  can be mediated through the qubus. Compare this with the previous entanglement distributor
  in Fig.~3.2.2.}
  \end{columnfigure}
\end{center}

Let us finally note that the two-qumode $C_Z$ gate,
as discussed in Sec.~2.3.1, of course, can be implemented directly
using beam splitters and squeezers, independent of a supposed
asymptotic scheme based on Eq.~(\ref{lloydtrick}) with Eq.~(\ref{comm2}).
Such a simple realization, however, is typically not available
for qubits. Therefore, the exact qubus-medi\-ated two-qubit entangling gate
construction here may turn out to be very useful. In Sec.~3.3.5,
we shall discuss optical qubus schemes, similar to Fig.~3.3.2,
utilizing controlled rotations
obtainable from strong or weak nonlinear interactions.

\subsubsection{Nonlinear resources and linear operations}

The GKP scheme \cite{GKP01} is in some sense the CV
counterpart to the DV KLM scheme \cite{KLM01}.
The essential technique exploited in both schemes is to induce a nonlinearity
by means of measurements and to create a highly
sophisticated, nonlinear ancilla state offline.
This saves one from directly applying nonlinearities online.
Conceptually, the essence of KLM and GKP is to
combine an abstract model of measurement-based quantum computation
with the notion of measurement-induced nonlinearities.

Similar to the discussion in Sec.~2.3, the measurement-based approaches
can be divided into those based on quantum teleportation
involving a nonlocal measurement and into cluster-based schemes
requiring only local measurements with all entangling
operations done offline prior to the actual computation.
In this sense, KLM is teleportation-based, whereas
GKP is cluster-based (though the original GKP scheme was not
presented as a cluster-based scheme, but can be recast correspondingly
\cite{Gu}).

In order to obtain the necessary non-Clifford gate
on the logical qubit through a non-Gaussian operation
on the physical qumode
(recall the discussion around Fig.~3.3.1), GKP propose
to produce an approximate version of the cubic phase state offline,
$D_3(\kappa)|p=0\rangle = e^{i\kappa \hat x^3}|p=0\rangle =
\int dx e^{i\kappa x^3} |x\rangle$,
using Gaussian two-mode squeezed state resources and photon number
measurements (see Fig.~3.3.3).
The resulting cubic phase state
is then sufficient to accomplish the cubic phase gate $D_3(\kappa)$
and to apply it to an arbitrary input state $|\psi\rangle$
through linear operations including homodyne detections (see Fig.~3.3.4).
The circuit in Fig.~3.3.4 with all gates performed offline
can be interpreted as a CV cluster computation on a non-Gaussian
cluster state, in which case homodyne detections are sufficient
for universality \cite{Gu}.

\begin{center}
  \begin{columnfigure}\label{fig16}
    \centering%
    \includegraphics[width=180pt]{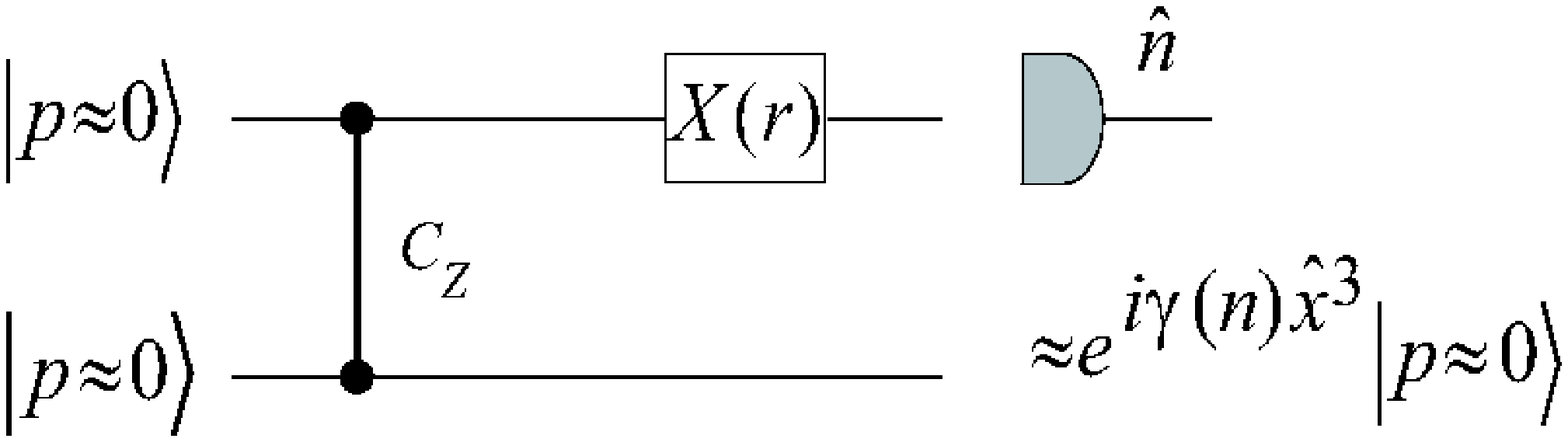}
  \figcaption{{\bf 3.3.3} The GKP approach for creating a cubic phase state \cite{GKP01}
  using linear, Gaussian resources and DV photon counting measurements.
  The Gaussian resource state, emerging from two momentum-squeezed states
  and the CV version of the controlled $Z$ gate, $C_Z = e^{2i\hat x_1\otimes\hat x_2}$,
  is a two-mode squeezed state up to a local Fourier rotation; $X(r) = e^{-2ir\hat p}$
  is a WH operator with fixed and sufficiently large $r$ (much larger than the
  resource squeezing parameter). The circuit as drawn
  is reminiscent of an elementary step in a CV cluster computation involving
  a nonlinear measurement onto the displaced number basis $\{X^\dagger(r) |n\rangle\}$ \cite{Gu}.}
  \end{columnfigure}
\end{center}

The approximate cubic phase state in Fig.~3.3.3 will depend on the
measurement result $n$, $e^{i\gamma(n) \hat x^3}|p=0\rangle$,
and so for the desired cubic phase state,
phase-free squeezing corrections are needed, $\hat S^\dagger[t(n)]e^{i\gamma(n) \hat x^3}
\hat S[t(n)] = e^{i\kappa \hat x^3}$ with $t(n) = [\kappa/\gamma(n)]^{1/3}$ \cite{Gu}.

\begin{center}
    \begin{columnfigure}\label{figGKPofflinecubic}
    \centering%
    \includegraphics[width=200pt]{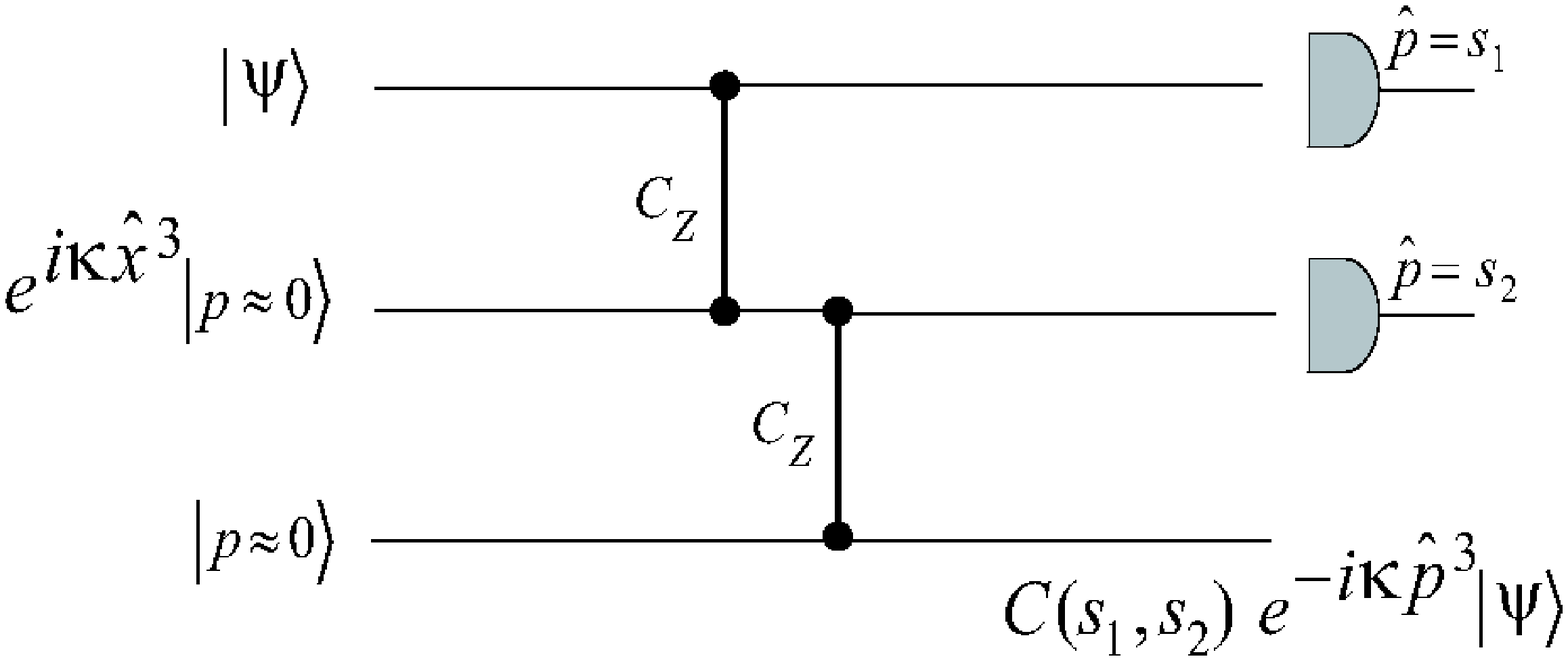}
  \figcaption{{\bf 3.3.4} Using an offline prepared cubic phase state
  $e^{i\kappa\hat x^3}|p=0\rangle$ in order perform the cubic phase gate $e^{-i\kappa\hat p^3}$
  on an arbitrary state $|\psi\rangle$ using linear, Gaussian operations.
  The final, Gaussian correction step to undo the operator $C(s_1,s_2)$ depending
  on the homodyne results $s_1$ and $s_2$ involves displacements, rotations,
  and squeezers because of the commuting properties of the cubic gate
  which is non-Gaussian/non-Clifford and so does not preserve the WH group
  under conjugation \cite{BartlettMunro}.}
  \end{columnfigure}
\end{center}

There is actually a CV scheme which appears to be an even closer
analogue of KLM, as it is also teleportation-based using CV Bell
measurements \cite{SamKimble} on non-Gaussian resource states \cite{BartlettMunro}.
In fact, we could as well interpret the scheme in Fig.~3.3.4
as a teleportation-based scheme. In this case, only the entangling gate $C_Z$
acting upon the second and third rails of the quantum circuit
in Fig.~3.3.4 would be performed offline prior to the $C_Z$ on the first and second rails.
This way the $C_Z$ on the upper two rails together with the two homodyne detections
can be interpreted as a collective, homodyne-based CV two-mode Bell measurement
on an input state $|\psi\rangle$ and one half of
a non-Gaussian resource state which is the cubic phase state
coupled to a momentum-squeezed state through a CV $C_Z$ gate.
This resource then is actually equivalent to a two-mode squeezed state
with one mode subject to a cubic phase gate
(up to a local Fourier transform), as the $C_Z$ gate and the cubic gate
$D_3(\kappa)$ commute.

Eventually, we may describe the protocol
in terms of standard CV quantum teleportation where
a nonlinearly transformed offline two-mode squeezed state
is used as an EPR channel for CV quantum teleportation.
In the case of a cubic offline resource, the online correction operations
during the teleportation process will be quadratic containing squeezers
and displacements; a quartic offline resource such as
a self-Kerr transformed two-mode squeezed state leads to cubic corrections
\cite{BartlettMunro}. However, there are various ways
how to incorporate a desired gate operation
into a CV quantum teleportation scheme (see Fig.~3.3.5).
As long as only linear (but collective two-mode),
homodyne-based measurements on an input mode and one mode of a nonlinear,
non-Gaussian resource state are permitted, the degree of the correction
operations is always one order less than the order of the desired gate.
As a result, only the cubic gate can realized using cubic resources
and Gaussian measurements and corrections; a quartic gate
requires linear measurements, but cubic corrections which, of course,
may also be implemented using cubic offline resources.

\begin{center}
    \begin{columnfigure}\label{figCVgatetelepscheme}
    \centering%
    \includegraphics[width=200pt]{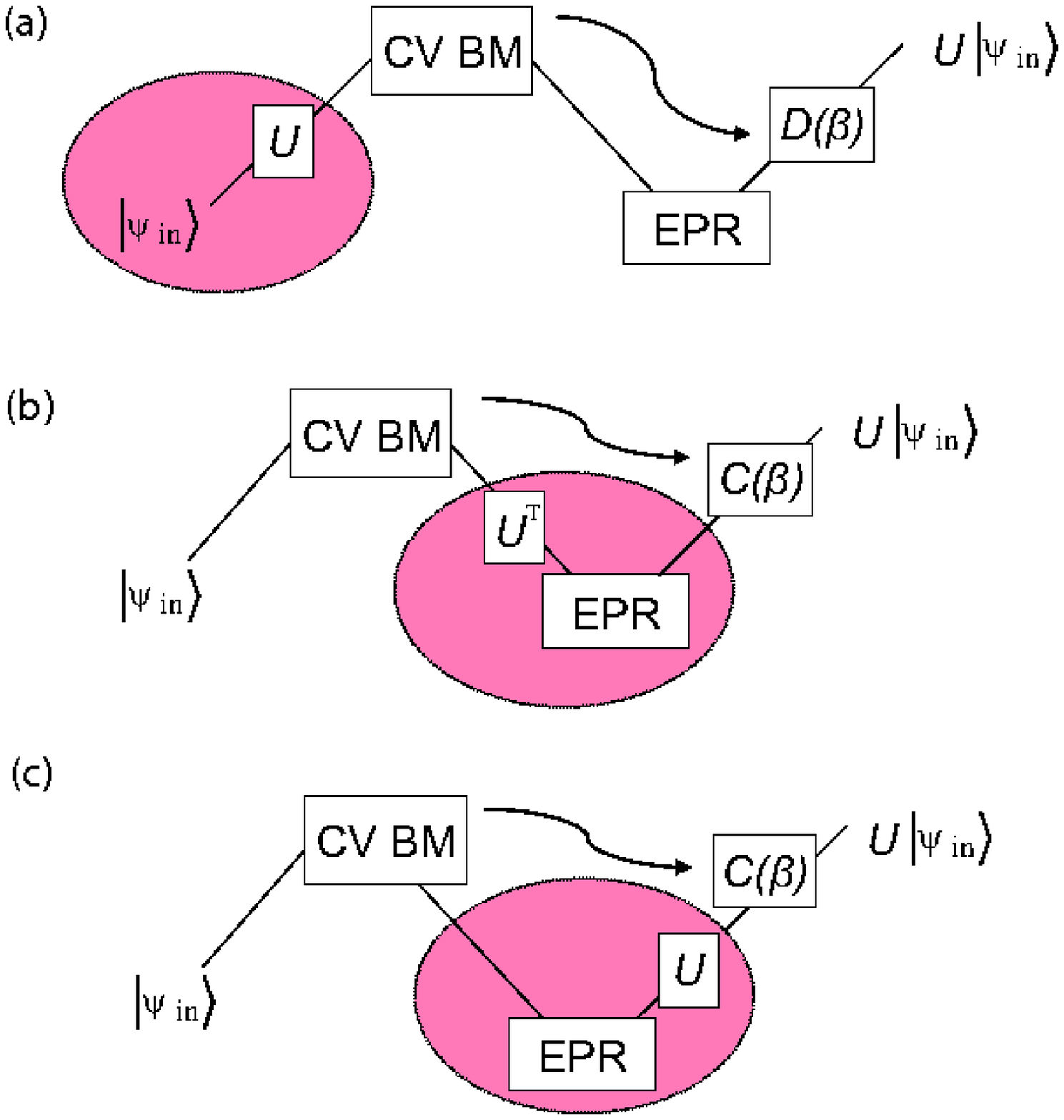}
  \figcaption{{\bf 3.3.5} Teleportation-based implementation of
  a nonlinear, unitary gate $U$ such as the cubic phase gate
  or a quartic Kerr-type gate. In all schemes, the online
  operations are: a two-mode, homodyne-based CV Bell
  measurement `CV BM' and the measurement-dependent corrections using
  displacements $D(\beta)$ or displacements and additional operations
  $C(\beta)$ with a Hamiltonian degree of one order
  lower that the nonlinearity order of $U$;
  the offline resource state is a nonlinearly transformed,
  Gaussian two-mode squeezed `EPR' state, except in the trivial,
  teleportation-based scheme (a) where the
  input state $|\psi_{\rm in}\rangle$ is first transformed according to the desired
  gate $U$ and then teleported; as $|\psi_{\rm in}\rangle$ may be arbitrary
  and unknown, the scheme (a) is not a valid offline-scheme and would require
  an online gate $U$ acting upon $|\psi_{\rm in}\rangle$. The schemes (b) and (c)
  are similar, only differing in the particular EPR-mode to which
  the nonlinear gate $U$ is applied offline. The nonlinear resources
  are always indicated by a red circle; only for the case of a cubic gate $U$
  are all online operations Gaussian.
  Up to local Fourier transforms and squeezers,
  the scheme of Fig.~3.3.4 is a special example of that in (b).
  Of course, we may also consider combinations of
  (a), (b), and (c).}
  \end{columnfigure}
\end{center}

Once a quartic self-Kerr-type gate can be implemented using CV quantum teleportation
[e.g. like in Fig.~3.3.5 (b) or (c)],
such a scheme could be applied to two qubits, as shown in Fig.~2.3.1.
Together with the two beam splitters (and taking into account
the finiteness of the resource squeezing for teleportation), this results in an {\it approximate,
unconditional, and thus deterministic} realization of a two-qubit $C_Z$ gate,
as opposed to the
perfect, nondeterministic implementation of KLM \cite{KLM01}.
However, note that the correction operations before the second beam splitter
are still cubic and so would require further nonlinear processing.

The scheme of Fig.~3.3.4 is a special example of the scheme (b) in Fig.~3.3.5.
In the general case of Fig.~3.3.5 (b), the gate $U$ to be implemented
can be arbitrary and need not be diagonal in the $x$-basis. For the
cluster-type circuit in Fig.~3.3.4, however, it is useful
that the entangling gates $C_Z$ and the desired cubic gate operation $D_3$
are all diagonal in $x$ and hence commute.\footnote{in the scheme of Fig.~3.3.5 (c),
an arbitrary gate $U$, instead of applying it at the very end
of CV quantum teleportation, can just be commuted through the final teleportation
displacement operation, $U D(\beta) = C(\beta)\, U$, with $C(\beta)$
a correction operation of one order lower than the order of $U$ and
$U$ applied offline to the EPR state \cite{BartlettMunro}.
Similarly, the scheme of Fig.~3.3.5 (b) may be understood
by rewriting the offline transformed, maximally entangled EPR state
of (c), $(\one\otimes U)|{\rm EPR}\rangle = (U^{\rm T}\otimes\one)|{\rm EPR}\rangle$,
in the limit of infinite squeezing;
for the finite-squeezing case, see main text.}
Starting with a two-mode squeezed state $\sum_{n=0}^\infty c_n \,|n, n\rangle_{1,2}$
with $c_n\equiv  \tanh^n r/\cosh r$ for a squeezing parameter $r$,
we may apply the transfer formalism for standard CV quantum teleportation \cite{Hofmann00}
and extend it to the present case of gate teleportation.
Then we obtain the conditional state after the Bell projection
of the input qumode and qumode 1
onto $\hat\Pi(\beta)\equiv |\Phi(\beta)\rangle\langle \Phi(\beta)|$ for the
CV Bell basis $|\Phi(\beta)\rangle\equiv [\hat D(\beta)\otimes\one]
\sum_{m=0}^{\infty} |m,m\rangle/\sqrt{\pi}$,
\begin{eqnarray}
\hat\Pi_{{\rm in},1}(\beta)
\left\{|\psi_{\rm in}\rangle \otimes \left[\left(U_1^{\rm T}\otimes \one_2\right)
\sum_{n=0}^\infty c_n \,|n, n\rangle_{1,2}\right]\right\}
\,,
\end{eqnarray}
corresponding to a conditional state of qumode 2 alone,
\begin{eqnarray}
\sum_n \frac{c_n}{\sqrt{\pi}}\, |n\rangle\sum_m U^{\rm T}_{mn}\langle m| \hat D^\dagger(\beta)
|\psi_{\rm in}\rangle =
\mathcal{D} \,U \hat D^\dagger(\beta) |\psi_{\rm in}\rangle,\nonumber\\
\end{eqnarray}
with the ``distortion operator" $\mathcal{D} \equiv \sum_n c_n |n\rangle\langle n|/\sqrt{\pi}$
and the matrix elements
$U^{\rm T}_{mn}\equiv \langle m| U^{\rm T} | n\rangle = \langle n| U | m\rangle$.
After a suitable correction operation $C(\beta)$, the input
state is ``transferred" onto the output state $\hat T_U(\beta) |\psi_{\rm in}\rangle$
with
\begin{equation}
C(\beta)\mathcal{D} \,U \hat D^\dagger(\beta)
\equiv C(\beta)\mathcal{D} \, C^\dagger(\beta) U\equiv
\hat T_U(\beta) \,,
\end{equation}
where the first equality defines the right correction operation
$C(\beta)$ depending on the gate $U$ and its commuting properties
with the displacement operator,
$U \hat D^\dagger(\beta) = C^\dagger(\beta) U$.
The degree of $C^\dagger(\beta)$ will always be
one order lower than the order of $U$ \cite{BartlettMunro}.
In the limit of infinite squeezing,
we obtain the desired gate teleportation.
For finite squeezing $r$, there will be a distortion
resulting in a nonunit-fidelity gate,
$F = \int d^2\beta |\langle
\psi_{\rm in}|U^\dagger \hat T_U(\beta) |\psi_{\rm in}\rangle |^2 < 1$.
We shall look at the scheme in Fig.~3.3.5 (b) as well as that
in Fig.~3.3.5 (a) from a different perspective in the following section.

Further refinements and proposals related with GKP can be found
in Refs.~\cite{GlancyKnill,Pirandola}.
An alternative approach to implementing a cubic phase gate relies upon
potentially more accessible non-Gaussian resources such as Fock-state ancillae
\cite{Gho06}. Similarly, nonlinear Fock-state ancillae may be exploited in order
to achieve nonlinear Fock-state projection measurements using linear,
Gaussian measurements in form of two homodyne detectors after a beam splitter
\cite{MarekFiurasek} (see Fig.~3.3.6). More specifically,
postselecting the homodyne results around zero leads to
$\hat\Pi_{{\rm in},1}(\beta\approx 0) |m\rangle_1 \propto \,_{\rm in}\!\langle m|$.
Projecting onto this two-mode basis corresponds to applying
a symmetric beam splitter transformation followed by $x$ and $p$
homodyne detections like for the CV Bell measurement in CV quantum teleportation
\cite{SamKimble}, as in Fig.~3.3.6.

\begin{center}
    \begin{columnfigure}\label{figMarekscheme}
    \centering%
    \includegraphics[width=80pt]{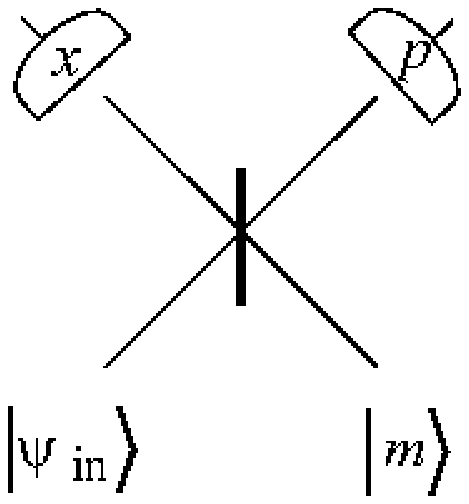}
  \figcaption{{\bf 3.3.6} Nonlinear (destructive) projection of an input state
  $|\psi_{\rm in}\rangle$ onto the Fock state $|m\rangle$
  through $x$ and $p$ homodyne detections and postselection using a beam splitter
  and an ancilla Fock state $|m\rangle$ \cite{MarekFiurasek}.}
  \end{columnfigure}
\end{center}

The proposal of Ref.~\cite{MarekFiurasek} can be used to realize
various quantum information tasks, including the implementation
of a NSS gate. However, the need for postselection renders
this approach again fairly inefficient. For example, for the NSS gate,
fidelities $F>0.9$
are obtainable at success probabilities $<10^{-4}$.
Nonetheless, in general, preparing nonlinear ancilla resource states
offline,\footnote{the degree of nonlinearity and non-Gaussianity of which
may even be tuned in a heralded state generation scheme \cite{Jain09}.}
possibly in a nondeterministic fashion, but with
reasonable fidelities, offers a promising approach to
universal quantum information processing, as the online operations
in this case can be restricted to only linear ones. These linear operations may
include squeezing corrections which could also be efficiently implemented,
provided high-quality CV cluster states are available.
Such linear, universal resource states will be part of the discussion
in the following section.

\subsubsection{Linear resources and nonlinear operations}

The circuit in Fig.~3.3.4 describes a protocol in which
a nonlinear cubic phase gate is implemented
using an offline cubic phase state and online Gaussian operations.
In order to accomplish the gate $D_3(\kappa)$, a cubic state
of the form $D_3(\kappa)|p=0\rangle$ is needed.
So there should be a sufficient
supply of cubic states such that after injecting these into
a CV Gaussian cluster state, sets of cubic gates can be applied
through homodyne detections whenever needed during
a computation \cite{Menicucci06,Gu,vanLoockJOSA2007}.

A canonical version of a Gaussian CV cluster state \cite{ZhangBraunstein,PvL07}
is shown in Fig.~3.3.7. An elementary cluster computation step
between two qumodes of the cluster corresponds to the circuit
in Fig.~3.3.8.

\begin{center}
    \begin{columnfigure}\label{figCVcluster}
    \centering%
    \includegraphics[width=220pt]{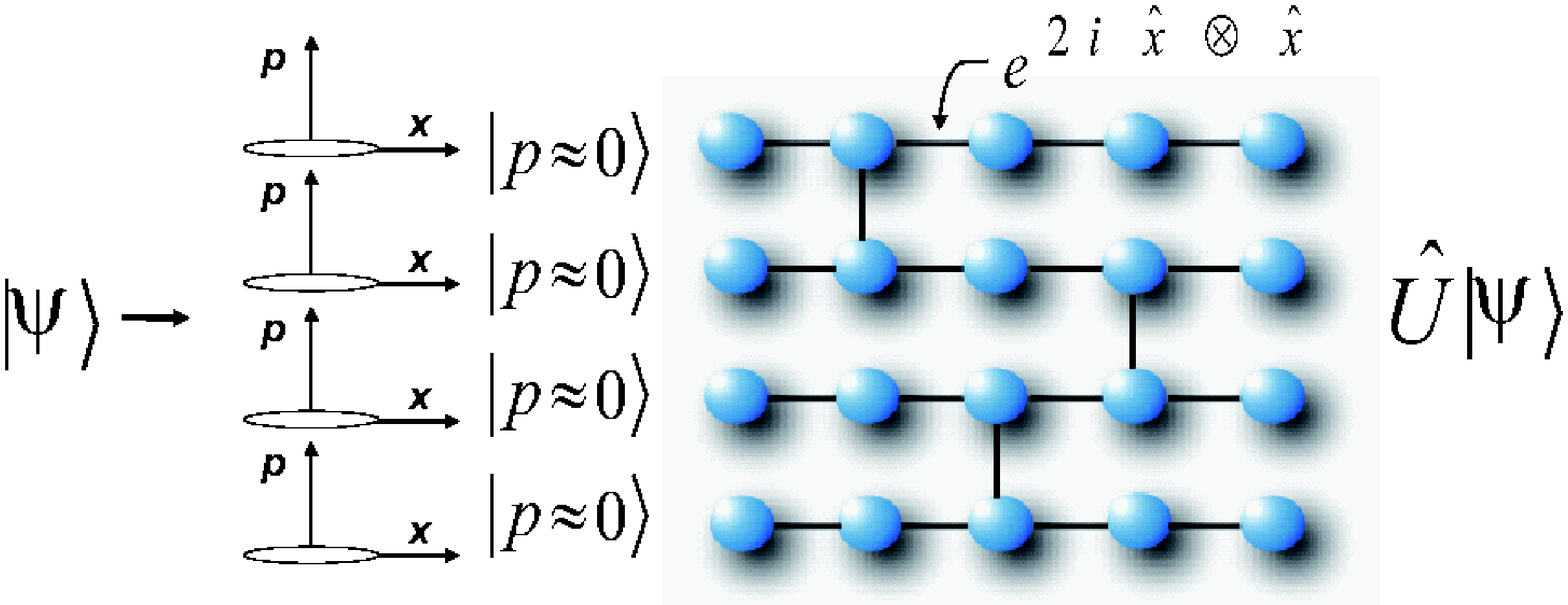}
  \figcaption{{\bf 3.3.7} An approximately universal,
  Gaussian CV cluster state built from
  momentum-squeezed states of light and Gaussian, CV versions of the
  $C_Z$ gate, $e^{2i \hat x\otimes \hat x}$.
  An arbitrary multi-mode state $|\psi\rangle$
  attached from the left can then be universally processed
  through local projection measurements of the individual
  qumodes of the cluster such as homodyne and photon number measurements
  \cite{Menicucci06}. The output state $\hat U|\psi\rangle$
  of the cluster computation will appear
  on the most right column in the remaining, undetected qumodes.}
  \end{columnfigure}
\end{center}

The gates in front of the $x$-homodyne detector in the circuit of
Fig.~3.3.8
can be absorbed into the measurement apparatus such that
instead of measuring the observable $\hat x$ the projection
is onto the rotated $p$-basis $\{D^\dagger |p\rangle\}$ measuring
the observable $D^\dagger \hat p D$.
This way we can apply any gate $D=e^{i f(\hat x)}$ to an arbitrary input state
teleported into the upper rail in Fig.~3.3.8, that is into
one or, in the multi-mode case, several qumodes of the most left column
in Fig.~3.3.7. Further application of such elementary steps,
by measuring out the other qumodes in the CV cluster state
beginning from the left in Fig.~3.3.7, may lead to, in principle,
universal quantum computation on multi-qumode states in the approximate,
asymptotic sense as discussed in Sec.~2.3.1 \cite{Menicucci06}.
A Fourier transform can be performed through the cluster
at any step in order to switch between $\hat x$ and $\hat p$ gates.
To complete the universal set in Eq.~(\ref{CVset}),
the two-qumode gate $C_Z$ is obtainable
through the vertical wires in Fig.~3.3.7, so
the cluster must be at least two-dimensional.
The total evolution of
the input is completely controlled by the measurements
with the cluster state prepared offline prior to the computation.

\begin{center}
    \begin{columnfigure}\label{figCVclusterstep}
    \centering%
    \includegraphics[width=180pt]{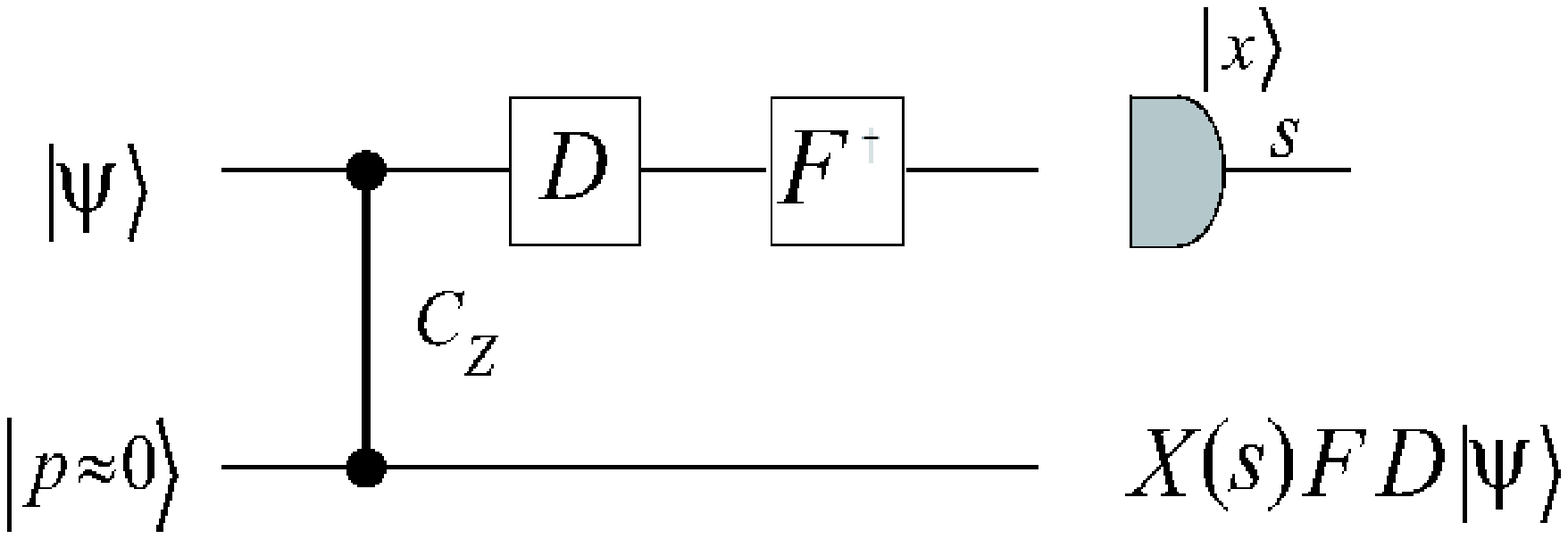}
  \figcaption{{\bf 3.3.8} Elementary step of a CV cluster computation.
  The desired gate operation $D=e^{i f(\hat x)}$ and
  the inverse Fourier transform $F^\dagger$
  can be absorbed into the measurement apparatus such that a projection
  of the upper qumode onto the basis $\{D^\dagger |p\rangle\}$ with result $s$
  leaves the lower qumode in the desired output state up to a Fourier transform
  and a WH correction $X(s)$. Compare this with the analogous qubit circuit
  in Fig.~2.3.5.}
  \end{columnfigure}
\end{center}

Any multi-mode LUBO as described by Eq.~(\ref{LUBO})
can be performed on an arbitrary multi-mode state
through homodyne detections alone. An additional {\it nonlinear measurement
such as photon counting is needed in order to be able to
realize gates of cubic or higher order.}
In this case, the basis choice of a measurement in one step
would typically depend on the outcomes of the measurements in the previous steps.
In contrast, in the all-homodyne-based scenario for LUBOs,
no such feedforward is required and all measurements may be conducted
in parallel -- a feature known as Clifford parallelism for qubit cluster computation.

It has been proven that a linear four-mode cluster state is sufficient
to achieve an arbitrary single-mode LUBO (see Fig.~3.3.9);
an arbitrary $N$-mode LUBO is possible
using a finite, two-dimensional CV cluster state of $\sim N^2$ qumodes
\cite{Ukai1}. In this case,
no more asymptotic evolutions with infinitesimal, elementary steps
must be considered, but rather combinations of beam splitter and single-mode
squeezing gates of appropriate strength. Such clus\-ter-based
LUBOs circumvent the complication of online squeezing
of, especially, fragile non-Gaussian states, since all squeezing gates
are performed offline on the Gaussian cluster state. Provided enough squeezing
is available to create the cluster states
\cite{PvL07,MenicucciFreqComb1,MenicucciFreqComb2,Gu},
this approach may also be used to {\it realize the necessary squeezing
corrections for nonlinear gate implementations,} as discussed in the preceding section.
The single-mode LUBO
scheme was recently implemented experimentally \cite{Ukai2,Miwa09}.
More experiments on the creation of various CV cluster-type states
and the offline implementation of CV $C_Z$-type gates
are presented in Refs.~\cite{Yukawa08,SuCluster} and Ref.~\cite{Yoshikawa08.prl},
respectively.

\begin{center}
    \begin{columnfigure}\label{figCVclusterLUBO}
    \centering%
    \includegraphics[width=200pt]{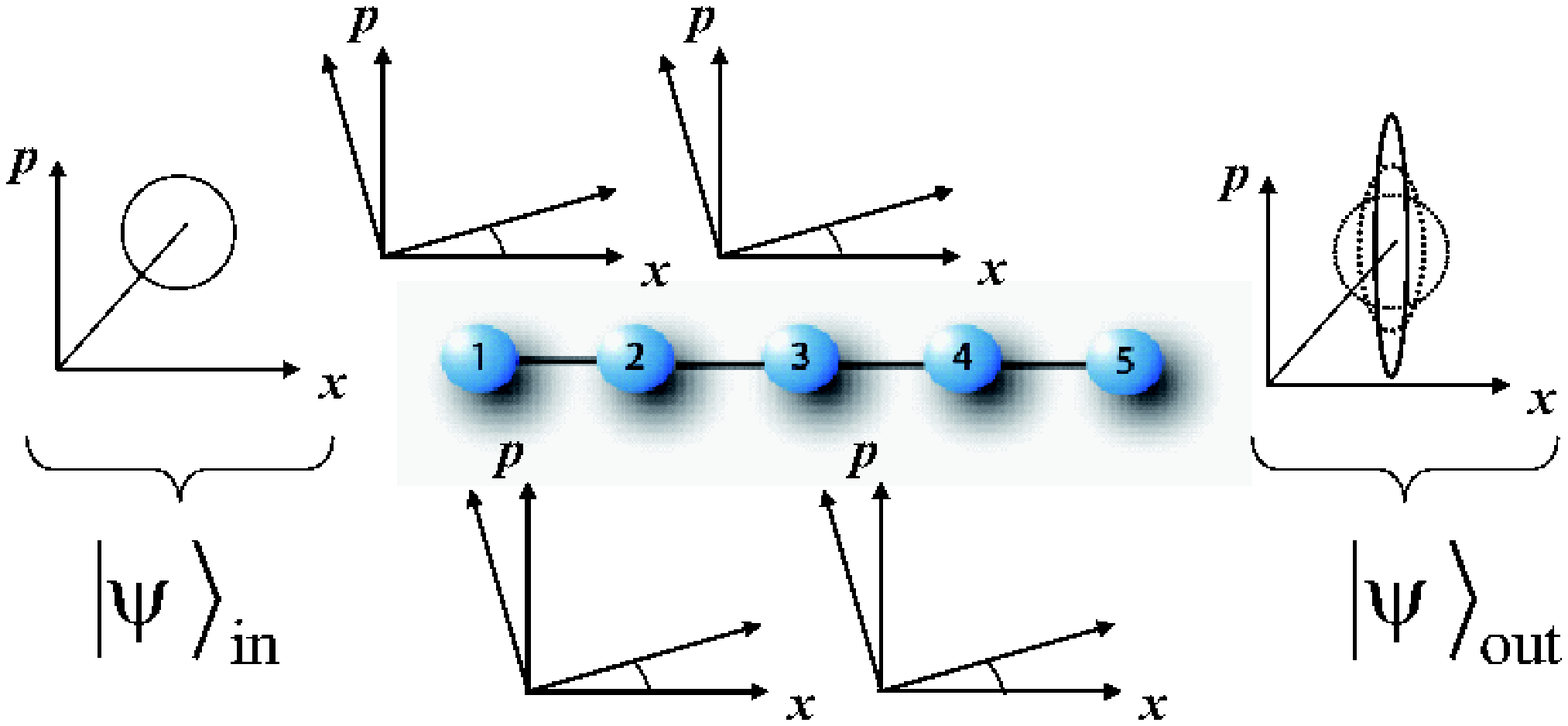}
  \figcaption{{\bf 3.3.9} Universal single-mode LUBO.
  After attaching an arbitrary single-qumode input state to
  a linear four-mode cluster state (qumodes 2-5), appearing in qumode 1,
  four elementary steps, involving four quadrature homodyne detections
  on qumodes 1-4
  with suitably chosen local-oscillator quadrature angles, are sufficient to
  obtain approximately any LUBO-transformed output state in qumode 5, provided the squeezing
  in the linear cluster state is sufficiently large. The $C_Z$-attachment of
  the input state and the two homodyne detections on qumodes 1 and 2 can be replaced
  by a CV Bell measurement on the input mode and the most left mode, qumode 2,
  of the cluster \cite{Ukai1}. The input state is depicted as a Gaussian coherent state
  only for illustration. Most significantly, the CV Bell measurement can also be used
  to teleport non-Gaussian input states into the four-mode cluster
  for universal, linear processing.}
  \end{columnfigure}
\end{center}

In comparison to the scheme in Fig.~3.3.4, which employs
a suitable nonlinear resource state and linear operations,
the GKP approach for realizing a cubic phase gate can also be
directly incorporated into a CV cluster computation \cite{Gu}.
In this case, the offline resource state remains Gaussian and
thus unconditionally producible, while some of the online operations
must then become nonlinear (see Fig.~3.3.10).
In order to obtain a desired cubic gate of any given strength $\kappa$,
as before, additional squeezing corrections are needed.
However, this time, also the squeezing corrections are performed
through cluster computation,
since any squeezing gate is available
by propagating the relevant state through a horizontal,
linear four-mode wire (see Fig.~3.3.11).

\begin{center}
  \begin{columnfigure}\label{fig17}
    \centering%
    \includegraphics[width=180pt]{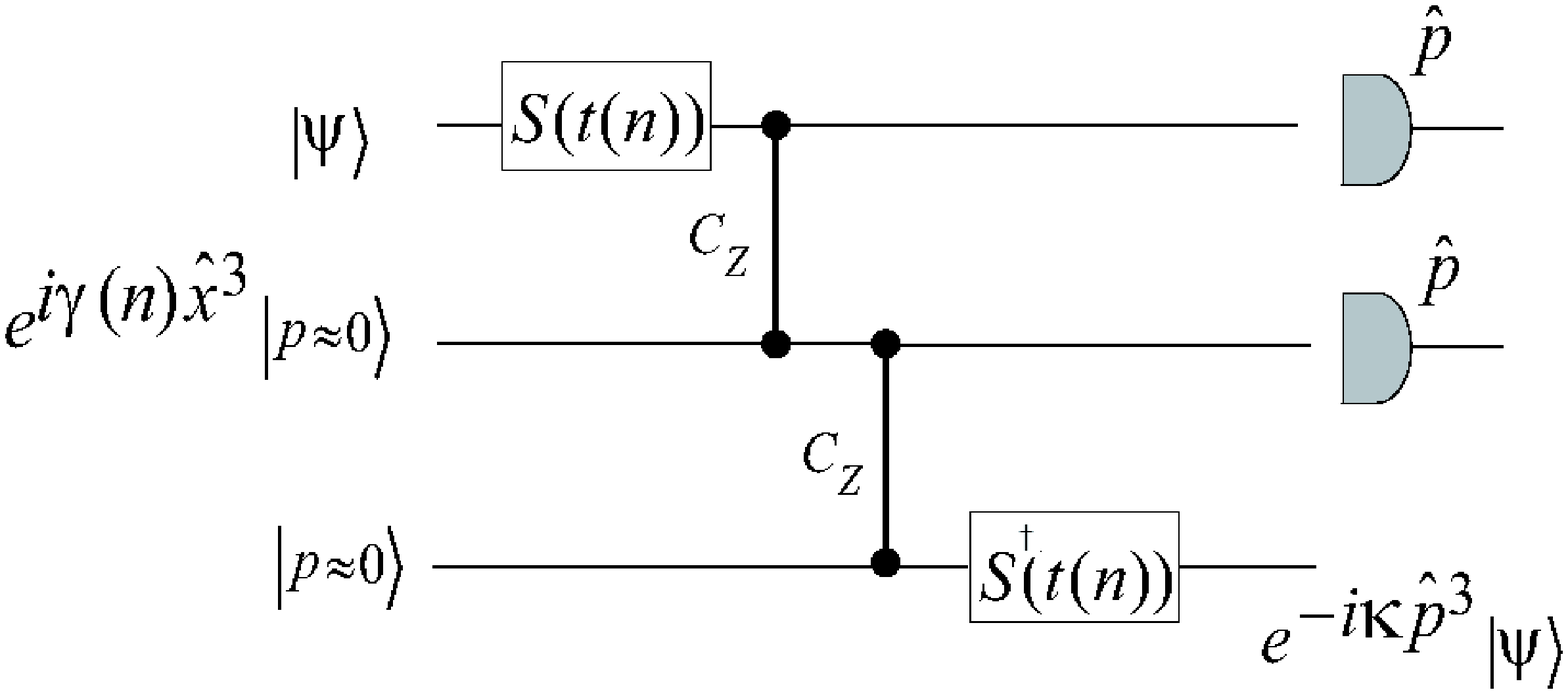}
  \figcaption{{\bf 3.3.10} Incorporating the GKP scheme into a CV cluster computation.
  The resource state remains Gaussian, the measurements
  are CV homodyne detections and DV photon number measurements
  to teleport the cubic phase state into the middle rail.
  The initial cubic phase state depends on the measurement outcome $n$
  and extra squeezing corrections $S$ are needed to obtain a cubic gate of
  any desired strength $\kappa$. Compared to Fig.~3.3.4, the $n$-dependent squeezing corrections
  $S$ are here to be done through the cluster (Fig.~3.3.11); the additional $p$-dependent
  squeezing and WH corrections $C(s_1,s_2)$ can also be incorporated into the cluster computation
  and are not shown for simplicity.}
  \end{columnfigure}
\end{center}

\begin{center}
  \begin{columnfigure}\label{fig18}
    \centering%
    \includegraphics[width=230pt]{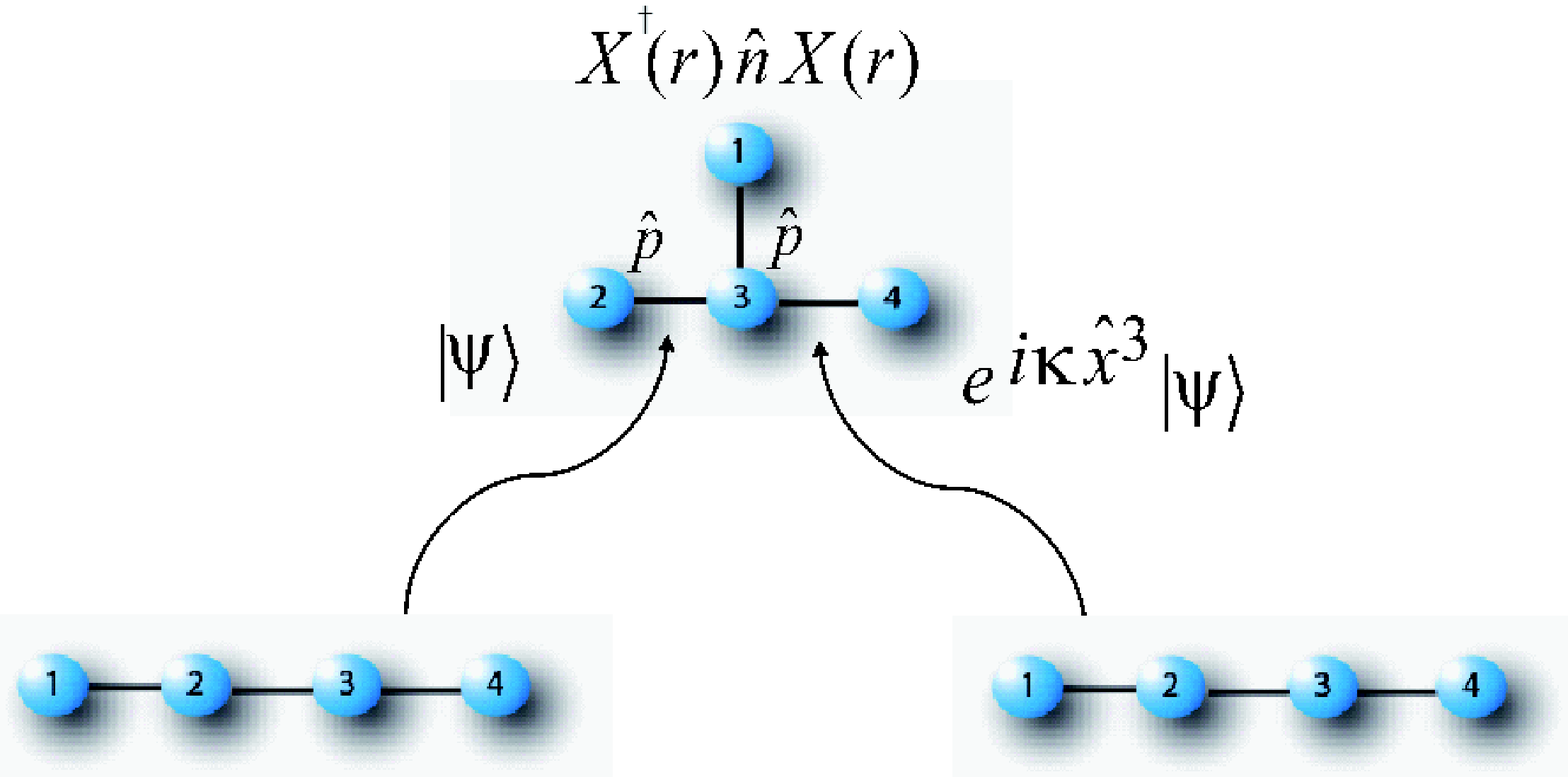}
  \figcaption{{\bf 3.3.11} Full CV cluster implementation of a cubic phase gate.
  The extra linear cluster states are needed to
  realize the necessary squeezing corrections which depend
  on the photon number measurement result. As the desired cubic gate
  is non-Gaussian,
  the order of the measurements matters and the number measurement
  has to be done first. This implementation
  using linear and nonlinear measurements on a Gaussian state
  is conceptually different from that illustrated in Fig.~3.3.4
  where the resource state is non-Gaussian and
  all online operations are linear.
  }
  \end{columnfigure}
\end{center}

\begin{center}
  \begin{columnfigure}\label{figCVgatetelepschemeNLmeas}
    \centering%
    \includegraphics[width=230pt]{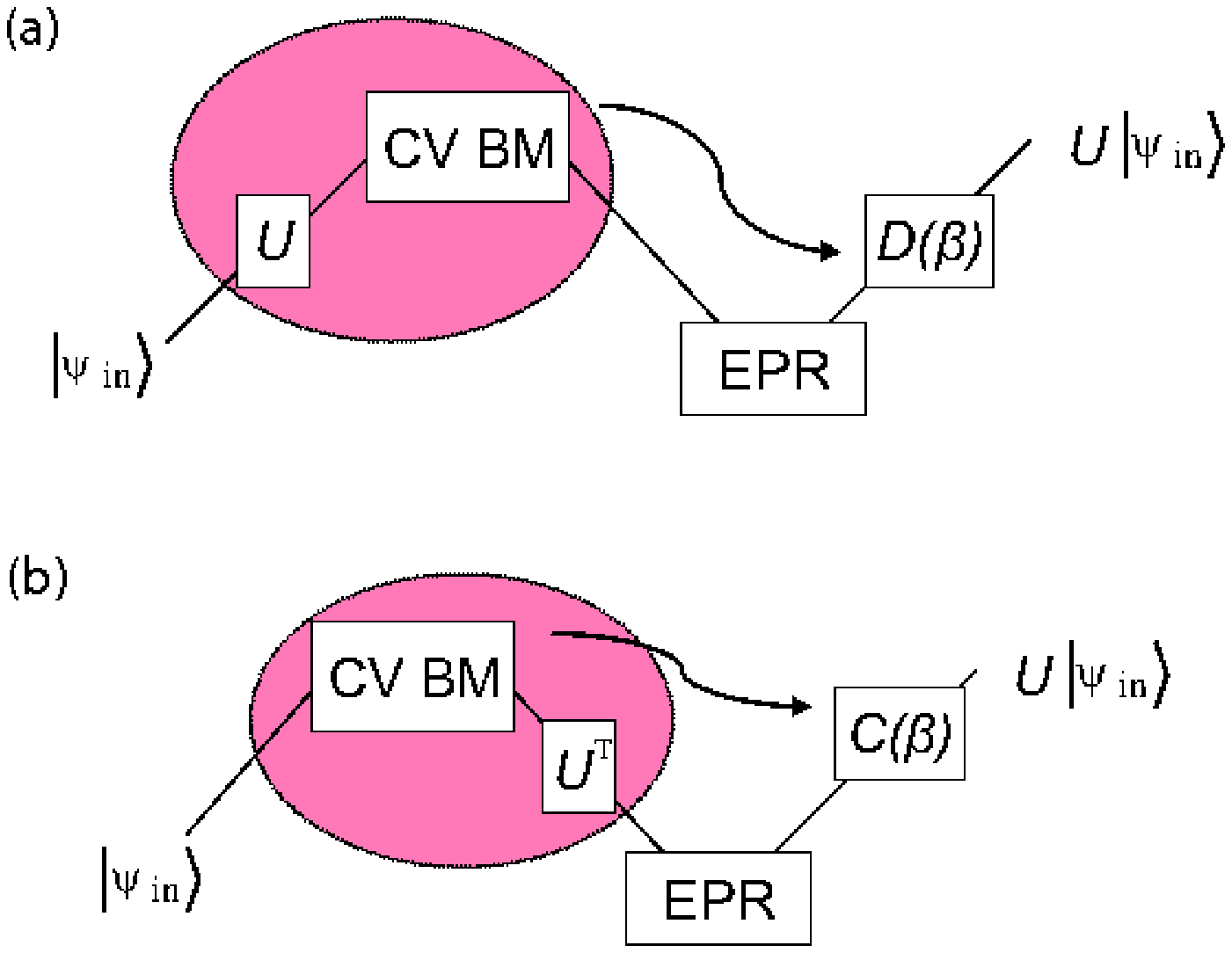}
  \figcaption{{\bf 3.3.12} Gate teleportation using collective,
  nonlinear projection measurements on an input state $|\psi_{\rm in}\rangle$
  and one half of a Gaussian two-mode squeezed state that serves
  as an EPR source; (a) compared to Fig.~3.3.5 (a), the nonlinear gate $U$
  is absorbed into the Bell measurement; similarly, (b)
  compared to Fig.~3.3.5 (b), the nonlinear gate $U^{\rm T}$
  is absorbed into the Bell measurement. The nonlinear elements
  are again indicated by the red circles, this time corresponding
  to suitable projective measurements.}
  \end{columnfigure}
\end{center}

Formally, we may also interpret some of the schemes in Fig.~3.3.5
as gate teleporations for arbitrary qumode gates $U$ using nonlinear operations on
linear resources, where
a Gaussian two-mode squeezed state is used as an EPR source and
a nonlinearly modified, two-mode Bell measurement is employed (see Fig.~3.3.12).
For example, the scheme in Fig.~3.3.12 (a) can be formally described as
\begin{eqnarray}
\hat \Pi^{(U)}_{{\rm in},1}(\beta)
\left(|\psi_{\rm in}\rangle \otimes
\sum_{n=0}^\infty c_n \,|n, n\rangle_{1,2}\right)=
\mathcal{D} D^\dagger(\beta) U |\psi_{\rm in}\rangle,\nonumber\\
\end{eqnarray}
using the definitions of Sec.~3.3.3
and $\hat \Pi^{(U)}(\beta) \equiv (U^\dagger\otimes\one)\hat \Pi(\beta)
(U\otimes\one)$, describing a projection onto the basis
$\{(U^\dagger\otimes\one)|\Phi(\beta)\rangle \}$ for a fixed $U$.
As such collective, nonlinear measurements are not known to be experimentally available,
there seems to be no advantage of the schemes in Fig.~3.3.12 besides the conceptual
insights they provide. The scheme in Fig.~3.3.12 (a) does not require
any correction operations other than displacements, because it is equivalent
to that in Fig.~3.3.5 (a), where the gate $U$ is trivially applied through teleportation
by first performing $U$ on $|\psi_{\rm in}\rangle$
and then teleporting $U|\psi_{\rm in}\rangle$ using standard CV quantum teleportation.

Concluding this section, we can say that it depends to a large extent
on the experimental implementability of nonlinear measurements such as photon number
resolving detections whether more advanced or ultimately universal, optical
quantum information protocols can be realized in the laboratory. If, similar to the
quantum-state tunability between offline
Gaussian and non-Gaussian resource states \cite{Jain09}, also the online
operations for measurement-based approaches, i.e., the quantum-state measurements,
could be tuned over a sufficient range of linear CV and nonlinear DV POVMs \cite{Puentes},
efficient experimental realizations
may then be possible in the near future. In this case, the offline resource
states may be, for example, Gaussian CV cluster states which can be built
unconditionally from squeezed light using beam splitters. However, finite-squeezing-induced
imperfections will then require some additional nonlinear element for quantum error correction.
Alternatively, instead of attempting to perform full computations over CV cluster states
at a precision that decreases linearly with the size
of the computation (number of measurement steps) for a given initial squeezing variance \cite{Gu},
one may just use Gaussian ancilla states, nonlinearly measured online or offline, for implementing
particularly difficult gates such as the NSS gate on DV photonic states and
do the simple gates on dual-rail encoded qubits directly in the standard circuit approach.

\subsubsection{Weakly nonlinear operations}

In this section, we shall now consider schemes where
a nonlinear element is part of an optical quantum protocol in form
of some nonlinear interaction, but this nonlinearity is no longer required to
be sufficiently strong. Recall the discussion in Sec.~3.3.2 on hybrid Hamiltonians.
Using an ancilla qumode as a quantum bus that mediates
a nonlinear interaction between two qubits, as depicted in Fig.~3.3.2,
we can construct universal two-qubit gates by means of
rotations or displacements of the qumode conditioned upon
the state of the respective qubit.
Sequences of such qubit-qumode interactions may then result
in an effective two-qubit entangling gate after the qumode is finally measured out
or when the qumode automatically disentangles from the two qubits,
as, for instance, described by the sequence in Eq.~(\ref{exact}).
In this example, though the ancilla qumode and each qubit do become entangled
during the gate sequence, an initially unentangled qumode
will be again unentangled at the end, as can be easily understood
from the right-hand-side of Eq.~(\ref{exact}).

In the quantum optical regime, using either dispersive light-matter interactions or
all-optical Kerr-type interactions,
qubit-controlled qumode rotations enable one
to perform various tasks from projecting onto the complete DV photonic Bell basis
to implementing photonic two-qubit entangling gates, employing either
DV (threshold) photon detectors \cite{Paris00}, CV homodyne detectors
\cite{Nemoto04,Barrett04,Munro05_2,Munro05}, or no detectors at all
\cite{Spiller06,vanloock07}.

In this section, we shall briefly discuss the {\it measure\-ment-free approach.}
Schemes where the qumode ancilla is eventually measured are particularly
useful for quantum communication and will be considered in the next section.
The measurement-based and measurement-free schemes can differ in their controlled,
qubit-qumode gate sequences and also in the scaling of the nonlinear element
with photon losses for the realistic case of imperfect,
lossy quantum gates and communication channels.

Most importantly, the qubit-qumode interactions, realizable, for instance,
through dispersive light-matter couplings in a CQED setting such as
the ``single-atom dispersion" discussed in
Sec.~3.1, are typically fairly weak.
Thus, the resulting controlled gates produce conditional phase angles $\theta$
which are initially too small to be useful, $\theta \sim 10^{-2}$ or smaller.
However, the hybrid qubit-qumode systems are designed such that
a sufficiently intense qumode beam leads to an effectively enhanced
interaction. On the other hand, qumode ancilla pulses carrying too many photons,
when temporarily entangled with a low-dimensional qubit, are very likely
to transfer photons containing which-path qubit information into the environment.
Therefore, the hybrid entangled states and the hybrid gates become
more sensitive to photon losses.

Let us write a sequence similar to that in Eq.~(\ref{exact})
in terms of quantum optical displacement operations, each depending
on a qubit Pauli operator,
\begin{eqnarray}\label{totalunitary}
\hat D\left[i\beta_2 \sigma_{z}^{(2)}\right] \hat D\left[\beta_1 \sigma_{z}^{(1)}\right]
\hat D\left[-i\beta_2 \sigma_{z}^{(2)}\right] \hat D\left[-\beta_1 \sigma_{z}^{(1)}\right]
\\
=
\exp \left[2 i\; {\rm Re}(\beta_1^* \beta_2)\;\sigma_{z}^{(1)}
\sigma_{z}^{(2)}\right].\nonumber
\end{eqnarray}
These controlled displacements, acting upon
the composite system of two qubits and one qumode,
generate a two-qubit entangling gate,
as discussed before. The two-qubit gate relies upon
a phase shift that is acquired by a qumode
whenever it goes along a closed loop
in phase space. This geometric phase depends on the area of the
loop and not on its form \cite{wan01}. It comes from the extra
phase factor in
$\hat D(\beta_1) \hat D(\beta_2)= \exp \left[ i\,{\rm Im}\left(\beta_1
\beta_{2}^*\right) \right] \hat D(\beta_1 +\beta_2)$. Note that
the qumode may start in {\it any} state.

In quantum optics, the controlled displacement operations of
Eq.~(\ref{totalunitary}) are not really available.
Two-qubit geometric phase gates directly implemented from controlled rotations
through dispersive light-matter interactions result in final states
with a qumode still entangled with the two qubits \cite{Ladd2006}.
The extra dephasing
from this effect, however, can be completely avoided by simulating every
controlled displacement in Eq.~(\ref{totalunitary}) through
yet another sequence of controlled rotations and uncontrolled displacements
\cite{vanloock07},
%(see Fig.~3.4.1),
\begin{eqnarray}\label{exactseries}
\hat D(\alpha\cos\theta)\hat R(\theta\sigma_z)
\hat D(-2\alpha) \hat R(-\theta\sigma_z) \hat D(\alpha\cos\theta) \nonumber\\
=
\hat D\left(2i \alpha \,\sin\theta\,\sigma_z\right)\,,
\end{eqnarray}
using Eq.~(\ref{controlledrotation}).

%\begin{center}
%  \begin{columnfigure}
%    \centering%
%    \includegraphics[width=110pt]{fig21.eps}
%  \figcaption{{\bf 3.4.1} bl.}
%  \end{columnfigure}
%\end{center}

In order to obtain a maximally entangling two-qubit gate,
we need $\beta_1\beta_2 = \pi/8$ (assuming real $\beta_1$ and $\beta_2$)
and so $2\alpha\sin\theta = \sqrt{\pi/8} \approx 0.6$ (assuming real $\alpha$).
For small $\theta$, this means that $\alpha$ must be sufficiently large such that
$\alpha\theta \sim 1$.
For instance, when $\theta\sim 10^{-2}$,
phase-space displacements corresponding to mean photon numbers
of the order of $10^4$ are required.
For stronger interactions $\theta$, correspondingly smaller displacements are enough.
Note that the strength of the displacements in Eq.~(\ref{exactseries}),
inserted into Eq.~(\ref{totalunitary}),
determines the effective enhancement of the nonlinear two-qubit gate,
as the qumode starts in an arbitrary state.

The deterministic, measurement-free gates described here could be, in principle,
used directly in a DV quantum computation, utilizing
either matter qubits dispersively interacting with optical ancilla qumodes
or photonic qubits weakly nonlinearly coupled to photonic ancilla qumodes.
Alternatively, these gates may provide a mechanism for growing DV qubit cluster states
in an efficient, deterministic fashion; for certain simple (though non-universal)
graph states such as linear or star (GHZ-type) graphs, the qumode may interact
as little as two times with each qubit to build up the graph \cite{Sebastien1}.
This is different from the nondeterministic, standard linear-optics
approaches \cite{Nielsen04,Browne05,Kieling1,Kieling2}, but similar to the
unconditional gate teleportations using Gaussian ancilla cluster resources discussed
in the preceding sections.

However, whereas the fidelity of even an ideal implementation
with CV cluster states
is fundamentally limited by the finite squeezing of the cluster states
and other imperfections in the extra non-Gaussian ancilla states or measurements,
the ideal weak-nonlinearity-based protocol would, in principle, achieve
unit fidelity at unit efficiency. Of course, this only holds provided that
the hybrid qubit-qumode interactions are perfect.
Realistically, these interactions are very sensitive to photon losses,
especially those with large $\alpha$-paths in phase space like in Eq.~(\ref{exactseries})
when $\theta$ is small, and so
for every gate protocol, a careful loss analysis is needed
\cite{Sebastien2,BarrettMilburn}. In general,
it will be useful to minimize the number of necessary interactions in a gate sequence.
One possibility then would be to employ measurement-based entangling
gates including postselection,
rendering the quantum computational routines again nondeterministic
\cite{Sebastien1,Sebastien3}. The measurement-based approach
is also preferred for quantum communication protocols when two spatially
separated memory qubits are to be entangled over a distance.
This will be discussed in the next section.

Let us finally mention that the concept of using a quantum bus
has other advantages too. In general, photonic qubit or qumode ancillae
may be employed to mediate interactions
between non-nearest-neighbour signal qubits in a solid-state system
\cite{cir97,bos99,dua03,bro03,lim05,barrett05}. In principle, this
allows for universality and scalability when arbitrarily many signal
qubits are added to the system. Two-qubit gates can then be
performed for any pair and there is no need for two qubits to be so
close together that individual addressing is no longer
possible. The first qubus proposals were based upon
both qubit signals and qubit ancillae, including, for instance,
the well-known ion-trap proposal \cite{cirzol95} with
a phononic ion qubit mediating
a gate between two internal two-level ion qubits.
The more recent hybrid approaches as described here would rather use
a photonic CV qumode ancilla serving as a quantum bus.

\subsection{Hybrid quantum communication}

Quantum communication over a distance as large as
1000 km or more is, in principle, possible
by means of quantum repeaters.
As discussed in detail in Sec.~2.4,
a quantum repeater operates by distributing
entangled states over sufficiently short
channel segments and connecting them through
teleportation in combination with entanglement distillation.
As the teleportation and distillation steps
require local Bell measurements and entangling gates,
with each repeater node
effectively functioning as a small quantum computer,
we shall in this section omit detailed discussions on the
connection and
distillation part of a repeater. For this purpose
(Figs.~3.4.3 and 3.4.4),
in principle, some of the quantum computational hybrid
protocols of Sec.~3.3 could be used.

Our focus in this section is on the distribution
of entangled qubit memory pairs between two neighboring
repeater stations. There are various non-hybrid approaches
for this of which
some were presented in Sec.~2.4.\footnote{those non-hybrid
approaches do not really combine DV and CV techniques
according to our definition of hybrid schemes. However,
recall that all these quantum repeater proposals
do rely upon local quantum memories and hence
local light-matter interfaces, rendering them
``hybrid" according to the supposed standard definition
of combined photon-atom systems.}
Typically, in these entanglement distribution protocols,
some form of local light-matter interaction is needed
in order to transfer the quantum correlations
encoded into the flying (optical) qubit/qumode systems
that travel along the channel onto the local static
qubit/qumode systems (which are typically either single electronic
spin qubits or collective atomic spin qumodes placed
in a cavity or just in free space, as discussed in Sec.~2.4).

A hybrid approach to entanglement distribution
may work as illustrated in Fig.~3.2.2 or, alternatively,
like the qubus scheme of Fig.~3.3.2. In either case,
the qumode(s) should propagate through the lossy communication
channel only once and any additional communication
will be classical, for instance, for confirming
a successful entanglement generation attempt among the
neighboring parties. Focussing on a qubus-based
repeater protocol like in Fig.~3.3.2, the measurement-free,
entangling gate sequen\-ces of Sec.~3.3.2 and Sec.~3.3.5
are better suited
for the local repeater operations, while a {\it measurement-based
approach} with only one interaction of the qumode ancilla
with each local qubit is optimal for the nonlocal
entangled-state preparation. In this case, the
mesaurement is needed in order to disentangle
the qumode from the tripartite entangled qubit-qubit-qumode
system and to project the two qubits onto a near-maximally
entangled state. When photon losses in the channel
are taken into account, the qumo\-de detection scheme
will typically include postselection of ``good"
measurement results.

Consider the qubus scheme of Fig.~3.3.2.
After a first dispersive interaction, i.e., a controlled
phase rotation of the qumode ancilla starting in a
coherent state $|\alpha\rangle$ ($\alpha$ real)
depending on the state of the first qubit [with initial state
$(|0\rangle + |1\rangle)/\sqrt{2}$],
the resulting hybrid qubit-qumode
state has the form of Eq.~(\ref{generalhybridentstate2})
with, for example, $|\psi_0\rangle\equiv|\alpha\rangle$,
$|\psi_1\rangle\equiv|\alpha e^{i\theta}\rangle$, and $\phi\equiv \alpha^2\sin\theta$
in the orthogonal qumode basis of Eq.~(\ref{qumodebasis2}) \cite{PvL06b}.
We may represent this hybrid state by $|\Phi^+(\mu)\rangle\equiv
\mu|u\rangle|0\rangle+
\sqrt{1-\mu^2}|v\rangle|1\rangle$.
Sending the qu\-mode through a lossy fiber channel with
amplitude transmission $\sqrt{\eta}$ then leads to a mixture
of two different hybrid entangled qubit-qumode states
\cite{pvlhybridrepeater08},
\begin{eqnarray}\label{qubitqubus}
F|\Phi^+(\mu)\rangle\langle\Phi^+(\mu)|+
(1-F)|\Psi^+(\mu)\rangle\langle\Psi^+(\mu)|\,,
\end{eqnarray}
with $|\Psi^+(\mu)\rangle\equiv
\mu|u\rangle|1\rangle+
\sqrt{1-\mu^2}|v\rangle|0\rangle$ and an attenuated amplitude
$\alpha\to\sqrt{\eta}\alpha$ throughout. Here the
``fidelity" is
$F\equiv [1+e^{-(1-\eta)\alpha^2(1-\cos\theta)}]/2$
and the Schmidt coefficient is
$\mu = \sqrt{1+e^{-\eta\alpha^2(1-\cos\theta)}}/\sqrt{2}$.
As this mixed entangled state is effectively written
in a two-qubit basis, its entanglement can be quantified.
Figure~3.4.1 shows the entanglement of formation \cite{Wootters98}
of the hybrid qubit-qumode state of Eq.~(\ref{qubitqubus})
as a function of the initial qumode photon number $\alpha^2$
for a 10 km fiber transmission. The trade-off between
good initial entanglement and loss-induced decoherence
for $\alpha$ too large ($\mu^2,F\to 1/2$)
and, on the other hand, vanishing initial entanglement and only little
decoherence for $\alpha$ too small ($\mu^2,F\to 1$)
is clearly reflected in the entanglement.

\begin{center}
  \begin{columnfigure}\label{figHQRenttransm}
    \centering%
    \includegraphics[width=220pt]{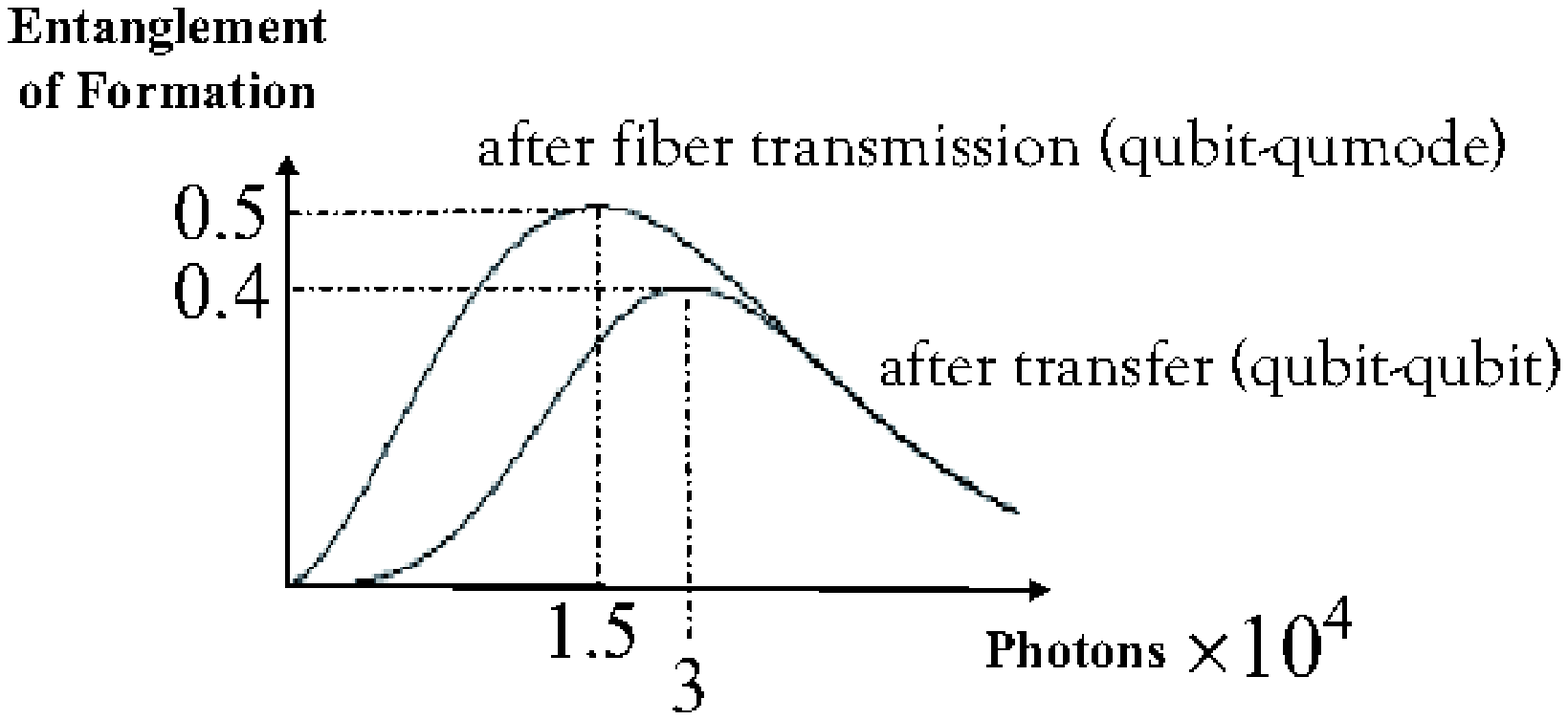}
  \figcaption{{\bf 3.4.1} Entanglement of formation for the hybrid qubit-qumode states
  as a function of the mean photon number in the initial qumode coherent state
  after optical-fiber qumode transmission. Similarly, the entanglement of formation
  for the qubit-qubit states after the entanglement transfer
  through homodyne detection. Losses are 1.8 dB corresponding to roughly 10 km
  fiber transmission. Entanglement transfer postselection efficiency is 36 \%
  and the local dispersive light-matter, qumode-qubit
  interactions are assumed to create phase shifts
  of the order of $10^{-2}$.}
  \end{columnfigure}
\end{center}

The essence of the qubus scheme in Fig.~3.3.2
is that the noisy qubit-qumode entanglement after the imperfect channel transmission
of the qumode is finally transferred onto the second qubit at the receiving
station. This can be accomplished through a second dispersive interaction
followed by a measurement on the qumode. Before the measurement, a
second controlled qumode rotation then
gives a tripartite entangled state of the two qubits and the qumode
\cite{pvlhybridrepeater08},
\begin{eqnarray}\label{qubitqubusqubit}
F|\Phi^+\rangle\langle\Phi^+|+
(1-F)|\Phi^-\rangle\langle\Phi^-|\,,
\end{eqnarray}
with the pure tripartite states $|\Phi^\pm\rangle$ equal to
%the desired and (for not too large losses) dominating state vector
\begin{eqnarray}\label{qubitqubusqubit2}
%|\Phi^\pm\rangle&=&
\frac{|\sqrt{\eta}\alpha\rangle
|\phi^\pm\rangle}{\sqrt{2}}
\pm\frac{e^{-i\phi}|\sqrt{\eta}\alpha e^{i
\theta}\rangle |10\rangle}{2}
%\nonumber\\
%&&\quad\quad\quad\quad\quad\quad\quad\quad\,\,
+\frac{e^{i\phi}
|\sqrt{\eta}\alpha e^{-i
\theta}\rangle |01\rangle}{2},\nonumber\\
\end{eqnarray}
and the maximally entangled Bell states $|\phi^\pm\rangle=
(|00\rangle \pm |11\rangle)/\sqrt{2}$.

For the final step of disentangling the qumode from the two
qubits through measurement and projecting the two qubits
onto an approximate version of $|\phi^+\rangle$ or
also $(|10\rangle + |01\rangle)/\sqrt{2}$, there are various choices
of which a homodyne detection is the experimentally
most efficient option \cite{PvL06b} (see Fig.~3.4.2).
In this case, a discrimination of the states $\{|\sqrt{\eta}\alpha\rangle,
|\sqrt{\eta}\alpha e^{\pm i \theta}\rangle\}$ in
Eq.~(\ref{qubitqubusqubit2}) along the $x$-axis will result in
overlap errors depending on the effective distance
of the Gaussian peaks that scales as $\sim\alpha\theta^2$ for small $\theta$
(see Fig.~3.1.3); hence $\alpha$ must be very large to compensate small
$\theta$ and suppress the overlap errors (e.g. photon numbers of
$\sim 10^{8}$ for $\theta\sim  10^{-2}$). As a result, the loss-induced
decoherence becomes too large.

A better option is homodyne detection along $p$
with peak distances $\sim\alpha\theta$. This scaling is the same as for the
measurement-free schemes and still feasible with weak interactions $\theta$.
However, in this case, only those outcomes consistent with
$|\sqrt{\eta}\alpha\rangle$ lead to an entangled two-qubit state
and the conditional states corresponding to
$\{|\sqrt{\eta}\alpha e^{\pm i \theta}\rangle\}$ must be discarded through
postselection. There will then be a trade-off between pair creation
efficiencies and fidelities, similar to the trade-off for the
hybrid entanglement after transmission; and for the optimal transmission,
the homodyne-based scheme will also not allow for a complete
entanglement transfer from the hybrid to the two-qubit states
(Fig.~3.4.1).

\begin{center}
  \begin{columnfigure}\label{fig20}
    \centering%
    \includegraphics[width=220pt]{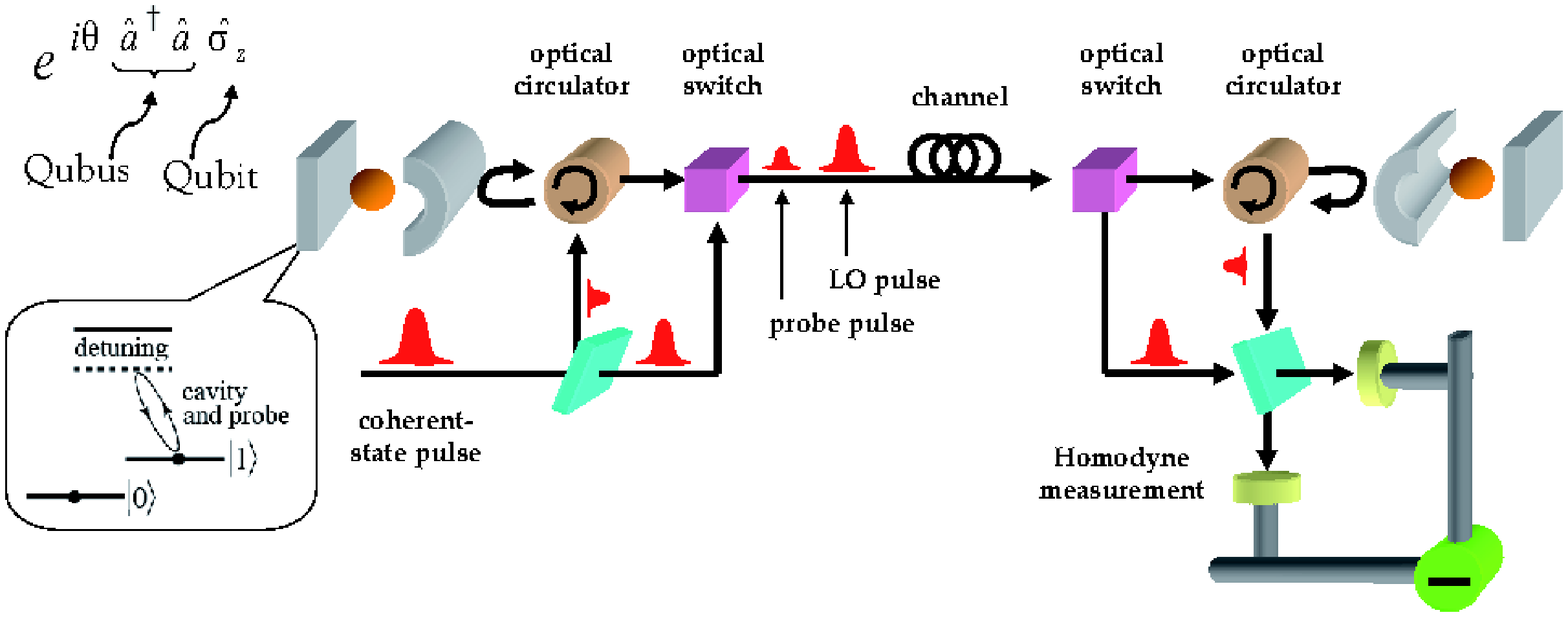}
  \figcaption{{\bf 3.4.2} Entanglement distribution in a hybrid quantum repeater using dispersive
  light-matter interactions and homodyne detection \cite{PvL06b}. ``LO"
  stands for ``local oscillator".}
  \end{columnfigure}
\end{center}

An interesting alternative is to replace the homodyne detection
by a non-Gaussian POVM for USD of coherent states (see Fig.~2.2.3).
In this case, an error-free identification of $|\sqrt{\eta}\alpha\rangle$
in Eq.~(\ref{qubitqubusqubit2}) is possible, completely eliminating
bit-flip errors \cite{pvlhybridrepeater08}. The remaining
phase-flip errors caused by photon losses may be minimized
and thus fidelities for given success probabilities maximized
through optimal USD
\cite{pvlhybridrepeater08}, attainable, for instance, by means of
photon number resolving detectors and using
two probe pulses \cite{Koashi09}.
Other non-Gaussian POVMs such as CSS state projections
can be considered \cite{MunroVanMeter08} as well as homodyne detections
on squeezed-state instead of coherent-state qumode ancillae \cite{Praxmeyer09}.
All these variations have in common that they achieve tunability
of the fidelity against the success probability including near-unit
fidelities at reasonable probabilities of success, a feature that is
very useful for a full quantum repeater architecture.

\begin{center}
  \begin{columnfigure}\label{fig9}
    \centering%
    \includegraphics[width=225pt]{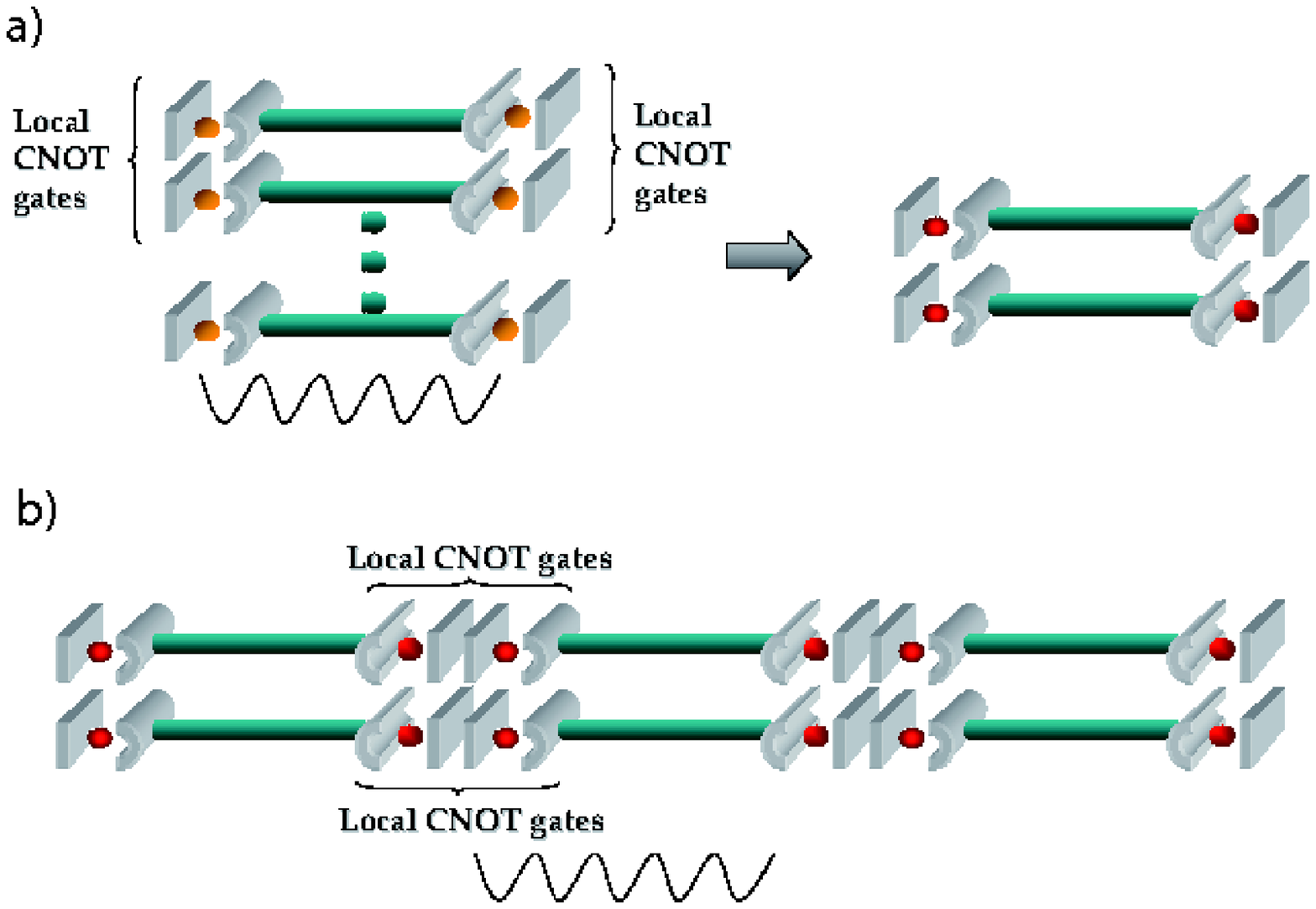}
  \figcaption{{\bf 3.4.3} Ingredients of a hybrid quantum repeater.
  (a) Distillation of better entangled states
   from a supply of noisy entangled states using local entangling gates
   such as $C_Z$ or $CNOT$, qubit measurements,
   and classical communication. (b) Connecting the distilled entangled states through
   entanglement swapping using local CNOT gates, qubit measurements, and classical communication.}
  \end{columnfigure}
\end{center}

The full hybrid quantum repeater is shown in Fig.~3.4.3.
The local entangling gates for entanglement distillation and swapping
could also be performed using dispersive light-matter interactions.
The local qumode ancillae in this case can be prepared independent
of the channel distances. Therefore arbitrarily intense qubus amplitudes
are possible, provided other imperfections such as local dissipations
and coupling inefficiencies in the CQED part of the repeater nodes
can be suppressed \cite{Ladd2006} (see Fig.~3.4.4). In this case,
CV homodyne measurements can result in sufficiently good
state discrimination for the local parity gates (in a measurement-based
implementation for the local gates), while the nonlocal qumode ancillae
that interact with two spatially separated qubits and hence must
be sufficiently weak could still be
detected through non-Gaussian USD by means of DV photon number mesaurements.

The distances between repeater stations are 10-30 km, almost directly compatible
with classical optical networks. A naturally given advantage
of the hybrid, photonic flying-qumode-based quantum repeater over the
photonic flying-qubit-based approaches is that it does not
require long-distance interferometry. Nonetheless, both on the experimental
and on the theoretical side there are still open issues
such as efficient local memory transfers (e.g. from electronic to nuclear spins,
recall the discussion in Sec.~2.4), efficient and error-resistant entanglement
distillations and connections, and complete optimizations
of hybrid repea\-ters in comparison to their non-hybrid counterparts.
Imperfections in the local repeater nodes as,
for instance, in a CQED-based scheme
might be circumvented through efficient
free-space qubit-qumode interactions \cite{4PIPAC}. Similarly,
the local CQED parts may also be highly improved through the use
of microtoroidal cavities \cite{KimbleAoki06}.

\begin{center}
  \begin{columnfigure}\label{figHQRlocalQubus}
    \centering%
    \includegraphics[width=220pt]{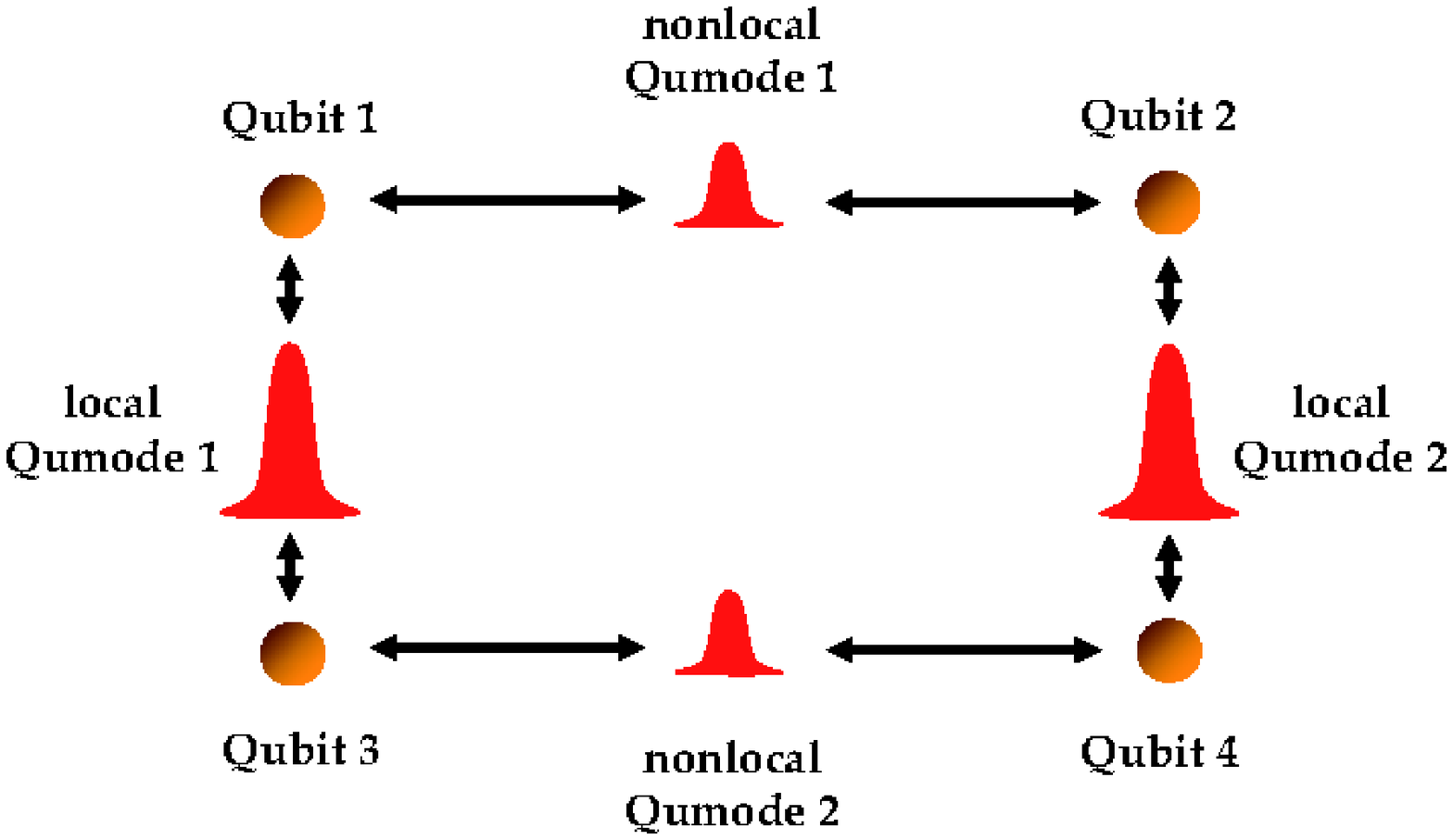}
  \figcaption{{\bf 3.4.4} A hybrid quantum repeater segment.
  The nonlocal qumodes travel through optical fibers and are subject
  to photon losses depending on the channel distance. Thus, error-free,
  non-Gaussian photon measurements may be used to eliminate
  measurement-based errors and maximize fidelities for sufficiently small,
  loss-resistant qumode amplitudes.
  The local qumode states, however, may have large
  photon numbers, as their photon losses are independent of
  the communication distance (neglecting local coupling inefficiencies
  and dissipations during the light-matter interactions). Therefore
  the local qumode states can be well discriminated through homodyne
  detection, allowing for efficient implementations of local
  two-qubit operations necessary for entanglement distillation.
  Alternatively, the measurement-free sequences from Sec.~3.3.5
  could be used for the local entangling gates.}
  \end{columnfigure}
\end{center}

Let us finally mention that there are indeed proposals for entanglement
distillation that could be incorporated into a full hybrid quantum repeater.
Since the flying quantum systems sent along the communication channel
are photonic qumodes, such distillations may directly act upon these qumodes,
for instance, in an entanglement distributor (Fig.~3.2.2)
rather than a qubus-based scheme. Besides those proposals listed in Sec.~3.1,
there are other very recent ideas achieving nondeterministic, noiseless
amplification \cite{LundRalph08,Ferreyrol09} with the help of DV Fock-state ancillae
and phase concentration through amplification
of attenuated coherent-state qumodes
utilizing thermal noise addition and photon subtraction \cite{MarekFilip09}.

\section{Summary and outlook}

Quantum information protocols can be formulated in terms of either
discrete or continuous degrees of freedom. In optical implementations,
the feasibility and the efficiency of a quantum information protocol,
including its basic subroutines such as state preparation, manipulation,
and measurements, depend on the type of variables employed in the scheme.
In particular, the use of continuous quantum variables leads to very efficient
implementations. However, recent studies have revealed that approaches
solely based upon Gaussian continuous-variable (CV) states and Gaussian operations
are ultimately limited:
universal quantum computation cannot be achieved in the Gaussian regime alone;
also entanglement distillation, a fundamental subroutine in quantum communication,
as well as more general forms of quantum error correction are
impossible in the realm of Gaussian states and Gaussian operations.

A possible way to circumvent these restrictions while maintaining some
of the advantages of the CV approach is to utilize
both continuous and discrete variables at the same time. For instance,
entangled Gaussian cluster states, unconditionally producible from
squeezed light sources, can still be used for universal quantum computation,
provided a non-Gaussian measurement (e.g. through photon counting)
is added to a cluster computation protocol.
Similarly, a nonlinear interaction of cubic or higher order,
otherwise hard to obtain on the level of single quanta,
can be effectively enhanced when intense optical pulses
in Gaussian states are used as a quantum bus to mediate
qubit-qubit interactions. Such a scenario is
particularly well sui\-ted to quantum communication schemes
where discrete-variable (DV) quantum information is locally encoded
into atomic qubit memories, but transmitted through an optical-fiber
channel using a CV quantum bus.

We gave a (certainly incomplete) review over existing hybrid proposals
for quantum information processing, including brief discussions of their
underlying principles and the tools that they use.
These hybrid schemes, based upon, for instance, linear resources,
together with the exploitation of measurement-induced or weak nonlinearities,
represent a promising route to efficient optical quantum computation and
communication. They can be seen as part of the current effort
to combine stable atomic and scalable solid-state systems
for reliable storage and fast processing of quantum information with
photonic systems for communication into hybrid devices.

The most attractive feature of those scheme
in which the necessary nonlinear element for universal quantum
information processing is solely provided through the measurement
apparatus is that in this case the linear resource states such as
Gaussian CV cluster states can be built
in an efficient, unconditional fashion. Nonetheless, this unconditionalness
comes at a price: the finite squeezing of the qumodes within a cluster state
leads to inevitable errors in a cluster computation. To suppress these
errors, some form of efficient quantum error correction will be needed
for which again a nonlinear element is required.
The most promising approach will be most likely based upon a combination
of different encodings, resources, and measurement techniques,
as envisaged by the hybrid schemes discussed in this review:
photonic qubits universally process\-ed on their own utilizing
polarization dual-rail encoding and efficiently coupled
with other qubits by employing nonlinearly transformed
or measured Gaussian photonic qumode ancillae.

\begin{acknowledgement}
I would like to thank the DFG for
financial support through the Emmy Noether programme.
Further I acknowledge useful inputs and assistance from
Akira Furusawa, Norbert L\"utkenhaus, Damian Markham,
Nick Menicucci, Bill Munro, Kae Nemoto, and Tim Ralph.
\end{acknowledgement}

\end{document}